\documentclass[11pt,a4paper]{article}
\pdfoutput=1

\usepackage{jheppub}
\usepackage{amsfonts,amscd,amsfonts,epsf,color}
\usepackage{caption}
\usepackage{subcaption}
\usepackage{lscape}
\usepackage{makecell}

\allowdisplaybreaks[4]

\usepackage[normalem]{ulem}

\newcommand{\be}{\begin{equation}}
\newcommand{\ee}{\end{equation}}

\DeclareMathOperator{\Tr}{Tr}

\subheader{LLNL-JRNL-824792, RIKEN-iTHEMS-Report-21, UTHEP-759, DMUS-MP-21/13}

\title{Confinement/deconfinement transition in the D0-brane matrix model -- A signature of M-theory?}

\collaboration{Monte Carlo String/M-theory Collaboration (MCSMC)}
\author[m]{Georg Bergner,}
\author[a]{Norbert Bodendorfer,}
\author[t]{Masanori Hanada,}
\author[a]{Stratos Pateloudis,}
\author[r]{Enrico Rinaldi,}
\author[a]{Andreas Sch\"{a}fer,}
\author[i]{Pavlos Vranas,}
\author[x]{and Hiromasa Watanabe}

\affiliation[m]{
University of Jena, Institute for Theoretical Physics,
Max-Wien-Platz 1, D-07743 Jena, Germany}
\affiliation[a]{
University of Regensburg, Institute of Theoretical Physics,\\
Universit\"{a}tsstrasse 31, D-93053 Regensburg, Germany}
\affiliation[t]{
Department of Mathematics, University of Surrey, Guildford, Surrey, GU2 7XH, United Kingdom}
\affiliation[r]{
Physics  Department,  University  of  Michigan,  Ann  Arbor,  MI  48109, United States\\
Theoretical Quantum Physics Laboratory, Cluster of Pioneering Research, RIKEN, Wako, Saitama 351-0198, Japan\\
Interdisciplinary Theoretical \& Mathematical Science Program (iTHEMS), RIKEN, Wako, Saitama 351-0198, Japan\\
Center for Quantum Computing (RQC), RIKEN, Wako, Saitama 351-0198, Japan}
\affiliation[i]{
Nuclear and Chemical Sciences Division, Lawrence Livermore National Laboratory,\\
Livermore CA 94550, United States\\
Nuclear Science Division, Lawrence Berkeley National Laboratory,\\
Berkeley, CA 94720, United States}
\affiliation[x]{
Graduate School of Pure and Applied Sciences, University of Tsukuba,\\
Tsukuba, Ibaraki 305-8571, Japan}

\abstract{
We study the confinement/deconfinement transition in the D0-brane matrix model (often called the BFSS matrix model) and its one-parameter deformation (the BMN matrix model) numerically by lattice Monte Carlo simulations.
Our results confirm general expectations from the dual string/M-theory picture for strong coupling.
In particular, we observe the confined phase in the BFSS matrix model, which is a nontrivial consequence of the M-theory picture.
We suggest that these models provide us with an ideal framework to study the Schwarzschild black hole, M-theory, and furthermore, the parameter region of the phase transition between type IIA superstring theory and M-theory.
A detailed study of M-theory via lattice Monte Carlo simulations of the D0-brane matrix model might be doable with much smaller computational resources than previously expected.
}
\begin{document}
\maketitle

\section{Introduction}\label{sec:introduction}

The thermodynamic features of the D0-brane matrix model~\cite{Banks:1996vh,deWit:1988wri,Itzhaki:1998dd} (often called the Banks-Fischler-Shenker-Susskind (BFSS) matrix model) and its one-parameter deformation known as the Berenstein-Maldacena-Nastase (BMN) matrix model~\cite{Berenstein:2002jq} are of great theoretical interest and, therefore, have been the subject of many investigations. Here we extend the insights gained by numerical simulations of these theories and provide a different viewpoint on their properties. Our main motivation is to give evidence that M-theory~\cite{Witten:1995ex} and Schwarzschild black holes can be described by these matrix models.

M-theory plays an important role in the web of string dualities.
It is the eleven-dimensional theory which has membranes as fundamental degrees of freedom.
It appears as the strong-coupling limit of type IIA superstring theory, and its low-energy limit should be eleven-dimensional supergravity~\cite{Cremmer:1978km}.
There exist a few proposals for the nonperturbative formulation of M-theory based on the holographic principle.
Among them, we consider the one based on the matrix-model approach~\cite{deWit:1988wri,Banks:1996vh,Itzhaki:1998dd}.~\footnote{
Another promising direction, to which numerical method similar to the one used in this paper can be useful, is to study gauge theories dual to D2-branes or M2-branes. Maximally supersymmetric Yang-Mills theory in $2+1$ dimensions is dual to D2 or M2, depending on the parameter region~\cite{Itzhaki:1998dd}, and hence, physics similar to the one considered in this paper would be seen. See Ref.~\cite{Catterall:2020nmn} for a lattice simulation of this theory.
}
The history of this approach dates back to the 1980s when the M-theory proposal had not been made yet.
At that time, the quantization of a supermembrane in eleven-dimensional spacetime was discussed, and a matrix model, which is called the BFSS matrix model today, was introduced as a natural regularization of the supermembrane in the light-cone gauge~\cite{deWit:1988wri}.
Later, this model was re-discovered as a candidate for the nonperturbative regularization of M-theory~\cite{Banks:1996vh}.

Further analysis using gauge/gravity duality~\cite{Maldacena:1997re} suggested that the BFSS matrix model can describe M-theory, type IIA superstring theory, and the phase transition between them~\cite{Itzhaki:1998dd}, based on the calculations on the string/M-theory side of the gauge/gravity duality.
The BMN matrix model~\cite{Berenstein:2002jq} is a one-parameter deformation of the BFSS matrix model discussed in Refs.~\cite{deWit:1988wri,Banks:1996vh,Itzhaki:1998dd}.
It has a rich phase structure and is often easier to analyze both analytically and numerically.
Thermodynamic features of the BFSS matrix model have been studied intensively, starting from  Ref.~\cite{Anagnostopoulos:2007fw}.
Agreement with type IIA superstring has been obtained with good precision using lattice simulations; for example, see Ref.~\cite{Berkowitz:2016jlq} for numerical results at large-$N$ and in the continuum limit.
However, the parameter region in which the BFSS matrix model is expected to be dual to M-theory has not been studied in the past.
Doing so was believed to be a very hard task because the stringy correction to the effective string coupling constant becomes large at temperature $T\sim N^{-10/21}$, which is a parametrically low temperature in the large-$N$ limit.
In this paper, we propose a much easier way to identify the M-theory parameter region.

Black hole thermodynamics~\cite{Bekenstein:1973ur,Hawking:1974sw} led to various deep insights into quantum gravity and quantum field theory, and gauge/gravity duality provides us with an ideal setup for the study of the quantum aspects of black holes and emergent geometry in holographic duality.
It has been successful for describing `large' black holes which have positive specific heat.
Detailed nonperturbative and quantitative tests based on Monte Carlo simulations for the BFSS matrix model were performed in Ref.~\cite{Anagnostopoulos:2007fw} and following papers, and good agreement was also observed for D1-brane and D2-brane theories~\cite{Catterall:2010fx,Kadoh:2017mcj,Catterall:2017lub,Catterall:2020nmn}.
Here, however, we are especially interested in small black holes with negative specific heat.
Holographic duality should also be applicable to small black holes, including usual Schwarzschild black holes.
For example, 4d SYM on S$^3$ should describe the 10d Schwarzschild black hole~\cite{Aharony:1999ti}.
The BFSS matrix model at very low energy should describe an 11d Schwarzschild black hole in M-theory~\cite{Itzhaki:1998dd}.
One of the goals of this paper is to give evidence that the BFSS matrix model and BMN matrix model do describe small black holes.

In principle, the real-time dynamics of the matrix model can be efficiently simulated on quantum computers~\cite{Gharibyan:2020bab}.
Hence, in the future, it might even become possible to study the entire life of a small black hole --- from the formation via gravitational collapse to eventual evaporation --- in a controlled manner in a fully quantum setup, in the context of superstring/M-theory.

\begin{figure}[htbp]
\begin{center}
\rotatebox{0}{
\scalebox{0.4}{
\includegraphics{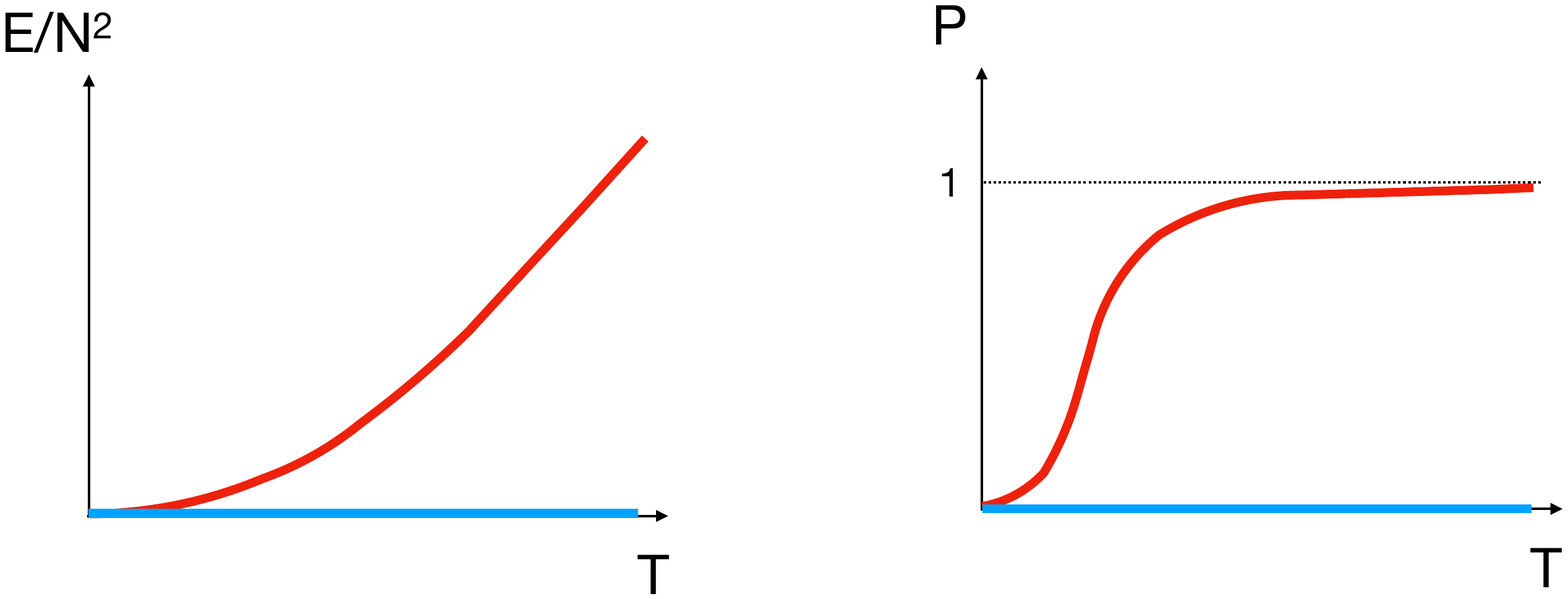}}}
\end{center}
\caption{
A cartoon picture of the (meta-)stable phases in the BFSS matrix model at finite temperature in the 't  Hooft large-$N$ limit ($\lambda=g^2N$ fixed).
Red and blue lines characterize the deconfined and confined phases, respectively. In the past, the existence of the confined phase was not pointed out.
Both are minima of the free energy, and the deconfined phase is the global minimum at any temperature.
The existence of the confined phase is a natural consequence of the dual M-theory description, hence, the numerical confirmation of the confined phase on the matrix model side would be an interesting clue towards demonstrating the validity of the M-theory description. 
}\label{fig:BFSS_minima}
\end{figure}

Another, but very closely related, motivation is that we want to understand the microscopic mechanism of the confinement/deconfinement transition.
Via gauge/gravity duality, deconfinement and the formation of a black hole are equivalent~\cite{Witten:1998zw}.
Strong-coupling results based on dual gravity analyses and weak-coupling results~\cite{Sundborg:1999ue,Aharony:2003sx} show striking similarity, and the latter show rather universal features regardless of the details of the theories~\cite{Aharony:2003sx}.
Such generic features can naturally be understood in the framework of partial deconfinement~\cite{Berenstein:2018lrm,Hanada:2018zxn,Hanada:2019czd,Hanada:2020uvt}:
Between the confined phase ($\sim$ thermal AdS) and the deconfined phase ($\sim$ large black hole) of SU($N$) gauge theory, there exists a partially-deconfined phase ($\sim$ small black hole) in which an SU($M$) subgroup ($0<M<N$) is deconfined.
For several theories, partial deconfinement has been demonstrated analytically at weak coupling~\cite{Hanada:2019czd, HanadaRobinsonPSB}, and numerically at strong coupling~\cite{Watanabe:2020ufk}.
In this paper, we provide evidence that the D0-brane matrix model provides us with an ideal setup to which a dual gravity analysis is tractable.
As a byproduct, we will see that the confined phase, which has not been considered before, should exist, and can be (meta-)stable up to rather high temperature (see Fig.~\ref{fig:BFSS_minima} for our conjecture).
Once knowing it to exist, it is not hard to find such a confined phase numerically.
Note that the existence of the confined phase is a consequence of the dual M-theory description and \textit{not} the type IIA string theory description.

This paper is organized as follows.
First, we give a short summary of our main claims in Sec.~\ref{sec:short_summary}.
Then we explain the details in the following sections.
In Sec.~\ref{sec:BMN_definition}, the BMN matrix model is defined.
The BFSS matrix model is obtained as a special case of the BMN matrix model, where the flux parameter $\mu$ is set to zero.
The BMN matrix model has many vacua: in Sec.~\ref{sec:backgrounds_definition} we specify the vacuum considered in this paper.
In Sec.~\ref{sec:conjecture_phase_diag}, we conjecture the phase diagrams of the BMN matrix model and BFSS matrix model, based on results from various numerical simulations (obtained in this paper, and also in previous studies) and dual gravity analyses.
In Sec.~\ref{numericalsection}, numerical results are presented and their consistency with the conjectures presented in Sec.~\ref{sec:conjecture_phase_diag} is discussed.
 
\section{Short summary of the main claim}\label{sec:short_summary}

In this section, we summarize the main claims of the paper.
The numerical evidence will be explained in later sections, together with various potential subtleties.

We study the thermodynamic features of the BFSS matrix model and BMN matrix model, whose definition is given in Sec.~\ref{sec:BMN_definition}.
These models can have various nontrivial backgrounds.
In this paper, we are interested in the `trivial' background.
We carefully study the properties of the configurations obtained in our simulations and extract the configurations corresponding to the trivial background.
The relevant parameters are the temperature $T$ and the flux parameter $\mu$.
The BFSS matrix model is obtained by setting $\mu=0$.

\begin{figure}[htbp]
\begin{center}
\rotatebox{0}{
\scalebox{0.2}{
\includegraphics{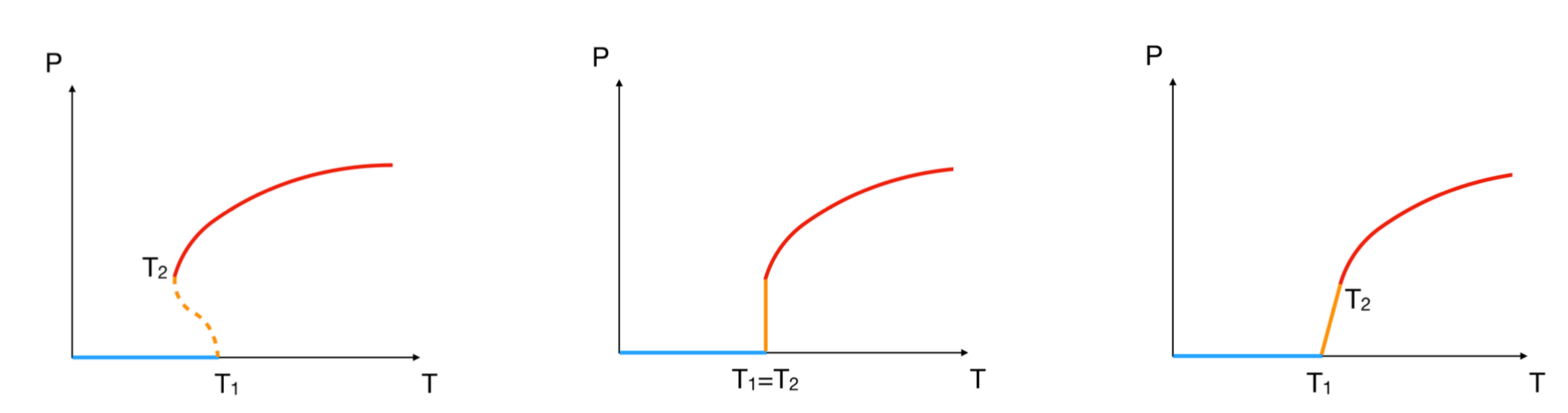}}}
\end{center}
\caption{
Typical shapes of confinement/deconfinement phase transition line in large-$N$ gauge theories.
The solid and dashed lines are minimum and maximum of the free energy at each fixed temperature.
Similar plots can be drawn by taking the vertical axis to be the energy $E$ instead of the Polyakov loop $P$.
(This figure appeared originally in Ref.~\cite{Hanada:2018zxn}.)
}\label{fig:P-vs-T-3-patterns}
\end{figure}

Based on lattice Monte Carlo simulations and dual gravity analyses, we propose the following phase structure in the $\mu$-$T$ plane.
If we vary $T$ at fixed $\mu$, there is a first-order phase transition with a strong hysteresis (see the left panel of Fig.~\ref{fig:P-vs-T-3-patterns}, and also Fig.~\ref{fig:Pol-1st-order-scenario}).
The high-temperature, high-energy phase is deconfined (the Polyakov loop $P$ is nonzero; see Appendix \ref{appendix:Polyakov} for the definition of the Polyakov loop), while the low-temperature, low-energy phase is confined (the Polyakov loop is zero).
These phases are the minima of free energy.
There is a local maximum of free energy separating these two phases, whose corresponding states are partially-deconfined.\footnote{The partially-deconfined phase is stable in the microcanonical ensemble.
}
There are three kinds of critical temperatures, $T_c$, $T_1$, and $T_2$ (see Fig.~\ref{fig:P-vs-T-3-patterns}, and also Fig.~\ref{fig:Pol-1st-order-scenario}).
$T_c$ is the temperature where the free energies of the confined and deconfined phases coincide.
At $T>T_c$ (resp., $T<T_c$), the deconfined phase (resp., confined phase) is the global minimum.
Temperature $T_1$ is the highest temperature for the confined phase, i.e, the confined phase is a local minimum of the free energy for $T_c < T <T_1$, and this minimum disappears for $T>T_1$.
Temperature $T_2$ is the lowest temperature of the deconfined phase which is a local minimum of the free energy at $T_2<T<T_c$.
As a function of $\mu$, $T_c$ is monotonically decreasing, as shown in Fig.~\ref{fig:tc_mu}.

Let us briefly summarize how this phase diagram is obtained and how it is related to M-theory.

\subsubsection*{BFSS matrix model ($\mu=0$)}
Let us start with the BFSS matrix model ($\mu=0$).
Ref.~\cite{Itzhaki:1998dd} studied the BFSS matrix model based on the dual gravity picture.
The authors found the phases corresponding to a type IIA black zero-brane and the M-theory black string or black hole.
It is natural to identify the type IIA and M-theory phases with the completely- and partially-deconfined phases, respectively.\footnote{
Here, we are assuming that the latter has negative specific heat and hence corresponds to the local maximum of free energy; see Sec.~\ref{sec:Schwarz11D} and Appendix~\ref{appendix:canonical-vs-microcanonical}.
On the large-energy, low-temperature side of the M-theory phase the dual picture is a black string,
while on the low-energy, high-temperature side it is a black hole.}$^,$\footnote{
From a different angle, Refs.~\cite{Banks:1997hz,Banks:1997tn,Banks:1997cm,Horowitz:1997fr} proposed that the BFSS matrix model describes an 11d black string or black hole boosted along the M-theory circle.}

\begin{figure}[htbp]
\begin{center}
\rotatebox{0}{
\scalebox{0.2}{
\includegraphics{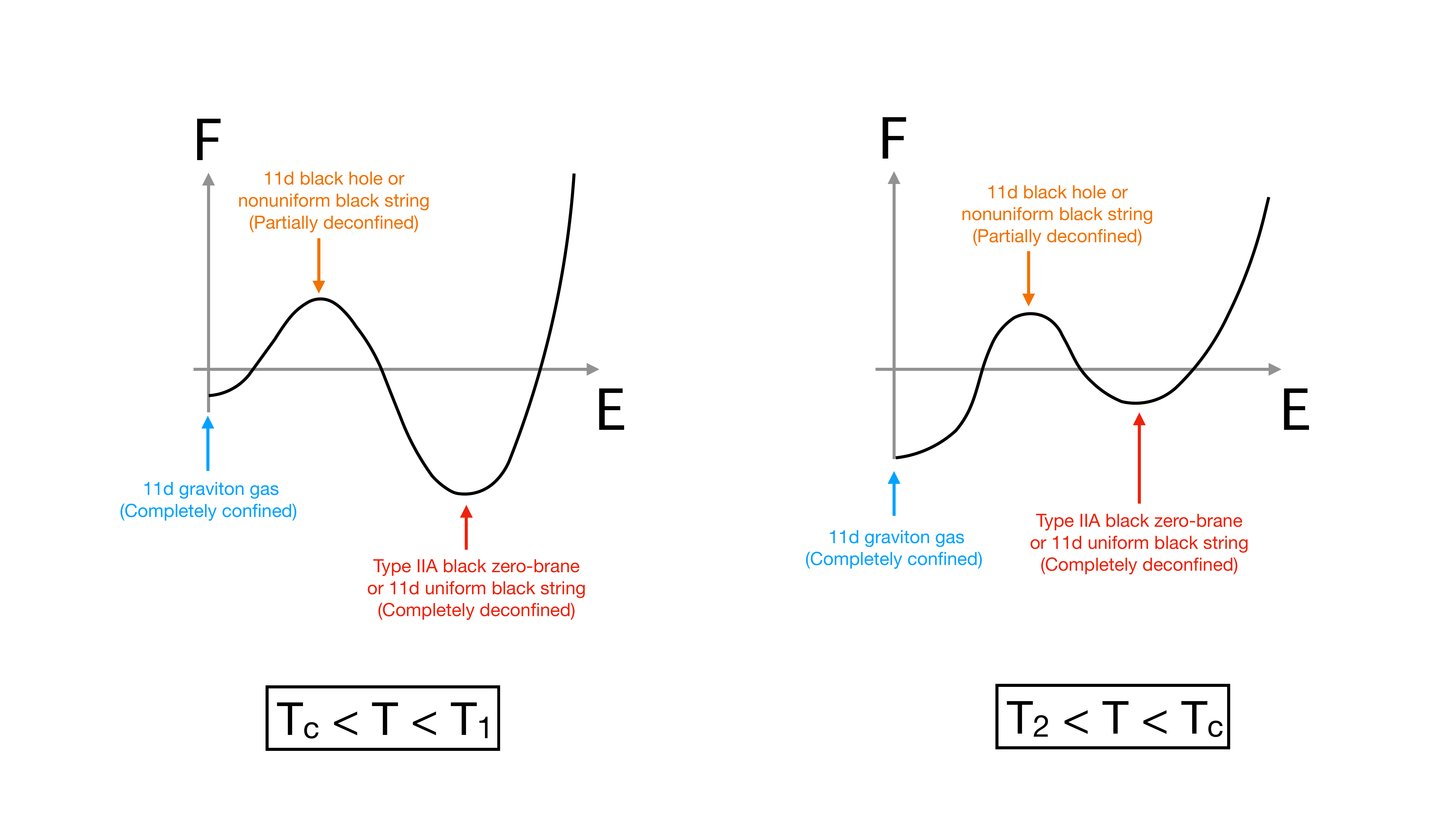}}}
\end{center}
\caption{
A conjecture for the relationship between free energy $F$ versus energy $E$ at fixed temperature $T$ in the BFSS matrix model.
An 11d Schwarzschild black hole gives the local maximum which separates the two minima.
(See Appendix~\ref{appendix:canonical-vs-microcanonical} for why negative specific heat implies a free energy maximum.)
}\label{fig:F-vs-E}
\end{figure}

Because of the existence of the partially-deconfined phase, we expect yet another, namely a completely-confined phase, which is likely to be dual to a graviton gas in eleven-dimensional spacetime; see Fig.~\ref{fig:F-vs-E}. (Such a phase was not discussed in Ref.~\cite{Itzhaki:1998dd}.)
On the gravity side, $T_1$ is the highest possible temperature of the eleven-dimensional black hole.
A very rough estimate for $T_1$ is $T_1 \gtrsim N^{2/9}$, as explained in Sec.~\ref{sec:gravity_BFSS}, and hence, the confined phase should survive up to a rather high temperature.
$T_2$ is interpreted as the temperature where the M-theory circle becomes large, and the transition from black string to black hole (Gregory-Laflamme transition~\cite{Gregory:1993vy}) takes place~\cite{Itzhaki:1998dd} and is estimated to be $T_2\sim N^{-5/9}$~\cite{Hyakutake:2015rqa}.
(See, e.g., Refs.~\cite{Harmark:2003eg,Kol:2004ww,Horowitz:2011cq,Kalisch:2018efd} for detailed calculations for Gregory-Laflamme transition, and Ref.~\cite{Hovdebo:2006jy} for those with a boost.)
Our numerical simulations confirm the existence of the confined phase.
Such a confined phase is tantalizing evidence for the M-theory phase in the matrix model.\footnote{
As far as we notice, the existence of the confined phase in the BFSS matrix model has not been appreciated in the past.
}

$T_c$ should be close to $T_2$, because otherwise the black zero-brane in type IIA supergravity (or boosted uniform black string in M-theory, which is essentially the same as a black zero-brane) is a good approximation, and we find sufficiently small free energy.
Specifically, both $T_2$ and $T_c$ are expected to be zero in the strict large-$N$ limit.
For $T_2<T<T_1$, we expect a two-state signal in the Monte Carlo simulations if $N$ is sufficiently large.
In the histograms of the energy or the Polyakov loop observables, there should be two peaks corresponding to two minima of free energy.
The local maximum of the free energy should be seen as a dip separating the two peaks.
However, in our simulation, we did not observe the two-state signal; the deconfined phase is too unstable for the small values of $N$ we could study with available computer resources ($N\le 16$).
As a consequence, we could not determine $T_2$ and $T_c$.

\subsubsection*{BMN matrix model}
\begin{figure}[htbp]
\begin{center}
\rotatebox{0}{
\scalebox{0.3}{
\includegraphics{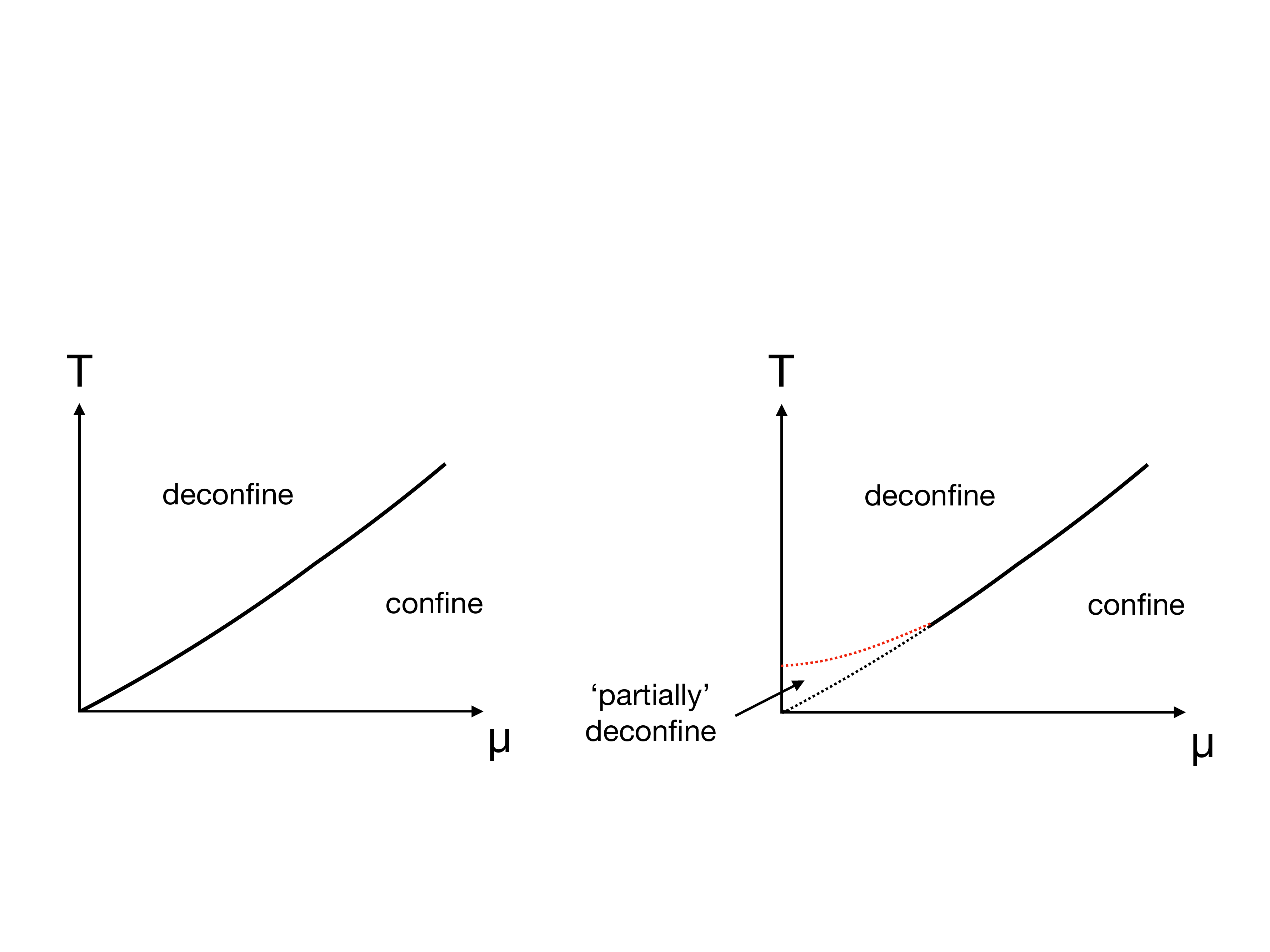}}}
\end{center}
\caption{Two kinds of conjectured phase diagrams of the BMN matrix model.
The large-$\mu$ region permits perturbative calculations~\cite{Furuuchi:2003sy,Spradlin:2004sx},
and the transition is found to be of first order.
The small-$\mu$ region has been studied by using the dual gravity description \cite{Costa:2014wya}
but the order of the transition has not been established.  We will argue that the left panel (first-order scenario) is likely to be true.
}\label{fig:conjectured-phase-diagram-BMN}
\end{figure}

We studied the BMN matrix model in order to confirm the two-state signal and give stronger support for the validity of the M-theory description in the BFSS limit.
The large-$\mu$ region is weakly-coupled, and hence perturbative methods are applicable.
The small-$\mu$ region (more precisely, small but order $N^0$) can be studied via dual type IIA supergravity.
By combining the large-$\mu$ and small-$\mu$ analyses, Ref.~\cite{Costa:2014wya} conjectured that the transition is of first order at any $\mu$, and the critical temperature $T_c$ decreases monotonically as a function of $\mu$ and becomes zero at $\mu=0$ (the left of Fig.~\ref{fig:conjectured-phase-diagram-BMN}).

We performed Monte Carlo simulations at various values of $\mu$ and observed quantitative agreement with the phase diagram conjectured in Ref.~\cite{Costa:2014wya}.
We observed clear two-state signals at $\mu\ge 0.5$, and even at $\mu=0.3$ our results support the existence of a first-order transition.
As shown in Fig.~\ref{fig:tc_mu}, the critical temperature $T_c$ agrees well with the conjecture in Ref.~\cite{Costa:2014wya} at $\mu\ge 0.8$.
Apparent deviation at $\mu<0.8$ can naturally be explained by taking into account finite-$N$ corrections on the gravity side.

\section{BFSS and BMN matrix models}\label{sec:BMN_definition}

The BFSS and BMN matrix models consist of nine $N\times N$ bosonic hermitian matrices $X_I$ ($I=1,2,\cdots,9$), sixteen
fermionic matrices $\psi_\alpha$ ($\alpha=1,2,\cdots,16$), and the gauge field $A_t$.
Both $X_I$ and $\psi_\alpha$ are in the adjoint representation of the $U(N)$ gauge group, and the covariant derivative $D_t$ acts on them as $D_tX_I = \partial_t X_I -i[A_t,X_I]$ and $D_t\psi_\alpha = \partial_t\psi_\alpha -i[A_t,\psi_\alpha]$.
The action is given by
\begin{equation}
S=S_b+S_f
\end{equation}
for BFSS and
\begin{equation}\label{eq:matrix-model-action-total}
S=S_b+S_f+\Delta S_b+\Delta S_f
\end{equation}
for BMN, where
\begin{equation}\label{eq:action-BFSS-bos}
S_b = \frac{N}{\lambda}\int_0^\beta dt\ \Tr\left\{
\frac{1}{2}\sum_{I=1}^9(D_t X_I)^2
-
\frac{1}{4}\sum_{I,J=1}^9[X_I,X_J]^2
\right\}
\ ,
\end{equation}
\begin{equation}\label{eq:action-BFSS-fer}
S_f = \frac{N}{\lambda}\int_0^\beta dt\ \Tr\left\{
i\bar{\psi}\gamma^{10}D_t\psi
-
\sum_{I=1}^9\bar{\psi}\gamma^I[X_I,\psi]
\right\}
\ ,
\end{equation}
\begin{equation}\label{eq:action-BMN-bos}
\Delta S_b =
\frac{N}{\lambda}\int_0^\beta dt\ \Tr\left\{
\frac{\mu^2}{2}\sum_{i=1}^3X_i^2
+
\frac{\mu^2}{8}\sum_{a=4}^9X_a^2
+
i\sum_{i,j,k=1}^3\mu\epsilon^{ijk}X_iX_jX_k
\right\}
\ ,
\end{equation}
and
\begin{equation}\label{eq:action-BMN-fer}
\Delta S_f =
\frac{3i\mu}{4\lambda}\cdot N\int_0^\beta dt\ \Tr\left(
\bar{\psi}\gamma^{123}\psi
\right)
\ .
\end{equation}
Here $\epsilon_{ijk}$ is the structure constant of $SU(2)$, and hence
$i\sum_{i,j,k=1}^3\mu\epsilon^{ijk}\Tr\left(X_iX_jX_k\right)=3i\Tr\left(X_1,[X_2,X_3]\right)$.
At $\mu=0$, the BMN matrix model reduces to the BFSS matrix model, which is obtained by dimensionally reducing 10d super Yang-Mills theory to 1 dimension.
The index $\alpha$ of the fermionic matrices $\psi_\alpha$ corresponds to the spinor index in ten dimension, and
$\psi_\alpha$ are Majorana-Weyl fermions in these ten dimensions.
$\gamma^I$ ($I=1,\cdots,10$) are $16\times 16$ sub-matrices of $32\times 32$ 10d Gamma matrices $\Gamma^{I}$.

In this paper, we will consider the thermodynamic features of the BFSS and BMN matrix models.
There are three parameters: the 't Hooft coupling $\lambda=g_{YM}^2N$, the temperature $T$ and the flux $\mu$.
The 't Hooft coupling has a dimension of $({\rm mass})^3$, and can be set to 1 by proper rescaling of time $t$ and matrices. In other words, all dimensionful quantities are considered in units of $\lambda$.
For example $T$, $\mu$, and the energy $E$ are actually $\lambda^{-1/3}T,\lambda^{-1/3}\mu$ and $\lambda^{-1/3}E$.
We mainly focus on the 't Hooft large-$N$ limit,\footnote{
	The only exception is when we discuss a relation to the M-theory region, which requires a different kind of large-$N$ limit.
}  in which $T$ and $\mu$ are fixed and the energy is proportional to $N^2$.

\subsection{Trivial vacuum in the BFSS matrix model}
In the BFSS matrix model, the potential term $-\frac{1}{4}[X_I,X_J]^2$ vanishes if the matrices commute with each other.
Therefore, at the classical level, there are flat directions.
Due to supersymmetry, a remnant of these flat directions survives at the quantum level.
Namely, the energy of almost-diagonal configurations $X_I\simeq{\rm diag}(x_I^1,x_I^2,\cdots,x_I^N)$ can be asymptotically small when the distance between diagonal entries $\sqrt{\sum_I|x_I^i-x_I^j|^2}$ is large. These diagonal entries are regarded as the locations of D0-branes.

Despite the existence of flat directions, D0-branes can form a bound state.\footnote{For the precise meaning of this bound state, see Ref.~\cite{Hanada:2021ipb}.}
In this paper, we will consider the `trivial vacuum', in which all D0-branes are bound together.
The trivial vacuum becomes more and more stable as $N$ becomes large.

\subsection{Trivial and fuzzy-sphere vacuum in the BMN matrix model}\label{sec:backgrounds_definition}
In the BMN matrix model, the flat direction is lifted due to the deformation term.
Instead, it has multiple supersymmetric vacua characterized by the fuzzy-sphere classical solution\footnote{
	Strictly speaking, the configuration \eqref{eq:fuzzy-sphere-configuration} should be interpreted as the center of a wave packet.
	See~\cite{Hanada:2021ipb} for details.
}
\begin{align}\label{eq:fuzzy-sphere-configuration}
X_i=\mu J_i\ (i=1,2,3)\ , \qquad X_a=0\ (a=4,\cdots,9)\ , \qquad \psi=0
\ ,
\end{align}
where $J_i$ are the generators of an SU(2) algebra.
As long as they satisfy the standard commutation relation $[J_i,J_j]=\sqrt{-1}\epsilon_{ijk}J_k$, any representation is allowed.
A generic representation is characterized by the spin $s=0,\frac{1}{2},1,\frac{3}{2},\cdots$
and the number of spin-$s$ representations $n_s$.
The matrix size $N$ is $N=(2s+1)n_s$ and the classification of vacua is characterized by partitions of $N$ ~\cite{Berenstein:2002jq}. One more interesting point of view is that the classification of these vacua on the gravity side can be interpreted as an electrostatic problem \cite{Lin:2005nh,Lozano2017}.

The mathematical reason why these vacua are interpreted as fuzzy-sphere vacua is that the matrices $X_i$ can be interpreted as the embeddings $X_i:\mathbf{S}^2\hookrightarrow \mathbb{R}^3$.
Being assigned to generators of $SU(2)$ they generate non-commutativity on the surface of $\mathbf{S}^2$ justifying the fuzzy-sphere interpretation~\cite{Madore1992}.
On equal footing is the interpretation from string theory, where the diagonal elements of the matrices $X_{i,a}$ are interpreted as positions of D0-branes and non-diagonal elements as open strings between them.
In the case of fuzzy sphere configurations, D0-branes are in a non-commutative sense localized on the surface of $\mathbf{S}^2$ which furthermore lives in $\mathbb{R}^3\subset \mathbb{R}^9$.
The interpretation of D2-branes as consisting of D0-branes is well known in string theory where under the presence of fluxes ($\mu$ in our case) D0-branes polarize and form non-commutative structures~\cite{Myers:1999ps}.

Hence, fuzzy-sphere vacua can be interpreted from the gravity perspective as concentric fuzzy spheres consisting of D0-branes, placed at the center of $\mathbb{R}^6$ and living only in the three dimensional subpart of spacetime whose radii scale as
\begin{align}
r_s\simeq\mu s
\ .
\end{align}

\noindent
When $n_0=N$ and $n_{s\ge 1}=0$, the ground state is given by
\begin{align}
X_1=X_2=\cdots=X_9=0\ , \qquad \psi=0
\ .
\end{align}
This state is called the ``trivial" vacuum.
Despite its simple appearance, the trivial vacuum has rich, nontrivial features.
In this paper, we focus on this vacuum.\footnote{
Unlike us, Ref.~\cite{Asano:2018nol} focused on the transition between various vacua.
}

At the classical level, the radius of the spin-$s$ fuzzy sphere is $\mu\sqrt{s(s+1)}$.
As $\mu$ becomes smaller, the radii of the fuzzy spheres approach zero, and hence, tunneling from the trivial vacuum to nontrivial fuzzy-sphere vacua can happen more easily.
Hence, it is crucial to ensure that the combination $\mu N$ is large enough to avoid ending up in an undesirable vacuum.
At zero temperature, the quantum fluctuation of each matrix entry is roughly $0.6N^{-1/2}$~\cite{Hanada:2021ipb}.\footnote{
	The value $0.6N^{-1/2}$ is the standard deviation.
	The order-one factor can be determined numerically, from the expectation value of $\sum_{I=1}^9{\rm Tr}X_I^2$, which is approximately $3.5N$~\cite{Berkowitz:2016jlq}.
	There are $9N^2$ matrix entries and  hence $\sqrt{\frac{3.5N}{9N^2}}\approx 0.6 N^{-\frac{1}{2}}$.
}
The radius of the smallest nontrivial fuzzy sphere ($s=\frac{1}{2}$) is $\frac{\mu}{2}\sqrt{3}$.
The radius of the largest fuzzy sphere ($s=\frac{N-1}{2}$) is $\frac{\mu}{2}\sqrt{N^2-1}$.
Therefore, when $N$ is not so large, say $N=16$, all possible fuzzy spheres are buried in quantum fluctuations if $\mu\lesssim 0.018$.
On the other hand, if $\mu$ is fixed and $N$ is sent to infinity, all fuzzy-sphere vacua are distinguishable~\cite{Hanada:2021ipb}.

\begin{figure}[htbp]
	\begin{center}
		\rotatebox{0}{
			\scalebox{0.35}{
				\includegraphics{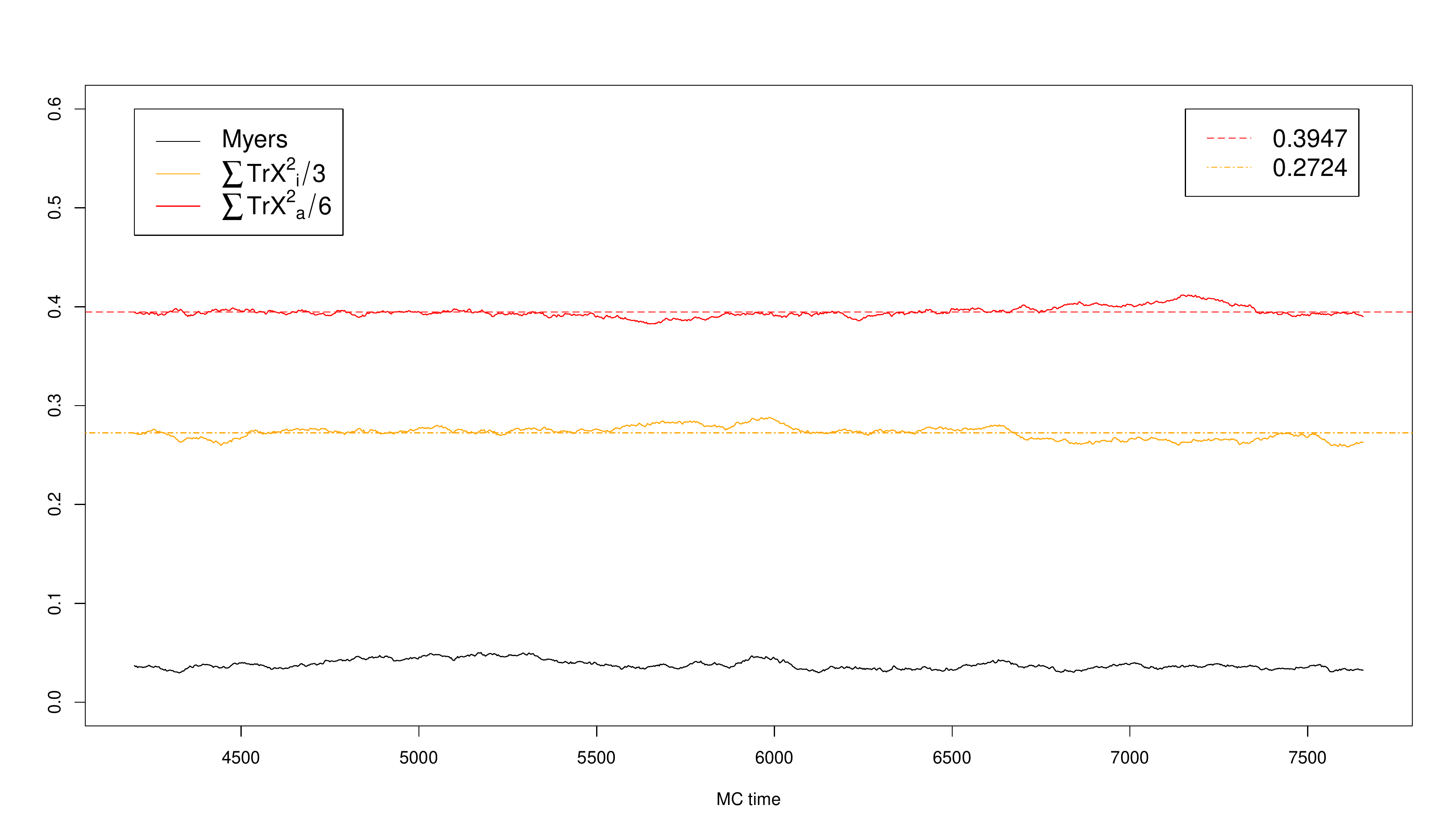}}}
	\end{center}
	\caption{
		Fluctuations of the first three matrices $\Tr{X_i^2/3}$ and  $\Tr{X_a^2/6}$ grouped together show an obvious breaking of the symmetries. The parameters shown are $N=16, S=24, T=0.23$ and $\mu=0.6$. We see that the Myers term is very small, indicating not a formation of a big fuzzy sphere. On the other hand, a priori there could be many small fuzzy spheres and in such a scenario the Myers term cannot distinguish these configurations.
	}\label{fig:fluctuations}
\end{figure}

The Myers term is a convenient order parameter for the formation of fuzzy spheres.
It is defined by
\begin{align}\label{eq:myers-term}
M=\frac{i}{3 N \beta}\int_{0}^{\beta} dt \sum_{i,j,k=1}^3\epsilon_{ijk}\Tr X^i X^j X^k
\ .
\end{align}
We monitor it during the simulations.
In addition, the different $\Tr (X^i)^2$ for each $i$ are measured to identify nontrivial vacua.
The classical value of $M$ is $\frac{\mu^3}{3N}\sum_s n_s s(s+1)(2s+1)$, which is 0 for the trivial vacuum, $\frac{\mu^3(N^2-1)}{12}$ for the largest fuzzy sphere (single fuzzy sphere of $s=\frac{N-1}{2}$),
and $\frac{\mu^3}{4}$ for $\frac{N}{2}$ fuzzy spheres of $s=\frac{1}{2}$.
We can detect a formation of large fuzzy sphere or many small fuzzy spheres, unless $\mu$ is too small, by monitoring $M$ or $R^2$ defined by
\begin{align}
R^2
=
\frac{1}{N\beta}\int_{0}^{\beta} dt \sum_{I=1}^9{\rm Tr}X_I^2.
\end{align}
However, it is difficult to detect the formation of a small number of small fuzzy spheres by monitoring $M$.
Still, when such ambiguity arises, it is unlikely that the properties of those vacua such as the deconfinement temperature are significantly different. Therefore, we neglect this subtlety in the present paper. To summarize all the above, we show the normalized $R^2_i$ and $R^2_a$ as well as the Myers term \eqref{eq:myers-term} in Fig.~\ref{fig:fluctuations}.

\section{Conjectured phase structure at finite temperature}\label{sec:conjecture_phase_diag}

In this section we present our conjecture regarding the phase diagram of the BMN and BFSS models. Numerical evidence will be shown in Sec.~\ref{numericalsection}.

First, let us discuss some generic features of the confinement/deconfinement transition at large $N$.

\begin{figure}[htbp]
\begin{center}
\rotatebox{0}{
\scalebox{0.2}{
\includegraphics{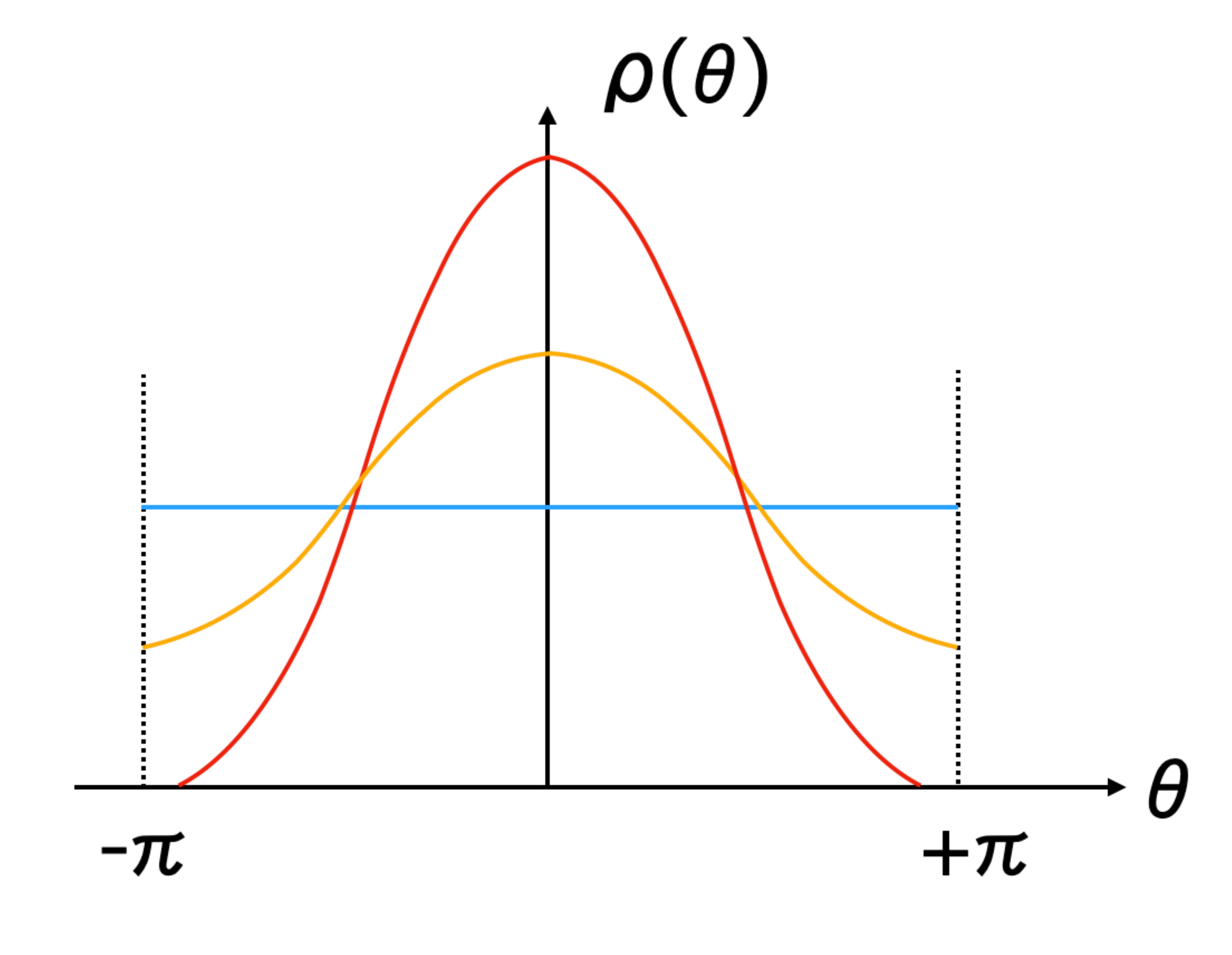}}}
\end{center}
\caption{
The uniform (blue), non-uniform and non-gapped (orange) and non-uniform and gapped (red) phases.
They can be understood as the confined, partially-deconfined and completely-deconfined phases, respectively.
}\label{fig:Pol-phase-cartoon}
\end{figure}

The distribution of the phases of the Polyakov line, which we denote as $\theta_1,\theta_2,\cdots,\theta_N$, provides us with a convenient way to understand the confinement/deconfinement transition.
In the large-$N$ limit, we can use a continuous function $\rho(\theta)$ defined at $-\pi\le\theta<\pi$ and
normalized as $\int_{-\pi}^{\pi}d\theta\rho(\theta)=1$ to describe the distribution of $\theta$.
Roughly speaking, there are three kinds of phases visualized in Fig.~\ref{fig:Pol-phase-cartoon}~\cite{Sundborg:1999ue,Aharony:2003sx}:\footnote{
Note that we are considering the thermal transition here. When the different boundary conditions such as the periodic boundary condition are used, other types such as the multi-peak distribution can appear as well.
Recently, an analogue of deconfinement in the index of 4d ${\cal N}=4$ SYM was found~\cite{Choi:2018vbz}, and multi-peak distributions were reported in, e.g., Ref.~\cite{ArabiArdehali:2019orz}.
Sometimes, such phases are called `partially-deconfined phase' as well.
Note also that, at strong coupling, the phase structure may be richer, and we may need to use other order parameters as well; in the dual gravity description, various black hole solutions can exist.
} the uniform (blue), non-uniform and non-gapped (orange) and non-uniform and gapped (red) phases.
They can be understood as the confined, partially-deconfined (equivalently, partially-confined) and completely-deconfined phases, respectively~\cite{Hanada:2018zxn,Hanada:2019czd,Hanada:2020uvt}.

For historical reasons, let us call
the transition between the uniform phase and the non-uniform, non-gapped phase
(resp., the non-uniform, non-gapped phase and the non-uniform, gapped phase) the Hagedorn-like transition (resp., the Gross-Witten-Wadia (GWW) transition).\footnote{
When the free-string picture is available, the transition between the uniform phase and the non-uniform, non-gapped phase is
actually the formation of the Hagedorn string~\cite{Hagedorn:1965st}, as long as the 't Hooft coupling is not too large.
Such an interpretation may not be valid when the gravity dual is M-theory.
The word ``GWW transition" is used to mean the formation of the gap in the eigenvalue distribution. In the weakly-coupled theory studied in Refs.~\cite{Sundborg:1999ue,Aharony:2003sx}, this transition is of third order, resembling the original Gross-Witten-Wadia model~\cite{Gross:1980he,Wadia:2012fr}.
}
Let $T_1$ (resp., $T_2$) be the transition temperature for the Hagedorn-like transition and the GWW transition, respectively.
In Fig.~\ref{fig:P-vs-T-3-patterns}, a sketch of the temperature dependence of the Polyakov loop $P$ is shown. From left to right, the panels correspond to $T_1>T_2$, $T_1=T_2$, and $T_1<T_2$ respectively.
The case with $T_1=T_2$ can be found, e.g., for the weak-coupling limit of 4d Yang-Mills theory on S$^3$. In this case, at the critical temperature $T=T_1=T_2$, states with different values of $P$ have the same free energy, and hence, all of them are equally important in the canonical ensemble. Such degeneracy in the free energy comes from the cancellation of the entropy factor and the Boltzmann weight, which is the same as the mechanism of the Hagedorn growth in string theory~\cite{Hagedorn:1965st}.
This gives a nice intuitive connection between deconfinement and the formation of a black hole~\cite{Sundborg:1999ue,Aharony:2003sx}.
When the interaction is introduced, depending on the details of the theory, the vertical orange line can be tilted toward the left ($T_1>T_2$) or right ($T_1<T_2$).
In the former case, the non-uniform, non-gapped phase has negative specific heat (i.e. $\frac{dE}{dT}<0$; note that the energy $E$ increases with $P$), and in the canonical ensemble, this is a maximum, rather than a minimum, of the free energy at each fixed $T$. Such phase resembles the so-called small black hole~\cite{Aharony:2003sx}.
Partial deconfinement gives a natural way to understand the negative specific heat~\cite{Hanada:2016pwv,Hanada:2018zxn}.
Note that, in the canonical ensemble, a first-order transition analogous to the Hawking-Page transition~\cite{Hawking:1982dh,Witten:1998zw} with hysteresis is expected when $T_1>T_2$.

\subsection{BFSS matrix model ($\mu=0$)}\label{sec:gravity_BFSS}
In this subsection, we argue that the BFSS matrix model should have a confined phase corresponding to a metastable vacuum.
The dual M-theory description plays a crucial role, and hence,
the confirmation of confinement is tantalizing evidence for the dual M-theory description.
\subsubsection{'t Hooft large-$N$ limit via dual IIA superstring theory}
Let us consider the 't Hooft limit of the BMN matrix model at $\mu=0$, corresponding to the BFSS model. The deconfined phase is dual to a black zero-brane in type IIA string theory \cite{Itzhaki:1998dd}. According to this, the dual gravity analysis predicts that the energy at sufficiently low temperature is $E \simeq 7.41N^2\lambda^{-3/5}T^{14/5}$ up to stringy corrections. No phase transition is expected; at any nonzero temperature, the energy is of order $N^2$, and hence, the system is in the deconfined phase.
This conjecture has been tested via Monte Carlo simulation by several groups, e.g., Refs.~\cite{Anagnostopoulos:2007fw,Catterall:2008yz,Hanada:2008gy,Catterall:2009xn,Hanada:2013rga,Kadoh:2015mka,Filev:2015hia,Berkowitz:2016jlq}.
In Ref.~\cite{Berkowitz:2016jlq}, the large-$N$, continuum limit has been taken in a wide temperature region $\lambda^{-1/3}T\ge 0.4$. Even at $\lambda^{-1/3}T=0.4$, the distribution of the Polyakov line phases $\rho(\theta)$ is clearly gapped, and the energy can be fit by the ansatz based on the weakly-coupled string theory.
As long as the dual type IIA description is valid, we expect only the deconfined phase.

As we will discuss below, a confined phase ($\frac{E}{N^2}\to 0$, $P\to 0$ as $N\to\infty$ at fixed $T$ and fixed $\lambda$) can also exist.\footnote{
This is not the 't Hooft large-$N$ limit, and standard 't Hooft counting does not hold.
} Except for the parametrically low temperature, the confined phase has larger free energy than the deconfined phase, and hence, it does not dominate the canonical ensemble.
Still, such a metastable phase can be identified in Monte Carlo simulations if we choose the initial configuration for the simulation carefully. We will argue that the existence of this confined phase constitutes strong evidence for M-theory on the gravity side.
\subsubsection{M-theory and confinement}\label{sec:Schwarz11D}
Let us continue the study of $\mu=0$, but depart from the 't Hooft limit
and go to parametrically small energy in the microcanonical ensemble.
As we will see, dual gravity picture suggests a first-order transition in the canonical ensemble.

Based on the gauge/gravity duality, it was conjectured~\cite{Itzhaki:1998dd} that there are two phases dual to the type IIA black zero-brane and M-theory black hole
(11d Schwarzschild black hole), based on the analysis of the gravity side.
Separately, Refs.~\cite{Banks:1997hz,Banks:1997tn,Horowitz:1997fr} studied the matrix-model description of the 11d Schwarzschild black hole
and discussed how thermodynamic relations might be explained.
More precisely~\cite{Itzhaki:1998dd,Horowitz:1997fr}:
\begin{itemize}
\item
At $E\gtrsim N^{2/3}\lambda^{1/3}$ and $T\gtrsim N^{-10/21}\lambda^{1/3}$, a type IIA black zero-brane is a good dual description.
The energy and entropy scales as $E\sim N^2\lambda^{-3/5}T^{14/5}$, $S\sim N^2\lambda^{-3/5}T^{9/5}$. The specific heat is positive.

\item
Below $E\sim N^{2/3}\lambda^{1/3}$, the M-theory circle is relevant in the thermodynamic analysis.
When the energy is not too small, the appropriate object in the dual picture becomes the uniform black string wrapped around and boosted along, the M-theory circle.
This object is a trivial M-theory uplift of black zero-brane, and hence thermodynamic properties are similar to those of a black zero-brane.

\item
When the energy is lowered further, the Gregory-Laflamme transition~\cite{Gregory:1993vy} to a localized boosted 11d black hole takes place at $S\sim N$ and $T\sim N^{-5/9}$~\cite{Hyakutake:2015rqa}.

\end{itemize}
Therefore it is natural to expect $T_2\sim\lambda^{1/3}N^{-5/9}$.
Note that this is outside of the 't Hooft scaling regime $T\sim \lambda^{1/3}N^0$ and we expect that $T_c$ is close to $T_2$.

The detailed properties of the M-theory black hole are not known. However, a few generic features can be inferred.
Firstly, we expect that the dual geometry is a black hole and graviton gas in a finite-size box, analogous to AdS times sphere.
In Refs.~\cite{Banks:1997hz,Banks:1997tn,Banks:1997cm}, it is assumed only a part of D0-branes (or more precisely, matrix degrees of frreedom) participate in the Shwarzschild black hole.
Other D0-branes can describe gravitons emitted from the black hole.
This can naturally be understood in terms of partial deconfinement~\cite{Hanada:2021ipb,Hanada:2016pwv}, i.e., the deconfined sector describes the degrees of freedom in the black hole.
By assumption all eigenvalues are bounded (i.e., we imposed such constraint in the numerical simulation), hence the degrees of freedom in the confined sector contribute to the bulk geometry as well and the geometry would not be too different from black zero-brane seen from an observer at sufficiently far from the black hole. Specifically, we expect a large M-theory circle near the center of the bulk geometry.
Secondly, the KK-momentum of the black hole along the M-theory circle can decrease as the energy is lowered, because the KK momentum corresponds to the number of D0-branes.
Therefore, we expect that the thermodynamic properties is not drastically different from a small black hole without boost.
Specifically, we conjecture that the M-theory black hole has negative heat capacity, namely the temperature increases as the energy decreases.\footnote{
We emphasize that this statement is speculative and it is desirable to have a more elaborate analysis.
}

In the canonical ensemble, such a phase corresponds to the local maximum of free energy separating the zero-brane phase (deconfined phase) and vacuum or graviton gas (confined phase).
To give a very rough estimate of $T_1$, we use the Schwarzschild solution in the noncompact space without boost, $S\sim \frac{T^{-9}}{G_{\rm N,11}}\sim N^3T^{-9}$, as an approximation. Then we obtain $T\sim N^{2/9}$ for the entropy at the Gregory-Laflamme transition, $S\sim N$. There is a jump of the temperature compared to the boosted uniform black string. It is likely that a non-uniform black string phase connects these two phases.
For complete evaporation to graviton gas, the black hole would have to be much smaller than the Gregory-Laflamme point. Therefore, we expect $T_1\gtrsim N^{2/9}$. A natural upper bound would be the 11d Planck scale $\sim N^{1/3}$.
We emphasize again that this is a very rough estimate; the important point is that $T_1$ can become parametrically large at large $N$.

It is natural to conjecture that such phases with negative specific heat, for which the nontrivial geometry along the eleventh direction is important,
are partially-deconfined.
See Fig.~\ref{fig:F-vs-E} for a sketch of the relationship between the free energy and energy at a fixed temperature.
The transition between the partially-deconfined phase and the completely-deconfined phase is the Gross-Witten-Wadia (GWW) transition where the gap is developed in the distribution of Polyakov line phases~\cite{Hanada:2020uvt,Hanada:2018zxn}.
In this sense, it would be natural to interpret the GWW transition as a `phase transition between string theory and M-theory'.

Our study in the 't Hooft limit suggests the phase diagram shown on the left of Fig.~\ref{fig:conjectured-phase-diagram-BMN}. Assuming the dual gravity analysis for the $\mu\to 0$ limit \cite{Itzhaki:1998dd} is valid, a cross-over from type IIA string theory to M-theory should be found as the energy is lowered if we zoom into the lower-left corner of this phase diagram.
Although the heat capacity is positive in the type IIA region, it can turn negative in the M-theory region.
Therefore, it is not unreasonable to expect that the phase transition we observe in this paper is connected to the typeIIA/M-theory phase transition.\footnote{This expectation may not be valid if there is a phase transition separating the $\mu\to 0$ limit and $\mu\sim N^0$ region.
A theoretically complicated feature of the limit of $\mu\to 0$ is that different fuzzy-sphere vacua become identical with the trivial vacuum of the BFSS matrix model unless the spin is increased as $\mu$ approaches zero.
}
Therefore, we might be able to learn about the M-theory black hole by studying the partially-deconfined phase in the BMN matrix model.

The highest temperature of the trivial-confined phase $T_1$ is the highest temperature of the partially-deconfined phase.
Because we expect that the partially-deconfined phase is dual to the eleven-dimensional Schwarzschild black hole,
we should find the temperature of the smallest possible black hole.
We can use the Gregory-Laflamme-transition temperature $\sim N^{2/9}$ as a lower bound for $T_1$.
A loose upper bound can be obtained by the Planck scale $\sim N^{1/3}$.

A parametrically large $T_1$ is probably a generic feature of string/M-theory.\footnote{
We thank Steve Shenker for drawing our attention to Ref.~\cite{Horowitz:1999uv} which is crucial in the argument provided below.}
As a well-known example, let us consider the case of the duality between 4d SYM and type IIB string theory on AdS$_5\times$S$^5$ with large fixed 't Hooft coupling $\lambda_{\rm 4d}=O(N^0)$~\cite{Aharony:1999ti,Aharony:2003sx}.
In this case, we have two scales: The Planck scale and the string scale. Supergravity is a good approximation when the black hole is bigger than the string length because strings can behave like point particles. Below this scale, the description as Hagedorn string is more appropriate. Therefore, $T_1\sim\ell_s^{-1}\sim\lambda_{\rm 4d}^{1/4}$ is expected, where $\lambda_{\rm 4d}$ is the 't Hooft coupling of 4d SYM. Note that this is the same as the Hagedorn temperature of the free string.\footnote{
The length of the string is related to the energy $E$ and entropy $S$ as $E\sim\frac{L}{\ell_s^2}$ and $S\sim\frac{L}{\ell_s}$.
The Hagedorn temperature $T_H$ is the temperature at which the free energy of the free string $F=E-TS$ becomes zero, and hence, $T_H\sim\frac{1}{\ell_s}$.
}
Note also that the mass of the 10d Schwarzschild black hole at $T_1$ is much larger than the Planck mass in this case.
Still, in the strong coupling limit ($\lambda_{\rm 4d}\to\infty$), $T_1$ becomes parametrically large.

Strictly speaking, the 10d black hole phase considered above is the equilibrium state of a 10d black hole plus graviton gas filling the rest of AdS$_5\times$S$^5$.
In the 't Hooft large-$N$ limit, the energy of the graviton gas is much smaller than that of the black hole. Hence it could be ignored in the thermodynamic analysis,
and the 10d black hole is stable in the microcanonical ensemble.
The situation can change away from the 't Hooft large-$N$ limit~\cite{Horowitz:1999uv}. A weakly-coupled string description is justified as long as $g_s\sim g_{\rm 4d}^2\ll 1$ and $\frac{\ell_s}{R_{\rm AdS}}\sim\lambda_{\rm 4d}^{-1/4}\ll 1$. If $\lambda_{\rm 4d}\gtrsim N^{8/17}$ (which is allowed while still satisfying $g_{\rm 4d}^2\ll 1$), the graviton gas can have larger entropy than the 10d black hole, even when the latter is not as small as $\ell_s$ and hence supergravity is a good approximation.
Therefore, at $\lambda_{\rm 4d}\gtrsim N^{8/17}$, a sufficiently small 10d black hole can evaporate completely to the gas of gravitons; it is unstable even in the microcanonical ensemble. This argument identifies the highest possible temperature of the 10d black hole, or more precisely, the system of graviton gas and the 10d black hole, which is equivalent to $T_1$.
The scaling with $N$ can be estimated by requiring the entropies of the black hole $\sim (\ell_{p,{\rm 10d}} E)^{8/7}$ and graviton gas $\sim (R_{\rm AdS}E)^{9/10}$ to be of the same order at $T=T_1$. By using $\ell_{p,{\rm 10d}}^9\sim N^{-2}$ for $R_{\rm AdS}\sim 1$, we obtain $E\sim N^{20/17}$ and $T_1\sim N^{2/17}$.
This is smaller than the Hagedorn temperature $\sim\lambda_{\rm 4d}^{1/4}$ if $\lambda_{\rm 4d}\gtrsim N^{8/17}$, but still it becomes parametrically large in the large-$N$ limit.

The same analysis~\cite{Horowitz:1999uv}  can be applied to the AdS$_4\times$S$^7$ geometry dual to the ABJM theory~\cite{Aharony:2008ug}.
A rough estimate of the highest possible temperature of the 11d black hole $T_1$ is obtained by requiring that the entropies of graviton gas $\sim (R_{\rm AdS}E)^{10/11}$ and black hole $\sim (\ell_{P,{\rm 11d}} E)^{9/8}$ are of the same order at $T=T_1$.
By using $\ell^{-1}_{P,{\rm 11d}}\sim \frac{N^{1/6}}{R_{\rm AdS}}$, we obtain $T_1\sim \frac{1}{R_{\rm AdS}}\left(\frac{\ell_{P,{\rm 11d}}}{R_{\rm AdS}}\right)^{-9/19}$ $\sim \frac{N^{3/38}}{R_{\rm AdS}}$.
This $T_1$ is smaller than the Planck mass, and hence the approximation by eleven-dimensional gravity is self-consistent.

The same logic could be applied to the BFSS matrix model if the gravity side were understood more precisely.
It would be nice if we could see the trivial-confined phase at a much larger $N$ and a higher temperature. That it was not observed in past studies is not necessarily a contradiction because a careful choice of the initial condition for the simulation is needed to see such a phase which has small free energy.
\subsection{BMN matrix model ($\mu>0$)}
Let us now consider the $\mu>0$ case with trivial vacuum, and gradually raise the temperature (canonical ensemble).
At sufficiently low temperature, the system is in the confined phase.\footnote{
	By `confined phase', we mean a phase in which the energy and entropy are of lower order in $N$ than $N^2$,
	while they become of order $N^2$ in the deconfined phase.
	Confinement is also indicated by a uniform distribution of the Polyakov line phases.
	Note that `confined phase' is sometimes used to indicate the phase with order $N^0$ energy and entropy.
}
In the same manner, if we take a nontrivial fuzzy-sphere vacuum and raise the temperature, the system is confined at a low temperature.
To distinguish them, we use the names ``trivial-confined phase" and ``fuzzy-sphere-confined phase".\footnote{
	We could also call them `confined phase on the trivial background' and `confined phase on the fuzzy-sphere background'.
}

As the temperature goes up, at some point, a transition to a deconfined phase can take place.
As long as the thermal excitation of each matrix entry is not too strong, we can still distinguish different fuzzy-sphere backgrounds.
Therefore, both a ``trivial-deconfined phase" and ``fuzzy-sphere-deconfined phase" can exist.\footnote{
	If we fix $\mu$ and $N$ and go to a very high temperature, the thermal excitation of each matrix entry becomes larger than the size of any fuzzy sphere,
	and the distinction between ``trivial-deconfined phase" and ``fuzzy-sphere-deconfined phase" disappears.
	}

As we will see shortly, all these phases --- trivial-confined phase, fuzzy-sphere-confined phase, trivial-deconfined phase, and fuzzy-sphere-deconfined phase --- can be minima of free energy in the canonical ensemble.
Multiple minima can coexist in certain parameter regimes, and the phase structure can be rather complicated.
We will take the trivial-confined phase and study the transition to deconfinement, but a priori, we do not know whether the trivial-deconfined or fuzzy-sphere-deconfined phase is obtained.
To make the analysis tractable, we split the discussion into two steps.
\begin{itemize}
	\item
	First, we ignore the possibility of a transition to the fuzzy-sphere-deconfined phase
	and consider the transition between the trivial-confined phase and the trivial-deconfined phase.
	We argue that there are three kinds of critical temperatures, $T_1,T_2$, and $T_{\rm c}$, that satisfy $T_2<T_{\rm c}<T_1$.
	The trivial-confined phase (resp.~trivial-deconfined phase) exists as a minimum of free energy only at $T\le T_1$ (resp.~$T\ge T_2$).
	The free energy of these two phases coincides at $T=T_{\rm c}$. The trivial-confined phase (resp.~trivial-deconfined phase) is favored at $T<T_{\rm c}$
	(resp.~$T>T_{\rm c}$).
	See Sec.~\ref{sec:trivial-conf-trivial-deconf}.

	\item
	Next, we take into account the fuzzy-sphere-deconfined phase.
	We show that a lower bound for $T_1$ is obtained if we observe the trivial-confinement/fuzzy-sphere-deconfinement phase transition.
	See Sec.~\ref{sec:trivial-conf-fuzzy-sphere-deconf}.

	\item
	When $\mu$ is very small and $N$ is not very large, we see yet another ``phase transition"; one or few of the eigenvalues escape and roll to very large values.
	This corresponds to a run-away behaviour toward the flat direction at $\mu=0$~\cite{deWit:1988xki,Anagnostopoulos:2007fw}.
	We can obtain a lower bound for $T_1$ also from this run-away behavior.
	See Sec.~\ref{sec:trivial-conf-run-away}.
\end{itemize}

\subsubsection{Trivial-confinement/trivial-deconfinement transition}\label{sec:trivial-conf-trivial-deconf}

Let us think about what kind of phase diagram can appear in the BMN matrix model.
At each fixed $\mu$, we expect one of the types of phase distributions shown in Fig.~\ref{fig:P-vs-T-3-patterns}.
The large-$\mu$ region permits perturbative calculations~\cite{Furuuchi:2003sy,Spradlin:2004sx},
and the transition is found to be of first order, i.e., the left scenario in Fig.~\ref{fig:P-vs-T-3-patterns} is realized.
Therefore, we expect one of the two phase structures depicted in Fig.~\ref{fig:conjectured-phase-diagram-BMN}, depending on the order of the phase transition at $\mu=0$.
Let us call them the first-order scenario and not-first-order scenario, respectively.

We expect that the first-order scenario is more likely to be realized for the following reasons:
\begin{itemize}
\item
We expect a first-order transition in the BFSS matrix model, and if the $\mu\to 0$ limit is smooth, this first-order character should persist for finite $\mu$.
(Note that the smoothness of the $\mu\to 0$ limit is nontrivial, and the order of $\mu\to 0$ and $N\to\infty$ can be important.)

\item
In the not-first-order scenario, the first-order transition at large $\mu$ splits into two transitions at small $\mu$ ($T_1$ becomes smaller than $T_2$, and $T_c$ disappears). Then the GWW transition should exist at some finite temperature  $T_2>0$ even at $\mu=0$.
It has to lie below $\lambda^{-1/3}T=0.4$, because $\rho(\theta)$ is gapped there~\cite{Berkowitz:2016jlq}.
But if this were the case a big puzzle would pop up:
why did the numerical simulations in the past agree well with dual gravity predictions,
although they have been performed at higher temperature which could be separated from the low-temperature region, where gravity is precise,
by the GWW transition?
Thus, a not-first-order scenario is disfavored.

\item
If the GWW transition lies in the type IIA string theory region,
there is yet another problem: below the GWW transition, it is natural to expect that the system is in the partially-confined phase, which corresponds to the RR-charge less than $N$; it does not agree with the charge counting in the gravity side.

 \end{itemize}

\subsubsection*{More on the first-order scenario}
\begin{figure}[htbp]
\begin{center}
\rotatebox{0}{
\scalebox{0.5}{
\includegraphics{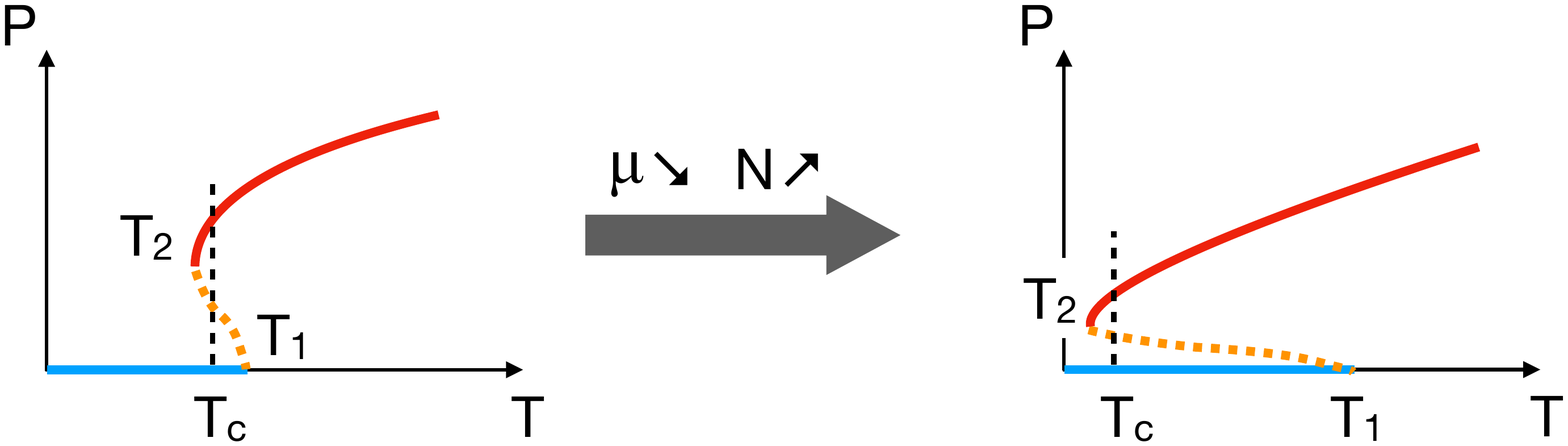}}}
\end{center}
\caption{
The first-order scenario.
The left panel is for larger $\mu$, and the right panel is for smaller $\mu$.
Both critical temperature and the size of the jump decrease as $\mu$ becomes smaller and $N$ becomes larger,
while $T_1$ can go up, as discussed in Sec.~\ref{sec:Schwarz11D}.
Strictly speaking, the orange part further splits into two phases: the black string and black hole phases.
}\label{fig:Pol-1st-order-scenario}
\end{figure}

\begin{figure}[htbp]
\begin{center}
\rotatebox{0}{
\scalebox{0.2}{
\includegraphics{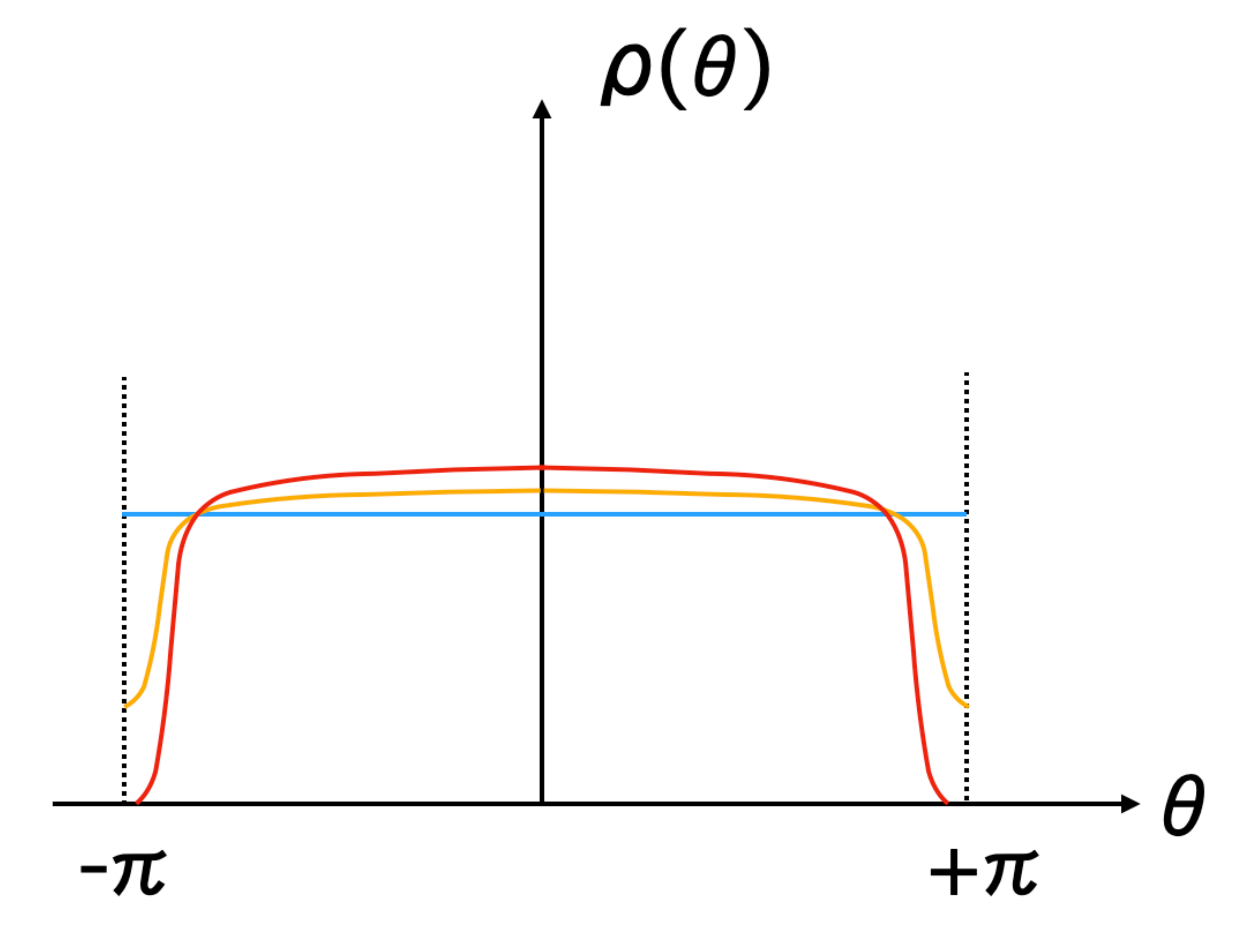}}}
\end{center}
\caption{
Possible form of the phase distribution around the phase transition at very small $\mu$.
}\label{fig:conjecture_GWW_strong_coupling}
\end{figure}

Suppose the transition is of first order, as shown in the left panel of Fig.~\ref{fig:conjectured-phase-diagram-BMN}.
Then, as $\mu\to 0$, the jumps of the energy and entropy have to approach zero
(more precisely, $\frac{E}{N^2}$ and $\frac{S}{N^2}$ have to approach zero), otherwise the $\mu=0$ limit disagrees with the dual gravity picture \cite{Itzhaki:1998dd} which has been tested throughly by numerical simulations.
It is natural to expect that the jump of the Polyakov loop also vanishes at $\mu=0$; see Fig.~\ref{fig:Pol-1st-order-scenario}.
Then, the distribution $\rho(\theta)$ has to become almost flat, while staying gapped; see Fig.~\ref{fig:conjecture_GWW_strong_coupling}.\footnote{
This is consistent with the claim in Ref.~\cite{Maldacena:2018vsr} that the phase distribution becomes flat at zero temperature. It is also consistent with the unpublished observation made by using the simulation data for Ref.~\cite{Hanada:2013rga}.
}

Let us summarize how the first-order scenario may work:
\begin{itemize}
\item
The transition remains of first order at small $\mu$.
As $\mu$ becomes small, the transition temperature approaches zero
(Fig.~\ref{fig:conjectured-phase-diagram-BMN} left).

\item
The jump of the energy and Polyakov loop at critical temperature becomes smaller.
(Fig.~\ref{fig:Pol-1st-order-scenario})

\item
The distribution of the Polyakov line phases at the GWW transition is distorted more and more at smaller $\mu$
and approaches the uniform distribution.
(Fig.~\ref{fig:conjecture_GWW_strong_coupling})

\end{itemize}

\subsubsection{Trivial-confinement/fuzzy-sphere-deconfinement transition}\label{sec:trivial-conf-fuzzy-sphere-deconf}
Suppose the first-order trivial-confinement/fuzzy-sphere-deconfinement transition is observed at $\tilde{T}_{\rm c}$ as $T$ is increased
before the trivial-confinement/trivial-deconfinement transition is observed.
Here we consider the situation that initial configuration for the simulation is taken from the trivial-confined phase, and temperature is gradually raised.
(We will encounter this situation later in Sec.~\ref{sec:SUSY-simulation}.)
Then, a natural expectation is that $\tilde{T}_{\rm c}<T_{\rm c}$.

More precisely: suppose that the two-state signal is observed between $\tilde{T}_2$ and $\tilde{T}_1$, where $\tilde{T}_2<\tilde{T}_1$. Then, it is natural to expect $\tilde{T}_1<T_2$.  Otherwise, we should see the trivial deconfined phase as well.

\subsubsection{Run-away behavior}\label{sec:trivial-conf-run-away}
The BFSS matrix model has flat directions. The BMN model shows remnants of these: When $\mu$ is small, scalars can be almost commutative, and eigenvalues can diverge as $\mu$ is sent to zero. All four phases under consideration can have an instability in this semi-flat direction. Let us denote such an instability as ``run-away behavior''. This effect is temperature-dependent: at sufficiently large temperatures, the flat directions are lifted.

In the trivial-confined phase, all D0-branes are sitting at the origin without any excitation~\cite{Hanada:2021ipb}. On the other hand, in the trivial-deconfined phase, each degree of freedom has an energy of order $N^0$. Because some amount of energy is needed for a D0-brane to move in the flat direction, it is natural to expect that the trivial-deconfined phase is more prone to the run-away behavior.
We expect the same for the fuzzy-sphere background, i.e., the fuzzy-sphere-deconfined phase is more prone to the run-away behavior.
It is also known that the problem with the flat direction is milder at high temperatures.
Hence, in the small-$\mu$ and not-so-large-$N$ region, it can be the case that the deconfined phase can be sufficiently stable only at higher temperatures. Consequently there can be a finite temperature range where we do not notice the metastable deconfined contribution in the simulations.\footnote{Note that it is hard to distinguish this possibility from a more standard scenario Fig.~\ref{fig:Pol-1st-order-scenario} with a very shallow meta-stable minimum; in Monte Carlo simulations, both would exhibit the run-away behavior.}

All of these observations suggest the following scenario. Suppose we start with the trivial-confined phase, gradually raise the temperature, and observe run-away behavior at some temperature $T_{\rm run\mathchar`-away}$. This temperature provides us with a lower bound for $T_1$, i.e., $T_{\rm run\mathchar`-away}\le T_1$. At sufficiently large $N$, $T_{\rm run\mathchar`-away}$ and $T_1$ should not differ much.

\section{Numerical determination of the phase transitions }\label{numericalsection}
After the discussion of the continuum physics and the conjectures about the phase transition, we now present our results from numerical simulations. The simulations have been done with the HMC algorithm. A careful analysis is needed to test the scenario described in Sec.~\ref{sec:conjecture_phase_diag}. Prior to this description, we consider the simpler case of the bosonic analogue of the BMN matrix model, where the transition point can be precisely identified even in the small $\mu$ limit. After this exercise, we
present the results of simulations for the full supersymmetric BMN model.
The BFSS matrix model is the limit of $\mu\to 0$ in the BMN matrix model.
Although our main interest lies in the limit of $\mu\to 0$ which has dual gravity descriptions, we will start from the large-$\mu$ region where the simulation is easier and gradually move toward smaller $\mu$.

The lattice action used for the simulations is discussed in Appendix~\ref{sec:lattice_regularization}.
We use $L$ to denote the number of lattice points. The lattice spacing is $a=\frac{\beta}{L}$.
The continuum limit is taken by sending $L$ to infinity.
 
\subsection{Phase transitions in the bosonic BMN model}\label{sec:bosonicbmn}

In this section we study the bosonic analogue of the BMN matrix model (bosonic BMN), whose action is only the bosonic part of the original, supersymmetric BMN matrix model. 
That is, we use the action $S=S_b+\Delta S_b$, where $S_b$ and $\Delta S_b$ are defined by \eqref{eq:action-BFSS-bos} and \eqref{eq:action-BMN-bos}, respectively. 
This model is much easier to simulate than the full, supersymmetric BMN model. 
This exercise is performed to confirm convergence to the bosonic BFSS model~\cite{Bergner:2019rca} in the $\mu\to 0$ limit.  

\begin{figure}[htbp]
	\begin{center}
		\rotatebox{0}{
			\scalebox{0.3}{
				\includegraphics{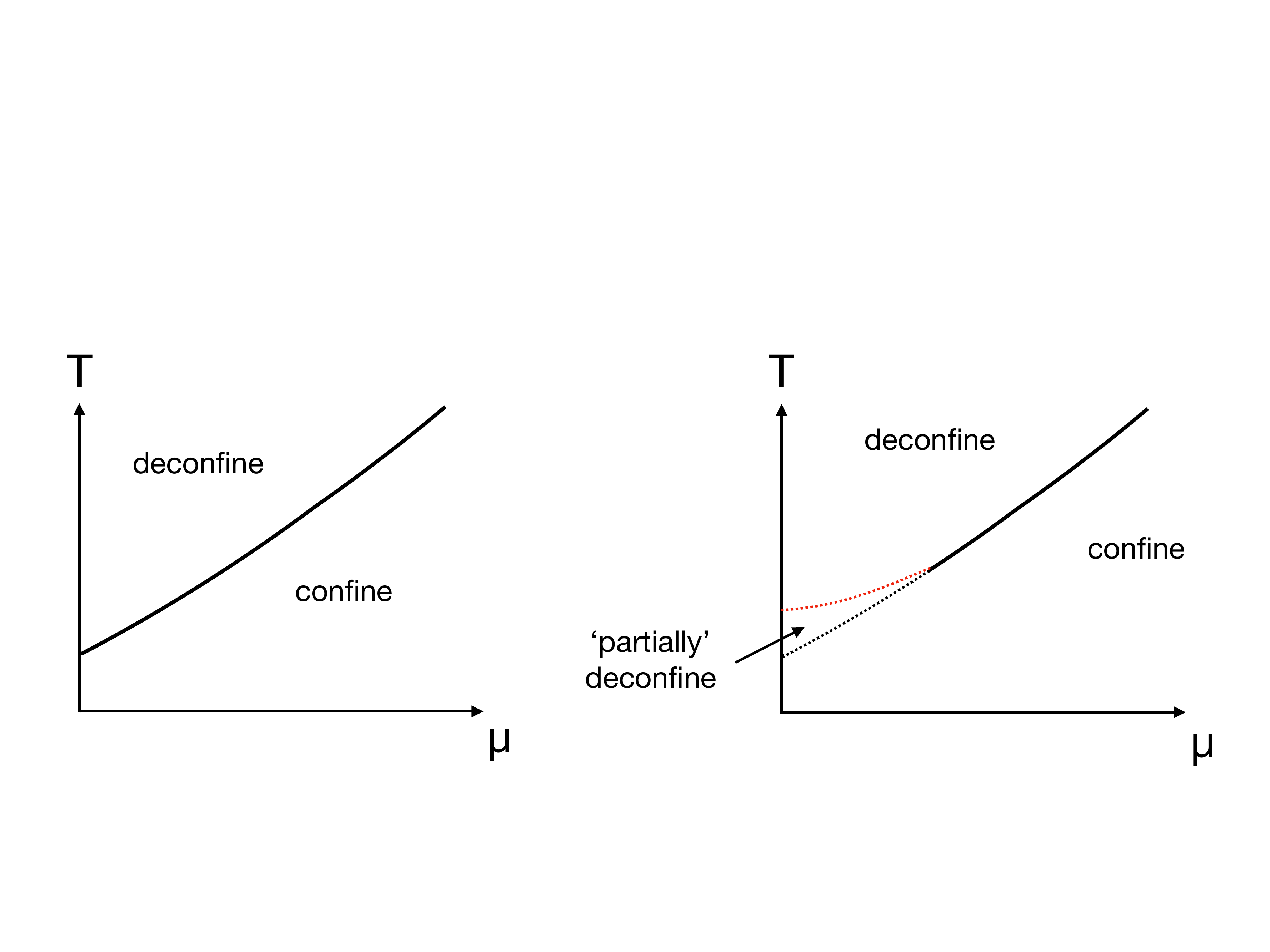}}}
	\end{center}
	\caption{Possible phase diagrams of the bosonic BMN model for the canonical ensemble.  
		Ref.~\cite{Bergner:2019rca} concluded that the left figure is the actual phase diagram.  
	}\label{fig:bosonic-BMN-phase-diagram}
\end{figure}

Historically, there was some confusion regarding the phase structure of the bosonic BMN model. 
At weak coupling (large $\mu$), perturbative analysis shows that the phase transition is of first order. 
At $\mu=0$, the first systematic study~\cite{Kawahara:2007fn} used the data with $N\le 32$ and suggested that the transition takes place at nonzero temperature and it is not of first order.
However, a later study with $N>32$ revealed that the transition is actually of first order after all~\cite{Bergner:2019rca}. 
From this fact, Ref.~\cite{Bergner:2019rca} concluded that the phase diagram of the bosonic BMN model looks like the left figure of Fig.~\ref{fig:bosonic-BMN-phase-diagram}.

Strictly speaking, to establish the phase structure unambiguously, it is necessary to confirm that the phase transition is of first order even in the intermediate-$\mu$ region. 
Below, we study a few intermediate values of $\mu$, and confirm this. In addition, we study the features of the transition in detail.

\begin{figure}[htbp]
	\begin{center}
			\scalebox{0.2}{
				\includegraphics{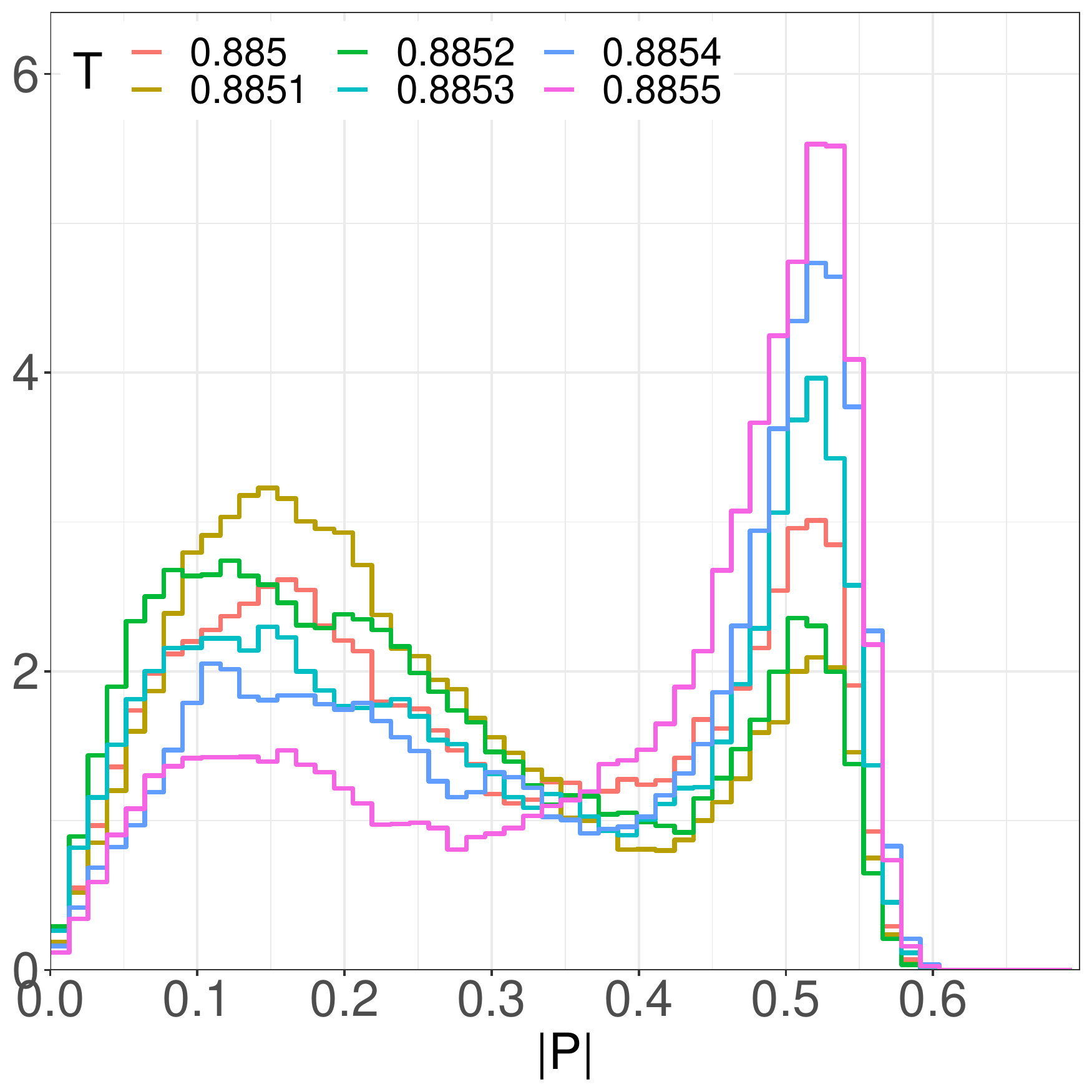}}
				\scalebox{0.2}{
				\includegraphics{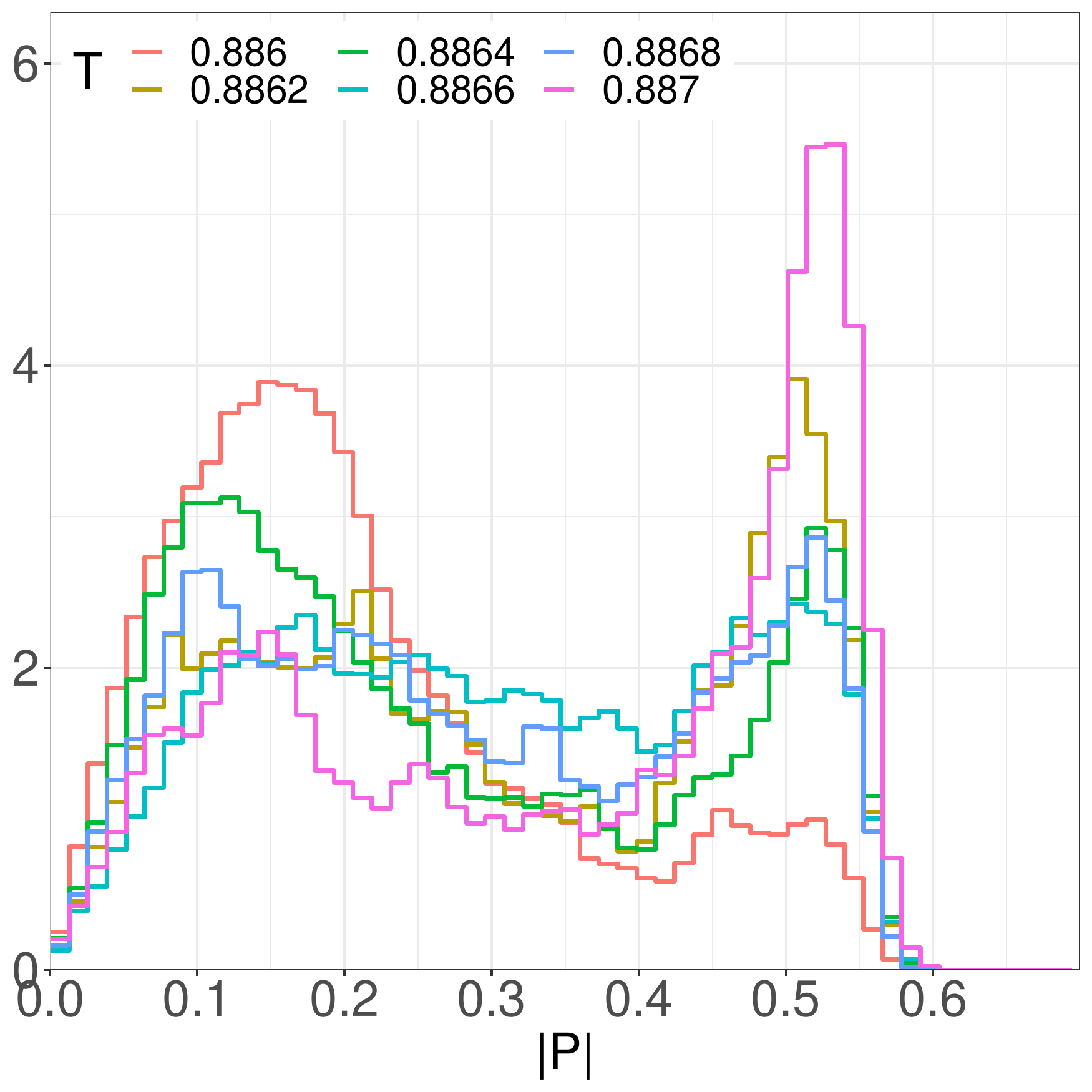}}
				\scalebox{0.2}{
				\includegraphics{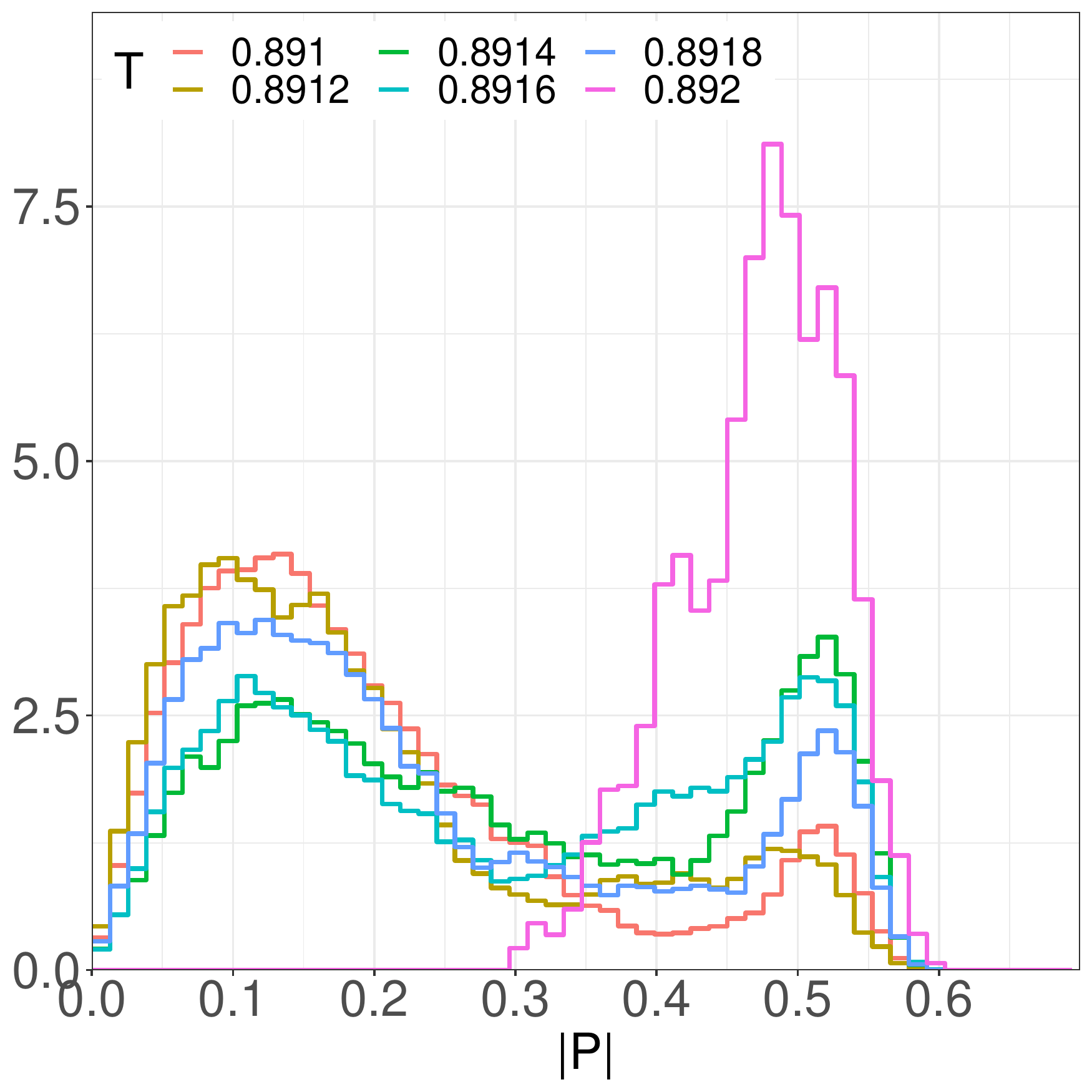}}
			\scalebox{0.2}{
				\includegraphics{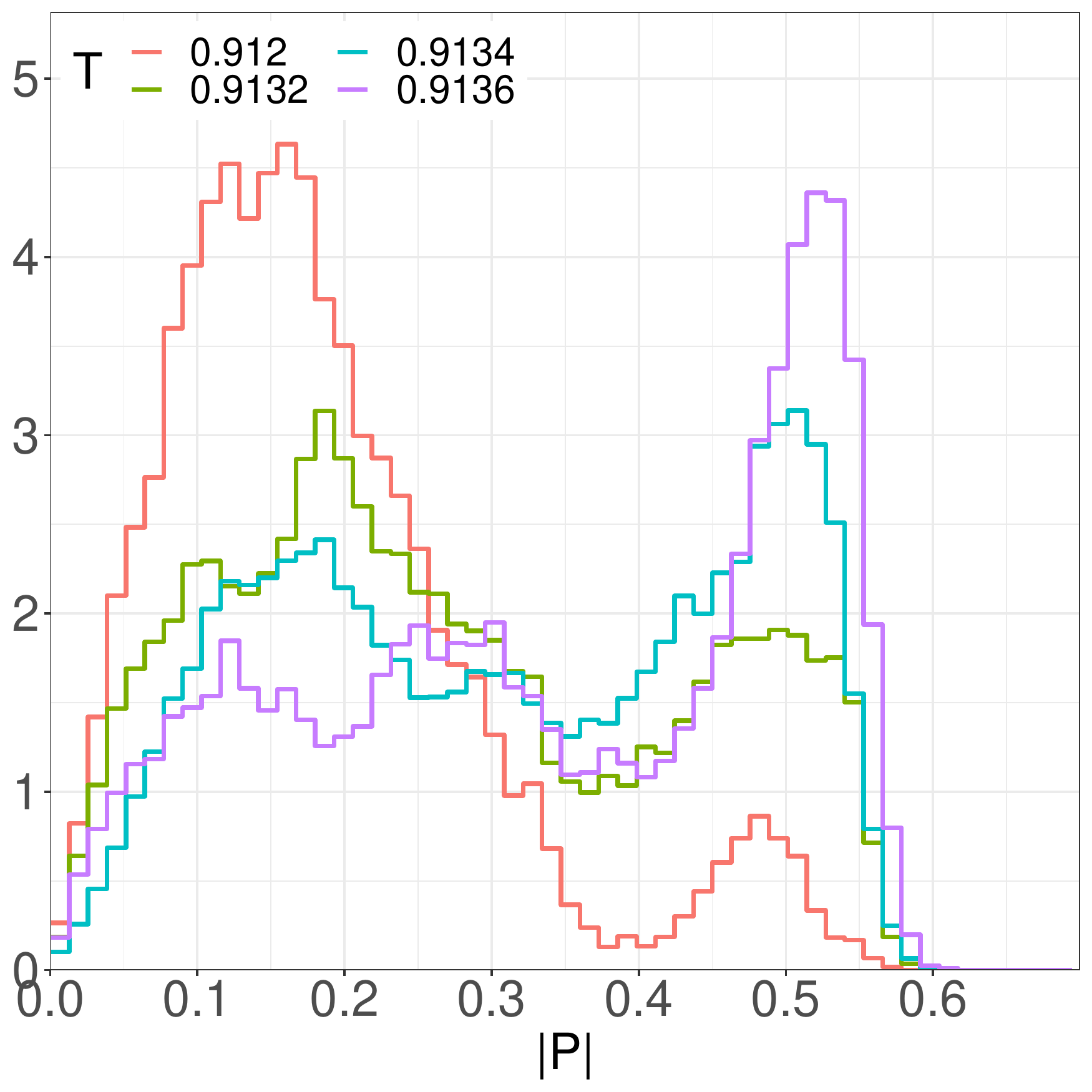}}
			\\
				\scalebox{0.2}{
					\includegraphics{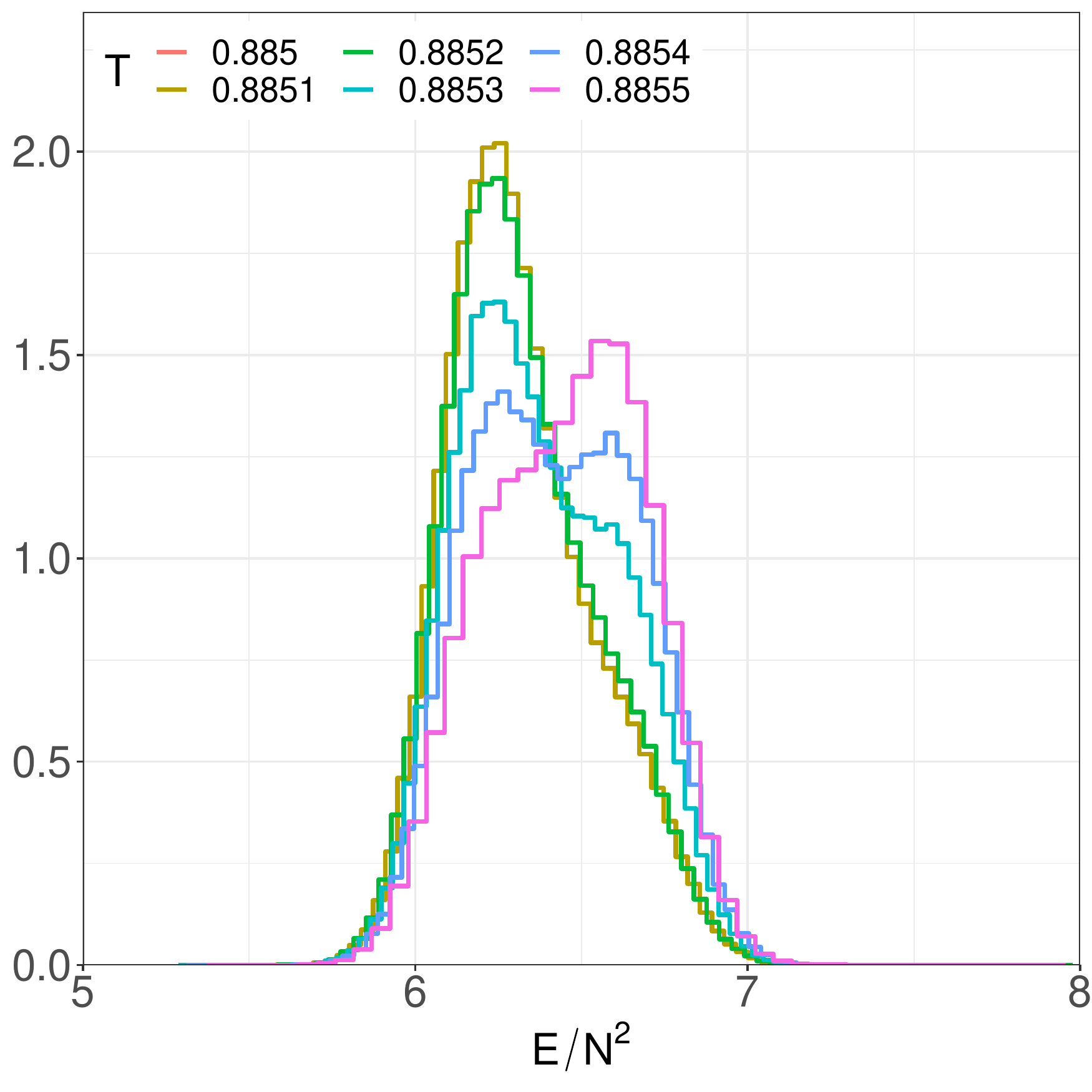}}
				\scalebox{0.2}{
					\includegraphics{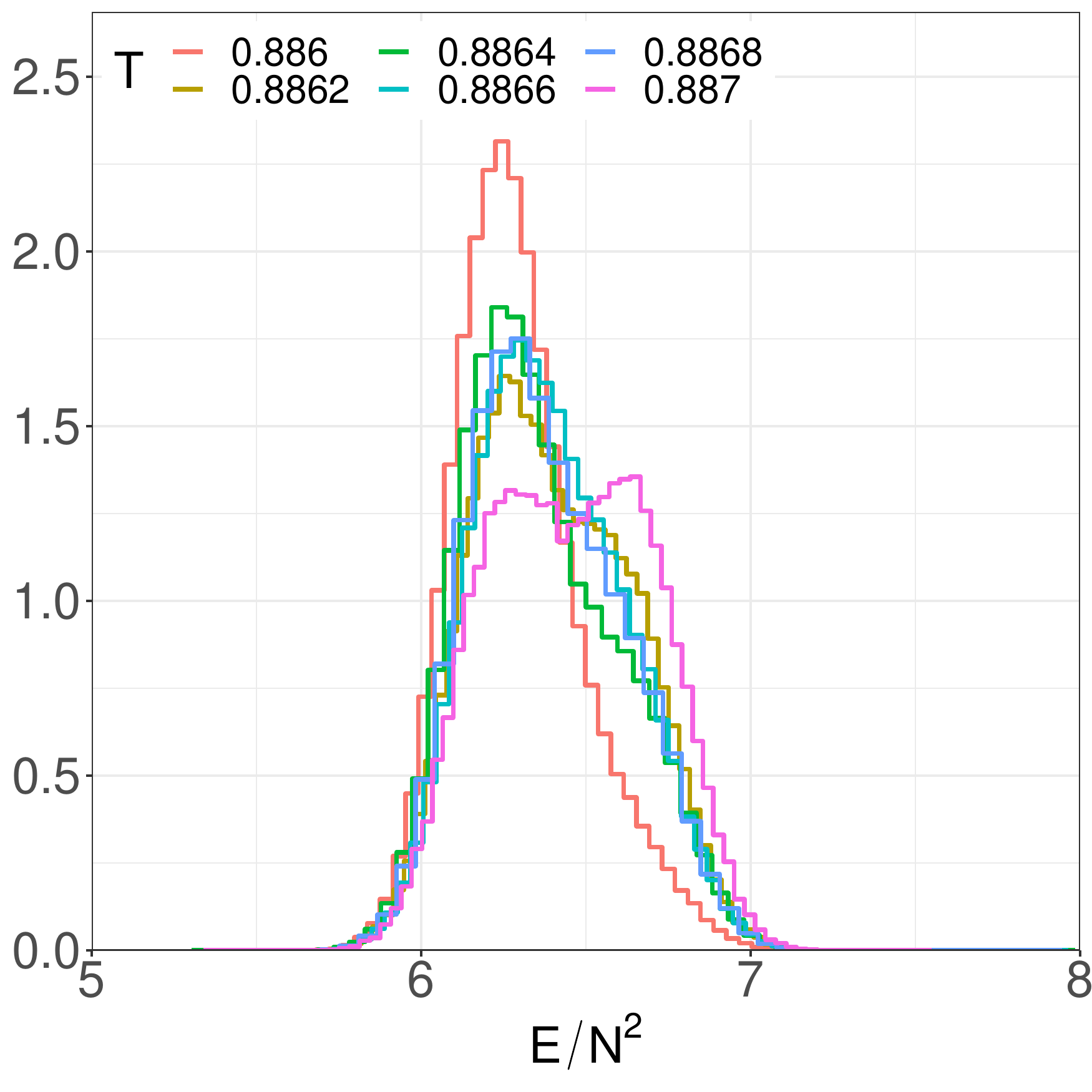}}
				\scalebox{0.2}{
					\includegraphics{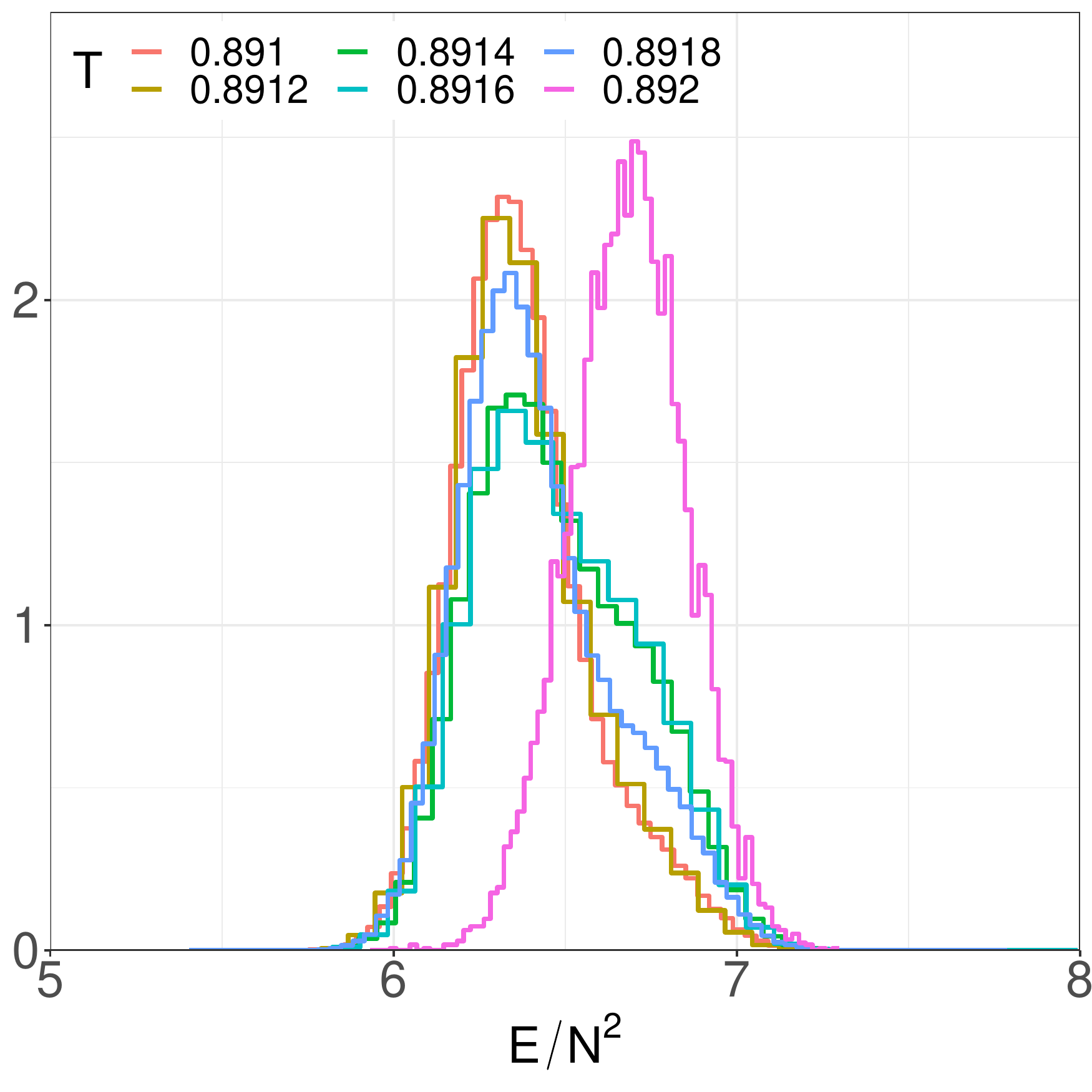}}
				\scalebox{0.2}{
					\includegraphics{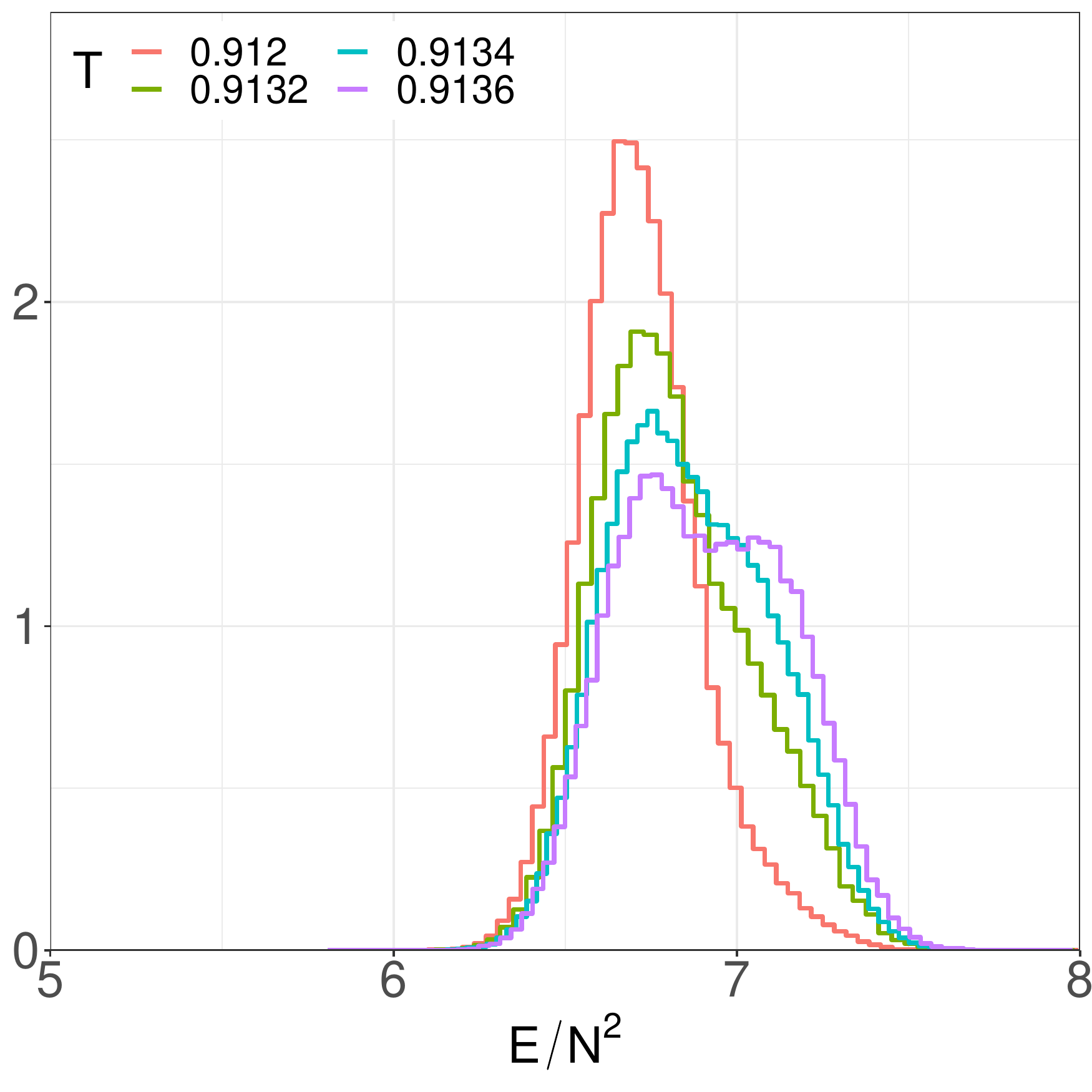}}\\
					\scalebox{0.2}{
					\includegraphics{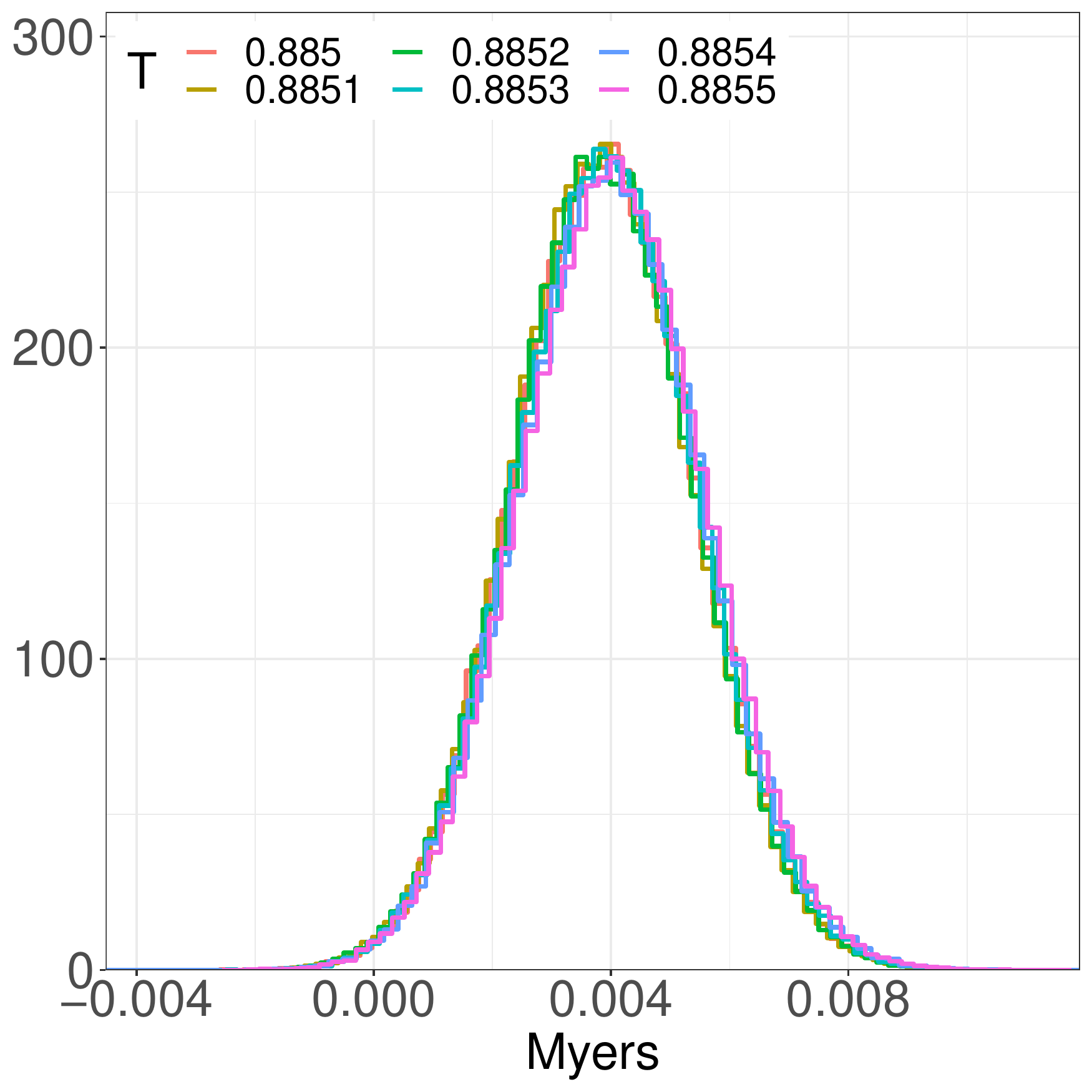}}
				\scalebox{0.2}{
					\includegraphics{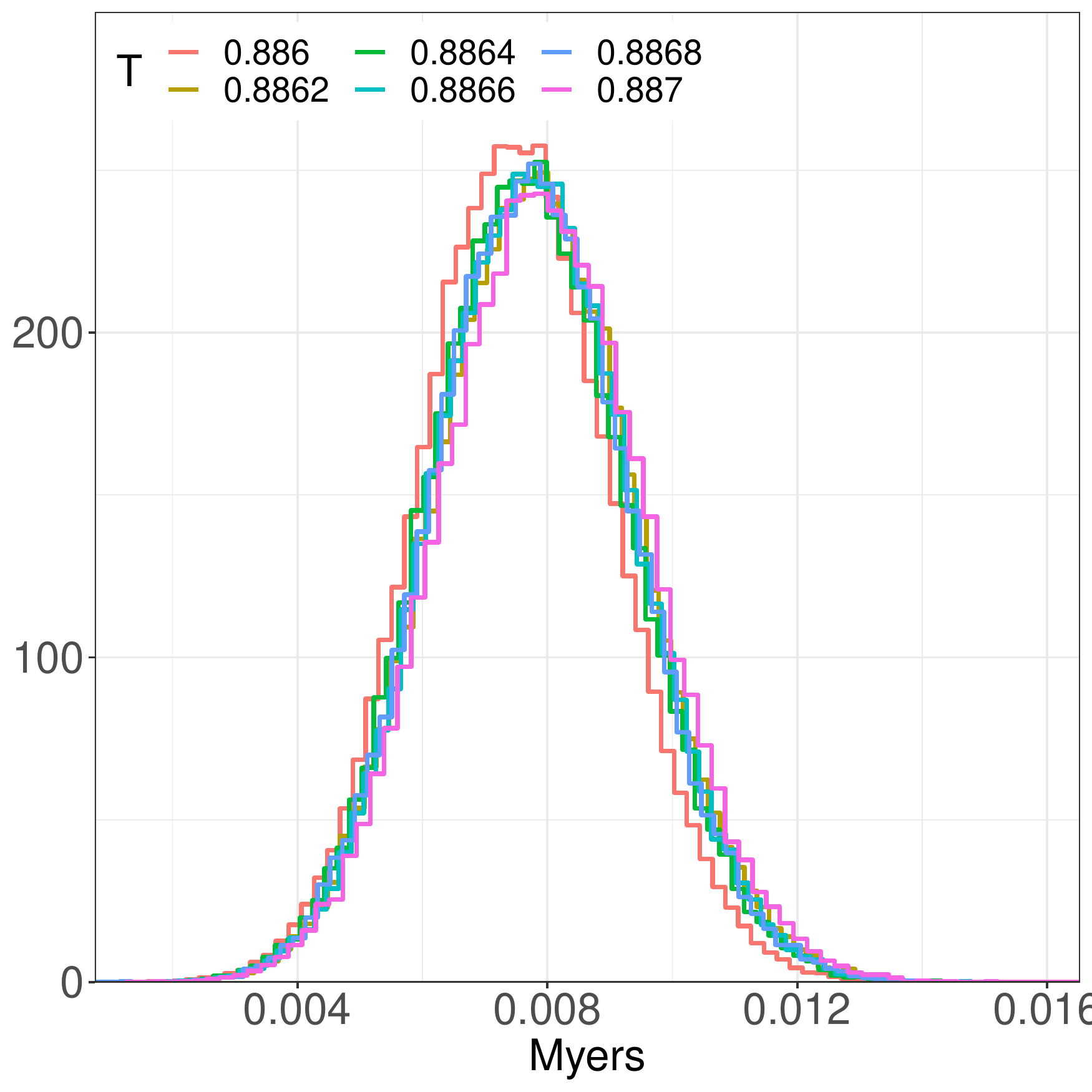}}
				\scalebox{0.2}{
					\includegraphics{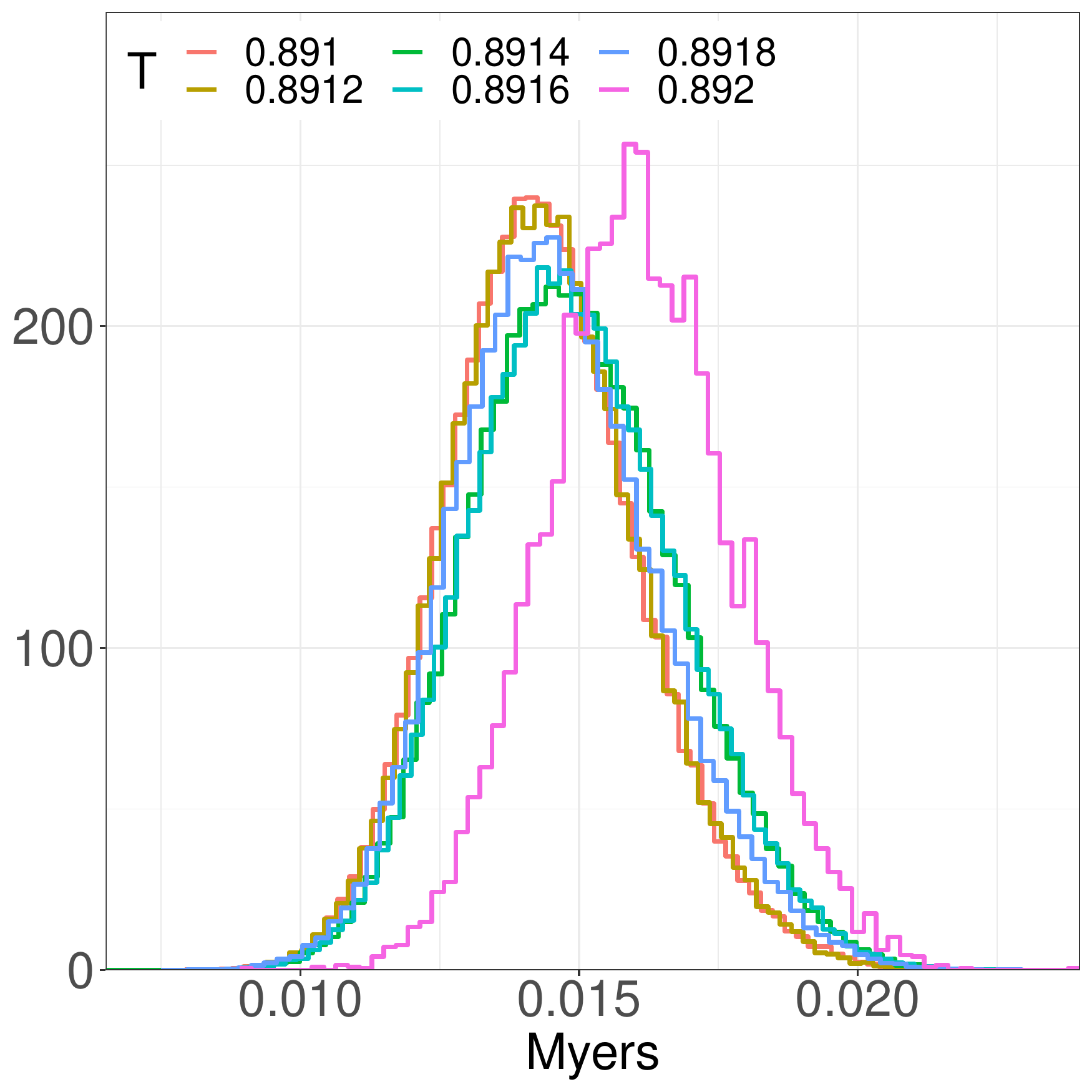}}
				\scalebox{0.2}{
					\includegraphics{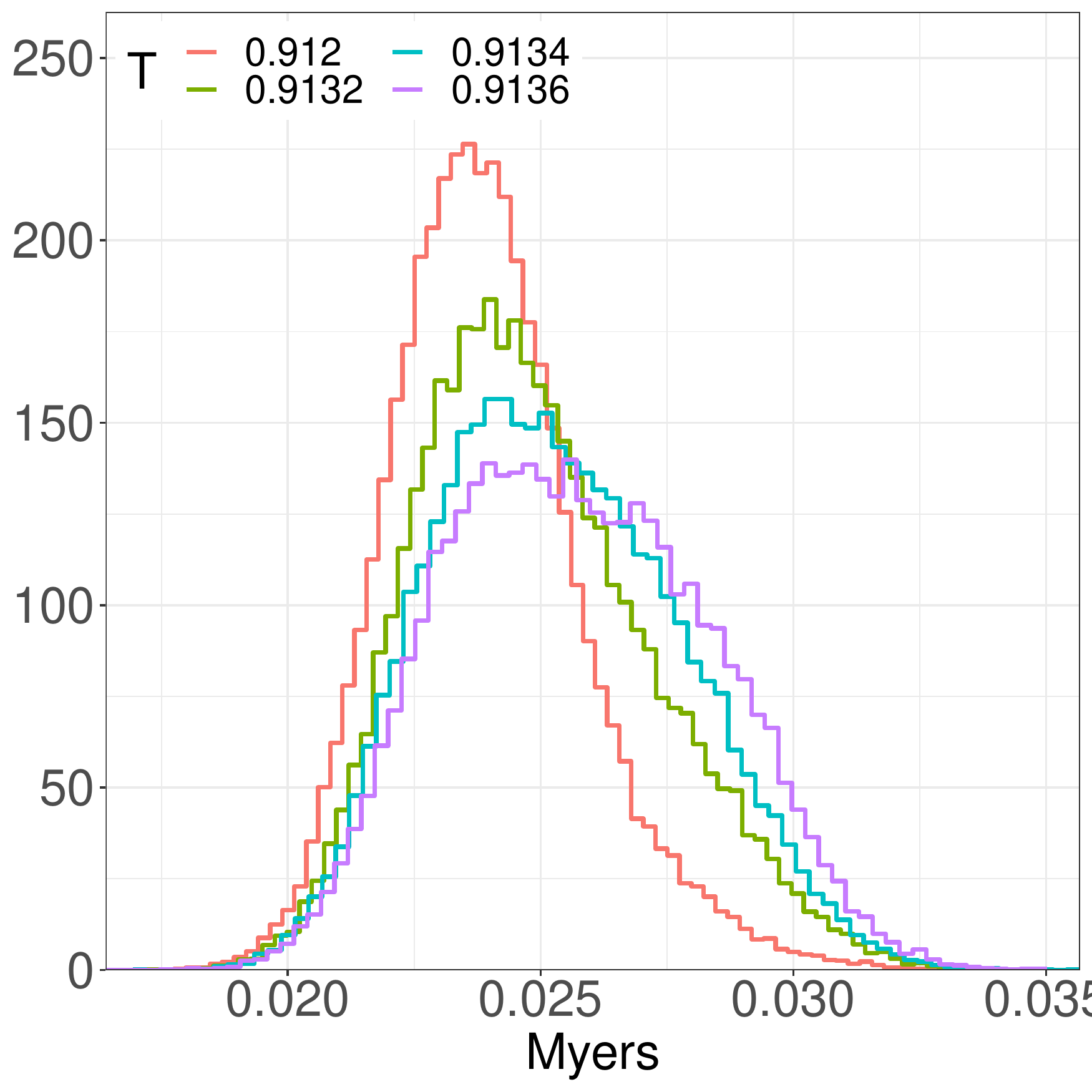}}
	\end{center}
	\caption{
	Histogram of the distribution of the Polyakov loop (first row), energy (second row), and Myers term (third row) during lattice Monte Carlo simulations for the bosonic BMN model at matrix size $N=64$ and lattice size $L=24$. 
	The values of $\mu$ are 0.125, 0.25, 0.5 and 1.0, from left to right. 
	A two-state signal near the transition temperature is observed at all values of $\mu$, particularly for the Polyakov loop $P$. 
	The jumps of energy and Myers terms are of order $N^2$,  but we do not observe the formation of a fuzzy-sphere background.
}\label{bosonic_bmn_lowmu}
\end{figure}

\begin{figure}[htbp]
	\begin{center}
	\rotatebox{0}{
		\scalebox{0.21}{
			\includegraphics{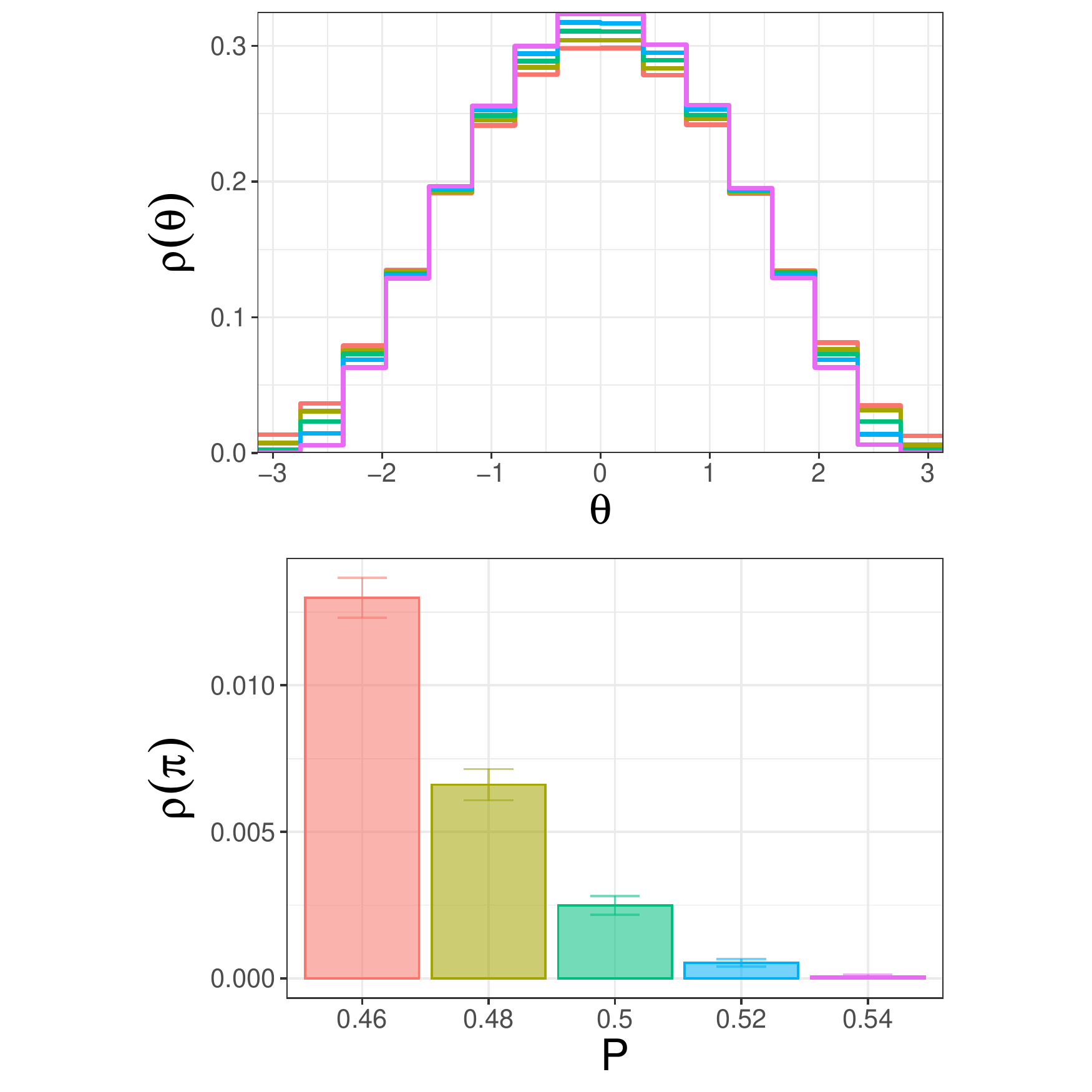}}
		\scalebox{0.21}{
			\includegraphics{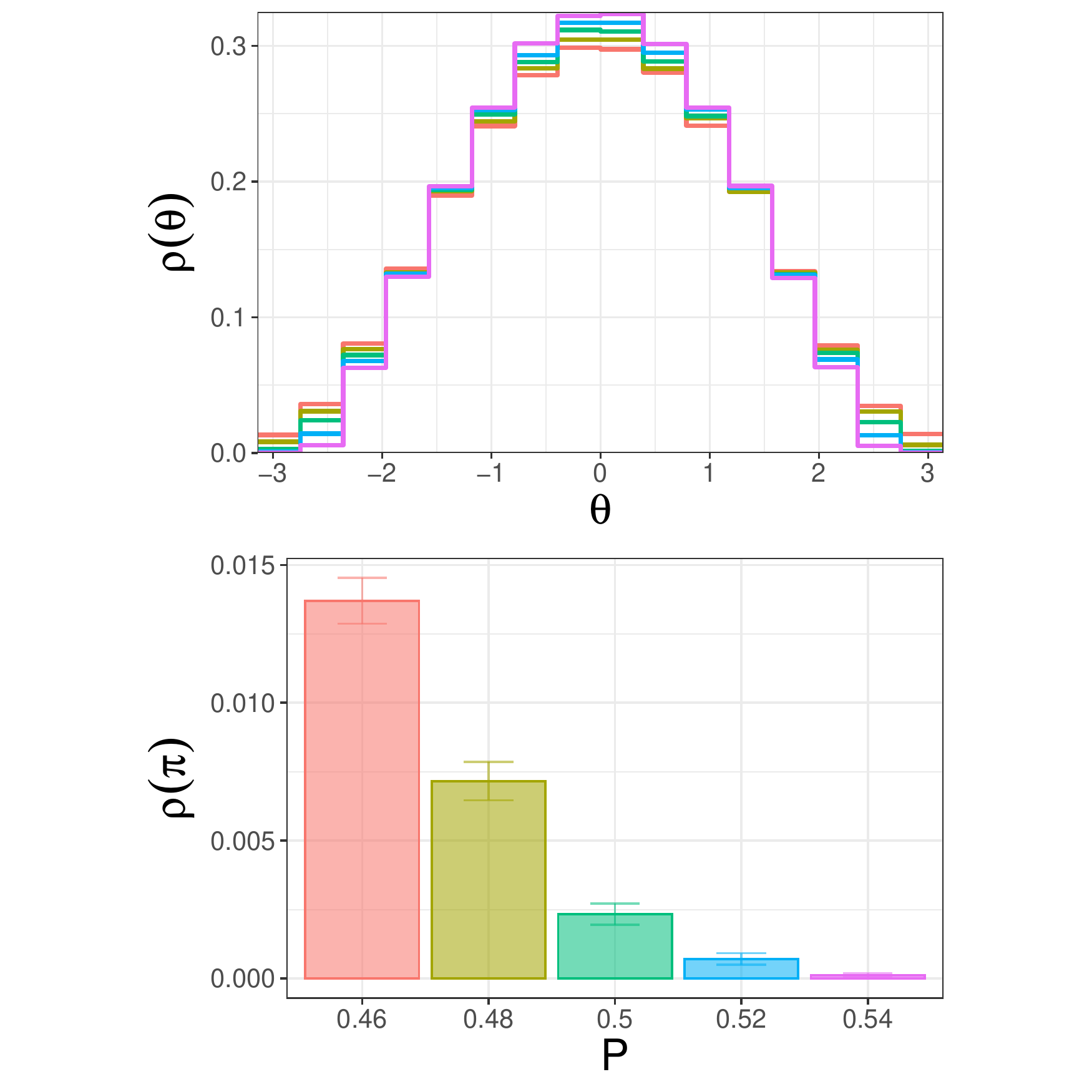}}
		\scalebox{0.21}{
			\includegraphics{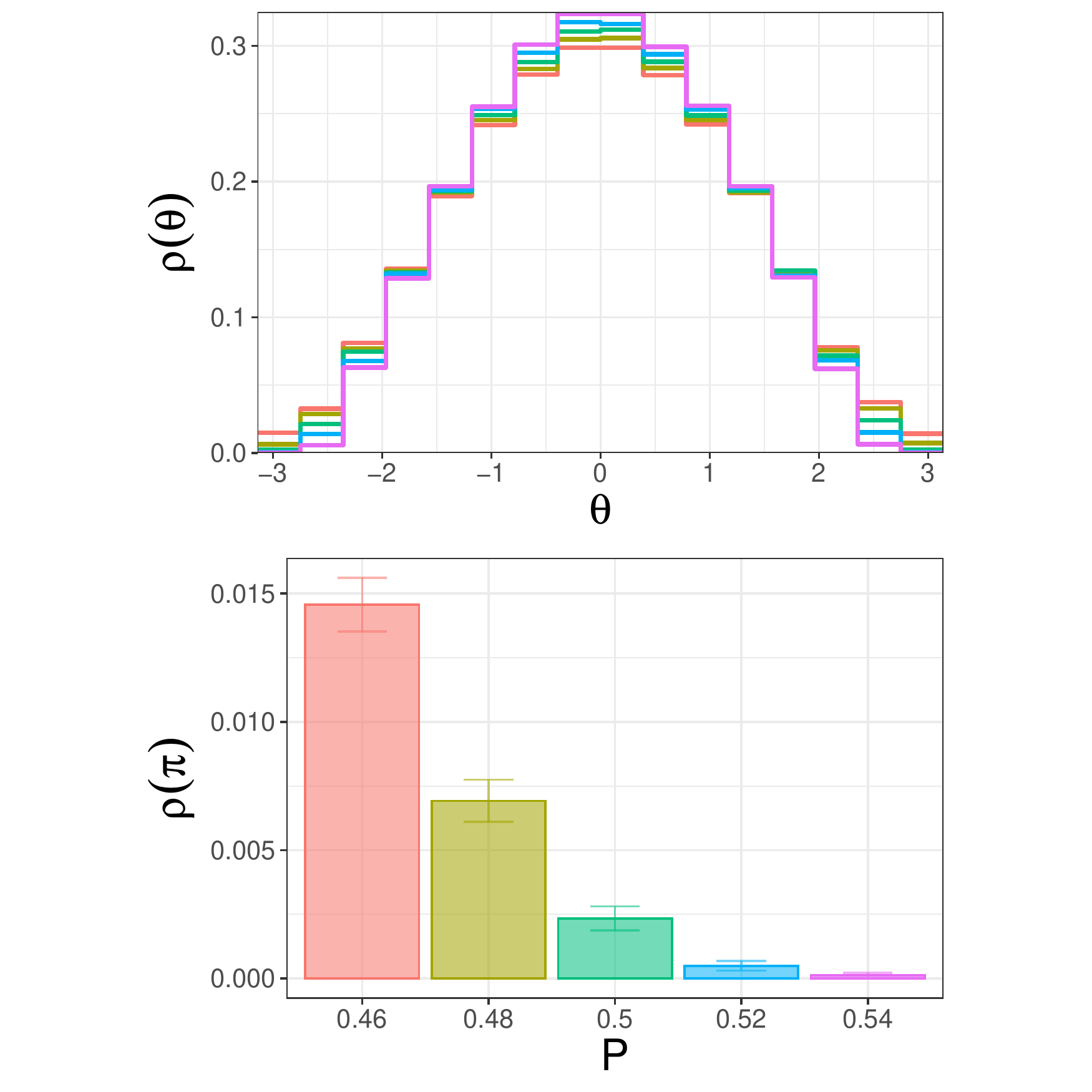}}
		\scalebox{0.21}{
			\includegraphics{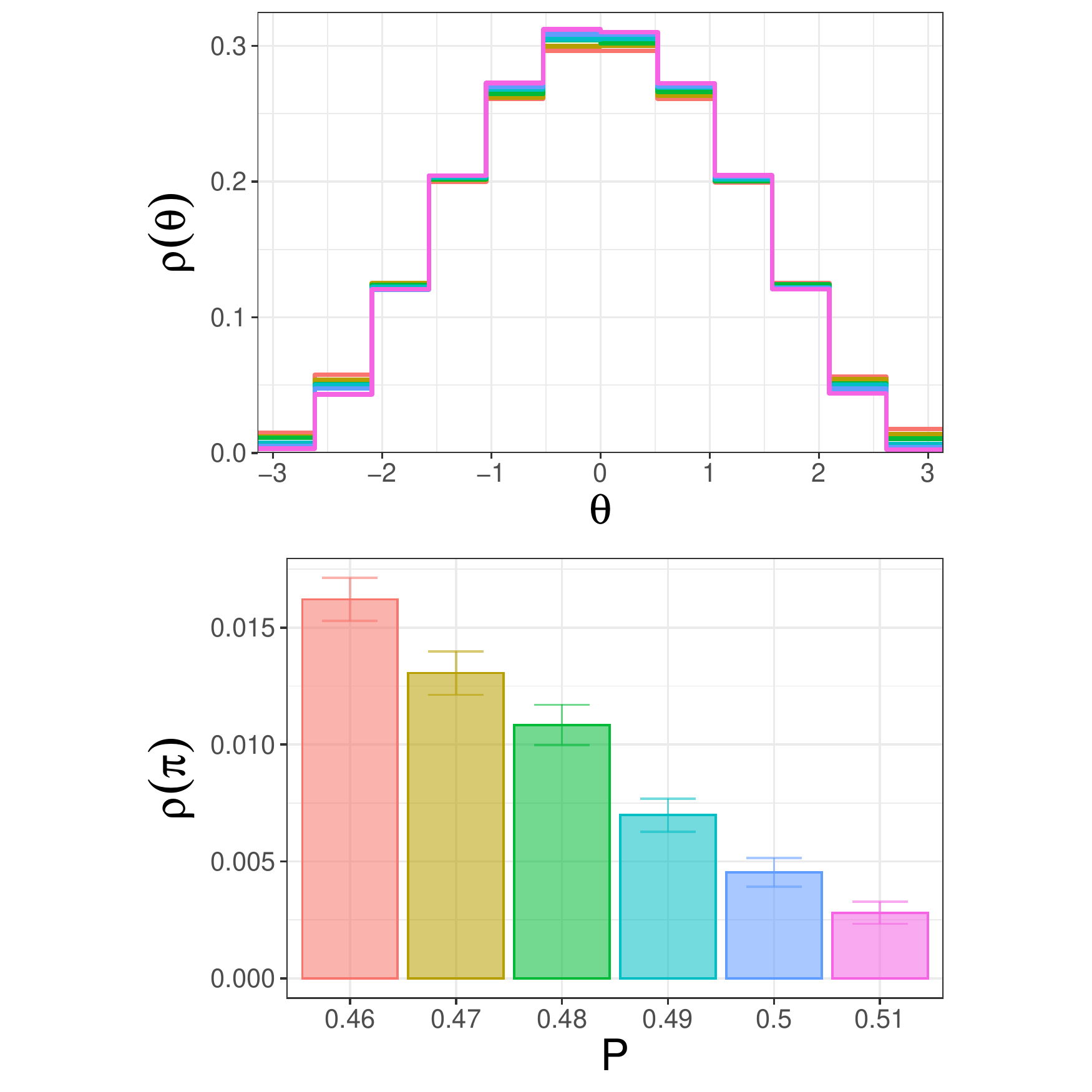}}
	}
	\end{center}
	\caption{Distribution of the phases of the Polyakov line near the transition temperatures for the bosonic BMN model. From left to right, $\mu=0.125, 0.25 ,0.5, 1.0$ with $T_c=0.8853, 0.8864, 0.8918, 0.9134$, respectively. The matrix size and the lattice size are $N=64$ and $L=24$, respectively. The gap opens at $P\simeq 0.5$. 
	}
	\label{fig:bosonic-BMN-phase-distribution}
\end{figure}

That the transition is of first order can be confirmed by observing the two-state signal in the Polyakov loop. 
Indeed, this is shown in Fig.~\ref{bosonic_bmn_lowmu}. 
We can see that 
the Polyakov loop jumps from $P\simeq \frac{1}{2}$ to $P\simeq 0$, at any $\mu$. (In Fig.~\ref{bosonic_bmn_lowmu}, the energy and Myers term are also shown. The two-state signal is not clearly visible for these quantities.)
We can see that
$P\simeq\frac{1}{2}$ corresponds to the GWW transition because the spectrum of the eigenvalues of the Polyakov line develops a gap as we can see at the phase distributions shown in Fig.~\ref{fig:bosonic-BMN-phase-distribution}.

\begin{figure}[htbp]
	\begin{center}
		\rotatebox{0}{
			\scalebox{0.35}{
				\includegraphics{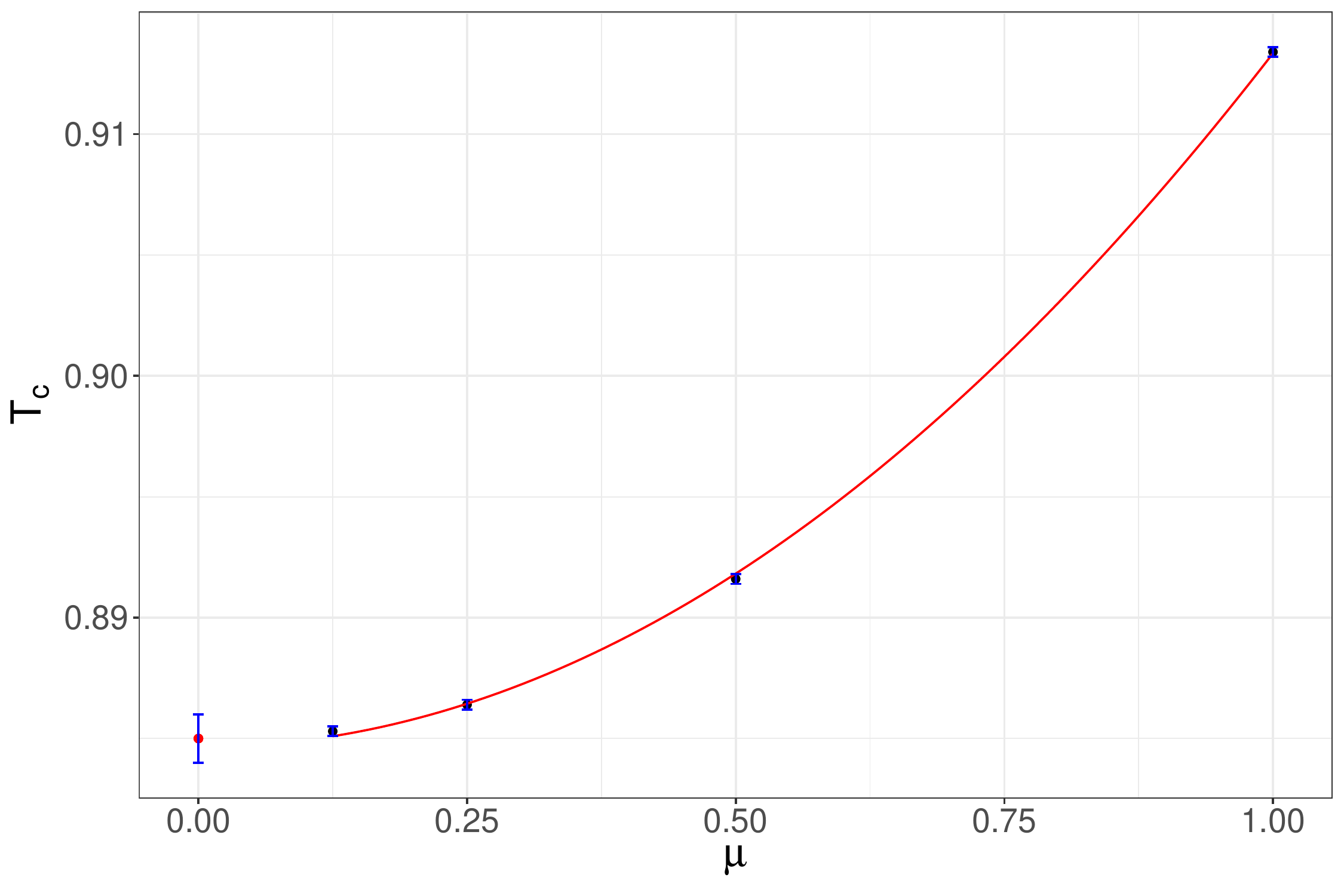}}
		}
	\end{center}
	\caption{Transition temperatures for the bosonic BMN model against  fluxes $\mu$. The parenthesis indicates the errors of the fit, and the left-most point is the extrapolated BFSS transition temperature (matrix size $N=64$, lattice size $L=24$) from \cite{Bergner:2019rca}. The fit is given by the equation $T_c=0.8846(01)+0.0297(02)\mu^2$.}
	\label{mu_extrapolation}
\end{figure}

The transition temperature, where the free energies of the confined and deconfined phases take the same value,
can be determined from the two-peak signal; this is the temperature at which two peaks contain the same number of configurations. 
The separation of two peaks is obtained by considering approximately same areas enclosed under two peaks. 
The numbers of trajectories in our simulations are large enough  to estimate the transition temperatures with errors of $0.02\%$. 
The critical temperatures determined for different $\mu$ in this manner are shown in Fig.~\ref{mu_extrapolation}. 
There is a symmetry $\mu\leftrightarrow -\mu$ and the $\mu$-dependence in $T_c$ is expected to be at least of quadratic order. $T_c$ can actually be well described by a quadratic fit.
The limit of $\mu\to 0$ is consistent with  the results for bosonic BFFS~\cite{Bergner:2019rca}, $\left.T_{c}\right|_{\mu=0}=0.885 \pm 0.001$ at $N=64, L=24$.

The dip in the histogram can be interpreted as the partially-deconfined phase, which sits at the maximum of the free energy in the canonical ensemble. At this point, $\rho(\theta)$ is expected to be nonzero everywhere between $-\pi$ and $+\pi$. This is consistent with the numerical observation: as mentioned above, if we look at configurations at each fixed $P$, then if $P<\frac{1}{2}$, $\rho(\theta)$ is nonzero everywhere between $-\pi$ and $+\pi$ as shown in Fig.~\ref{fig:bosonic-BMN-phase-distribution}.

The jumps of energy and entropy are of order $N^2$, at any $\mu$ (see Fig.~\ref{bosonic_bmn_lowmu}). The jump of energy goes down with $\mu$ but does not vanish even at $\mu=0$~\cite{Bergner:2019rca}. 
In the supersymmetric theory, on the other hand, we expect the jump to vanish at $\mu=0$, up to the $1/N$-suppressed corrections as we discuss
in Sec.~\ref{sec:conjecture_phase_diag} and in Fig.~\ref{fig:Pol-1st-order-scenario}.

\subsection{Phase transitions in the full BMN model}\label{sec:SUSY-simulation}

Our main interest is the thermal transition from confinement to deconfinement, while the configuration of the $X_i$ is kept in the trivial background with no relevant fuzzy sphere contributions.\footnote{Trivial (fuzzy-sphere) background means fluctuations around a configuration of the trivial (fuzzy-sphere) vacuum.} This is due to the fact that this scenario is assumed in the dual gravity predictions.
We want to verify a continuous transition line from perturbative predictions at weak coupling to the conjectures at strong couplings obtained with help from the dual gravity description.

A phase transition is a well-defined concept in the thermodynamic limit, which is the large-$N$ limit in the case of the matrix models.
In our numerical studies at finite $N$ and finite simulation time, several ambiguities concerning the exact definition of the transition have to be taken into account.

The first one is related to the common difficulty in identifying a first order transition.
In the ideal world, we would be able to study very large $N$ efficiently.
If $N$ were sufficiently large such that the separation of two phases is clear, and at the same time, the number of configurations collected in the simulation would be very large such that the tunneling between two phases can be captured, then the two-state signal would be very clean.
In this idealized case it would be possible to determine $T_c$, $T_1$, and $T_2$ precisely.
For the bosonic BMN model, we can actually perform such an analysis. However, for the full BMN model, such a detailed study is not possible for all parameter sets because of large simulation cost.
Compared to the bosonic BMN model, we have fewer data points, and the $1/N$-corrections are larger. Still, the large-$N$ behavior can be deduced by comparing the results for different values of $N$.

The second complication arises at finite $N$ due to tunneling to fuzzy-sphere configurations.
As we explained in Sec.~\ref{sec:conjecture_phase_diag}, there can be confinement or deconfinement with either a fuzzy-sphere or a trivial background of the $X_i$ fields.
At small values of $\mu$ or $N$, tunneling between different backgrounds can take place with a non-negligible rate.
At larger $\mu$, the tunneling is well suppressed even at the values of $N$ we study in this work, and we observe the trivial-confinement/trivial-deconfinement transition. However, at smaller $\mu$, deconfinement is associated with the transition to a fuzzy-sphere background.
Let $\tilde{T}_{\rm c}$ be the critical temperature at which the transition between the trivial-confined phase and fuzzy-sphere-deconfined phase is observed. If we do not see the transition to the trivial-deconfined phase, then we expect $\tilde{T}_{\rm c}\le T_{\rm c}$. $T_{\rm c}$ is the point where the free energies of the trivial-confined and trivial-deconfined phases coincide as explained in Sec.~\ref{sec:trivial-conf-fuzzy-sphere-deconf}.

We have tested different strategies to disentangle the influence of the tunneling to a fuzzy-sphere background and to obtain more information about the deconfinement transition in the trivial background. One strategy corresponds to constraining the Myers term, which forces the system to stay in trivial background.\footnote{
The full path integral contains multiple backgrounds, and we are interested in a particular background. Therefore, we have to restrict the path integral to the fluctuation around such a background, to obtain physically meaningful result~\cite{Banks:1996vh,Anagnostopoulos:2007fw,Hanada:2013rga}.
In this paper, we identify the background we are interested by specifying the value of the Myers term. See Sec.~\ref{sec:simsetup} for more details.
} Alternatively we have also investigated indications of metastability related to hysteresis effects. These investigations will be detailed below.

Our investigation led us to some expected and some unexpected observations, depending on the parameter region. The results are summarized in Fig.~\ref{fig:tc_mu}. (A complete list of simulation points is given in Appendix~\ref{appendix:simulation_parameters}, with short explanations regarding the simulation setups and outcomes.)
We split the parameter space into four regions:
\begin{itemize}
	\item
	For $\mu\ge 2.0$, the transition between the trivial-confined and trivial-deconfined phases can be easily identified.
	We observe convergence to the perturbative limit as $\mu$ becomes large.
	At intermediate $\mu$, the Pad\'e resummation based on the large- and small-$\mu$ behaviors obtained via perturbative calculation and dual gravity analysis gives a reasonable approximation to the simulation results.
	(See Refs.~\cite{Asano:2018nol,Schaich:2020ubh} for similar observations.)

	\item
	For $0.8\le\mu<2.0$, convergence to the dual gravity prediction is observed as $\mu$ becomes small. Below $\mu=0.8$, the deviations from the predicted line start to increase again.

	\item
	For $\mu< 1.6$, transitions to fuzzy-sphere configurations are observed frequently (see Ref.~\cite{Asano:2018nol} for a similar observation), which makes it in some regions impossible to observe the trivial-deconfined phase.
	In these cases, deconfinement and tunneling to a fuzzy-sphere background take place simultaneously, at least at the values of $N$ and $L$ we have studied.

	\item
	For $\mu< 0.8$, we observe pronounced metastable confined and deconfined phases. The system tends to remain in one of the phases for a longer simulation time, at least in the range of temperatures $T\in[0.26,0.27]$. This effect is independent of any constraint of the Myers term and even appears for fuzzy-sphere background. We observe indications that the metastable confined state persists even in the BFSS limit.
As we will see, this apparent deviation from the gravity prediction may be simply a finite-$N$ effect.

\end{itemize}
\begin{figure}[htbp]
	\begin{center}
		\rotatebox{0}{
			\scalebox{0.6}{
				\includegraphics{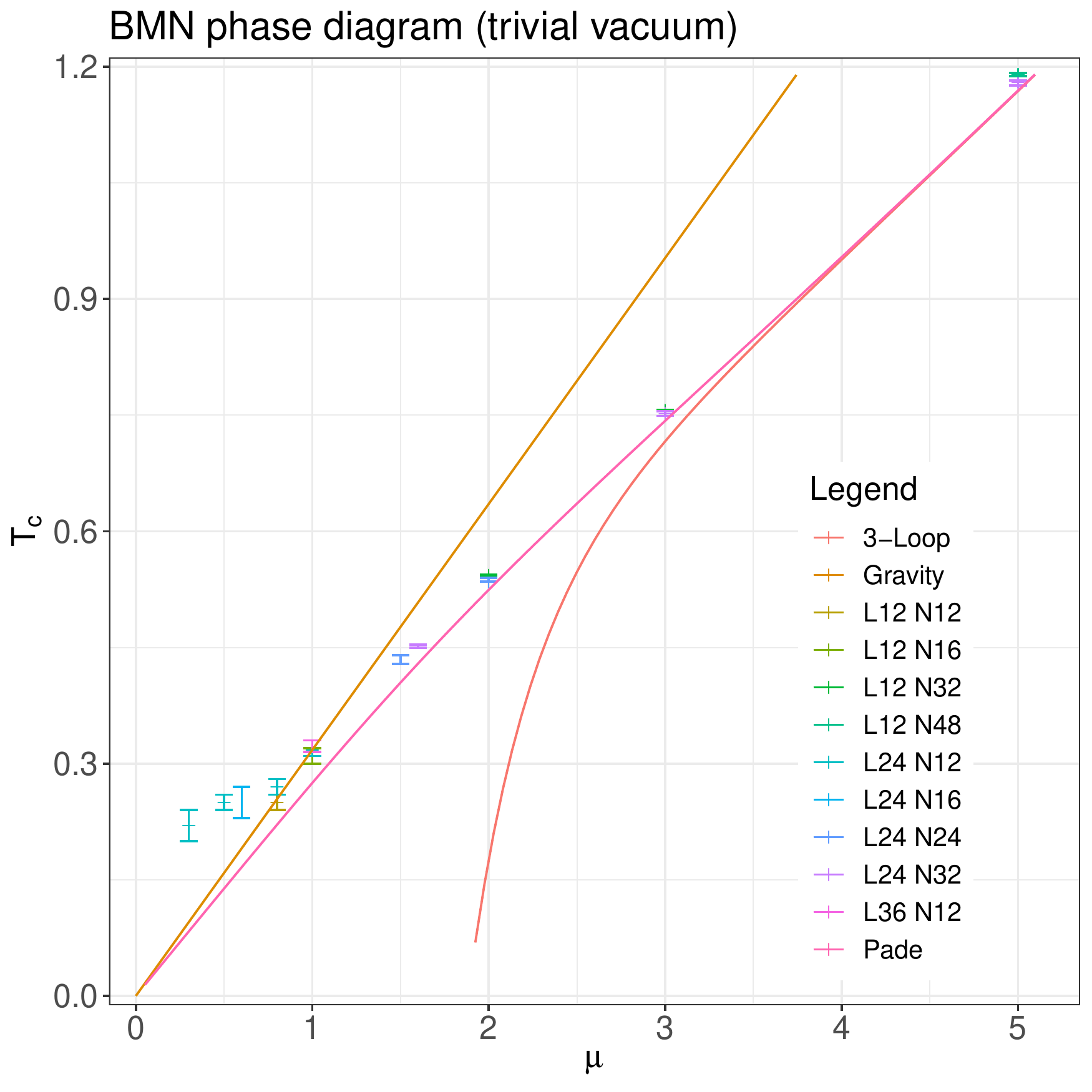}}}
	\end{center}
	\caption{
		The critical temperature $T_c$ vs. flux parameter $\mu$ for the full BMN model. The data points for the critical temperature correspond to a two-state signal with about equal probability for confined and deconfined phase, and their error bars are given by the neighboring simulated temperatures, for which these probabilities differ. The error bars without data points for $\mu=0.6$ indicate observation of a hysteresis, see Fig. \ref{Fig:mu06T1T2}. For $\mu=1.5$ and $1.6$, we also observed hystereses in the region given by the error bars but did not investigate the full extent of the hysteresis region. For comparison, the prediction of the perturbative analysis~\cite{Furuuchi:2003sy}, Pad\'e resummation~\cite{Asano:2018nol}, and gravity calculation~\cite{Costa:2014wya} are plotted.
In the main text, the way the transition temperature at each parameter set is determined is detailed, and the apparent deviation from the gravity predictions at small $\mu$ is explained. (Note that the latter can simply be a finite-$N$ artifact.)
In the BFSS limit $\mu\to 0$, although we could not identify the two-state signal, we still conclude that we observe the confined phase; see Sec.~\ref{sec:BFSSLimit}.
	}\label{fig:tc_mu}
\end{figure}

\subsubsection{Simulation setup}
\label{sec:simsetup}
We use the same simulation program as done in earlier simulations of the BFSS model~\cite{Berkowitz:2016jlq}. It is based on the RHMC algorithm~\cite{Clark:2003na}, neglecting the phase of the Pfaffian based on the arguments presented already in the BFSS case~\cite{Anagnostopoulos:2007fw,Catterall:2008yz}. The program includes a possible constraint for $R^2\equiv\frac{1}{N}\sum_{I=1}^9 \Tr X_I^2$, which has been used in some simulations of the BFSS model to stabilize the trivial background in Ref.~\cite{Berkowitz:2016jlq}, but not in the current paper. We have added in the same way constraints for the Myers term and the Polyakov loop. (See Ref.~\cite{Watanabe:2020ufk} as an example for a simulation with a constraint on $P$.)
Each constraint forces an observable $\mathcal{O}$ to lie in the interval $[x_{\rm min},x_{\rm max}]$ around a given value $x$ depending on the coupling strength $\gamma$. Inside the chosen interval the constraint has no effect. Expressed in terms of the Heaviside function $\Theta$, the additional contribution to the action is
\begin{align}
S_{\rm constraint}(x;x_{\rm max}, x_{\rm min},\gamma)=\left\{
\begin{array}{cc}
\gamma\cdot(\mathcal{O}-x_{\rm min})^2 & (\mathcal{O} <  x_{\rm min})\\
0 & (x_{\rm min}\le\mathcal{O}\le x_{\rm max})
\\
\gamma\cdot(\mathcal{O}-x_{\rm max})^2 & (x_{\rm max} < \mathcal{O})
\end{array}
\right.
\end{align}
where $\gamma$ is a sufficiently large positive number. Consequently, the value of $\mathcal{O}$ is approximately constrained between $x_{\rm max}$ and $x_{\rm min}$.
We have added this constraint contribution to the action in some of the simulations to stay as close as possible to the trivial background in certain parameter ranges. It prohibits tunneling to nontrivial backgrounds, which is expected to be suppressed in the large-$N$ limit, even at small $N$. We will discuss in detail in which cases the constraints have been included.

In all our simulations, we have kept the trajectory length of the molecular dynamics constant and adjusted the stepsize to achieve a similar acceptance rate. Statements about tunneling time and probability are hence always in reference to this trajectory length.

In addition to a transition indicated by the Myers term, we observe a run-away behavior of the scalar fields in the BFSS limit, remnants of which show up already at small $\mu$.  We monitored $\Tr X_I^2$ for each $I=1,2,\cdots,9$ to keep it under control.
We observed that this effect appears preferably in the $X^9$-direction, presumably due to a lattice artifact specific to our regularization.
The run-away behavior is generally enhanced by lattice artifacts and consequently reduced in the large $L$ limit.

\subsubsection{Convergence to perturbative limit at $\mu\ge 2.0$}
\begin{figure}[htbp]
	\centering
		\rotatebox{0}{
			\scalebox{0.25}{
				\includegraphics[trim={1cm 0cm 0cm 0},clip]{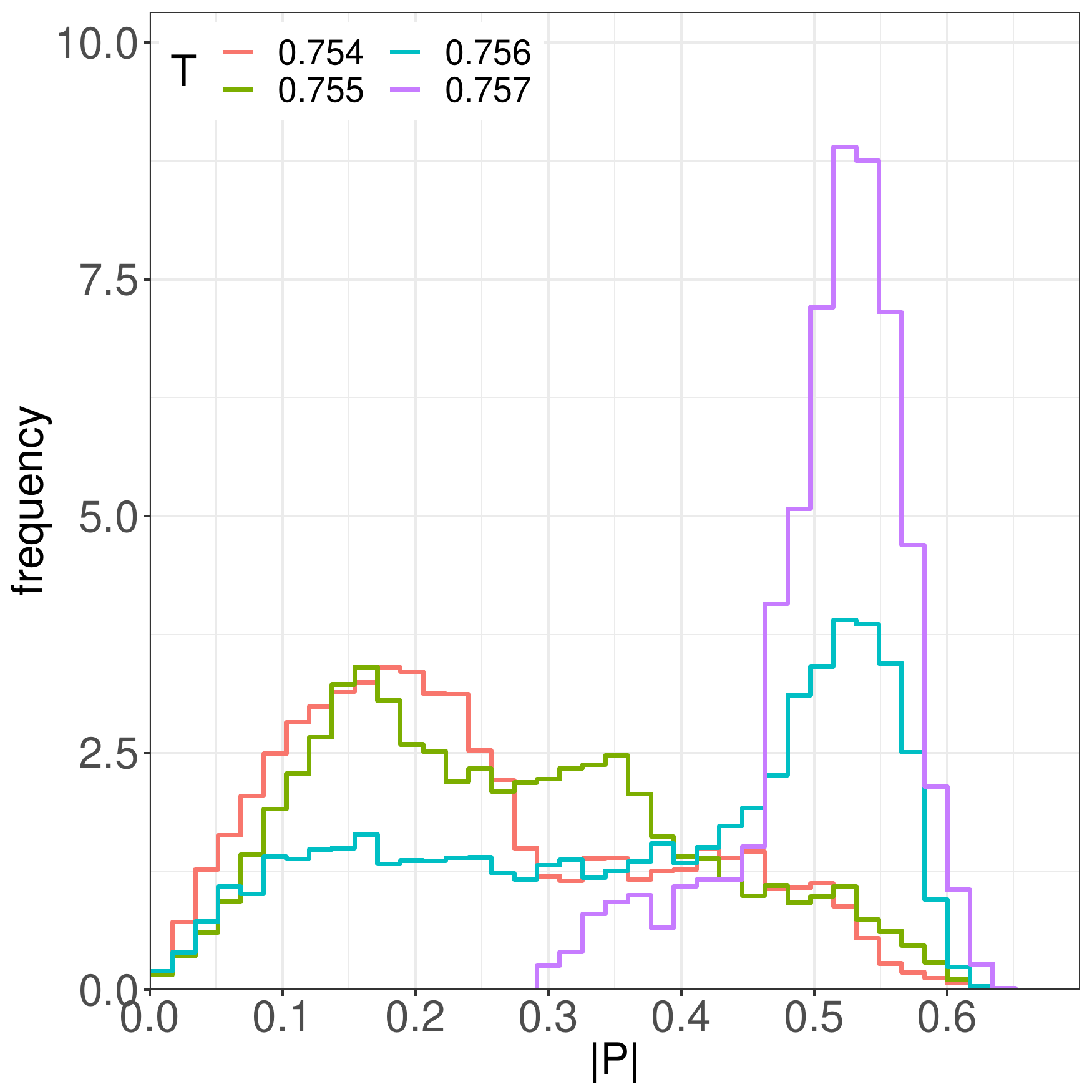}}}
	\rotatebox{0}{
	\scalebox{0.25}{
\includegraphics{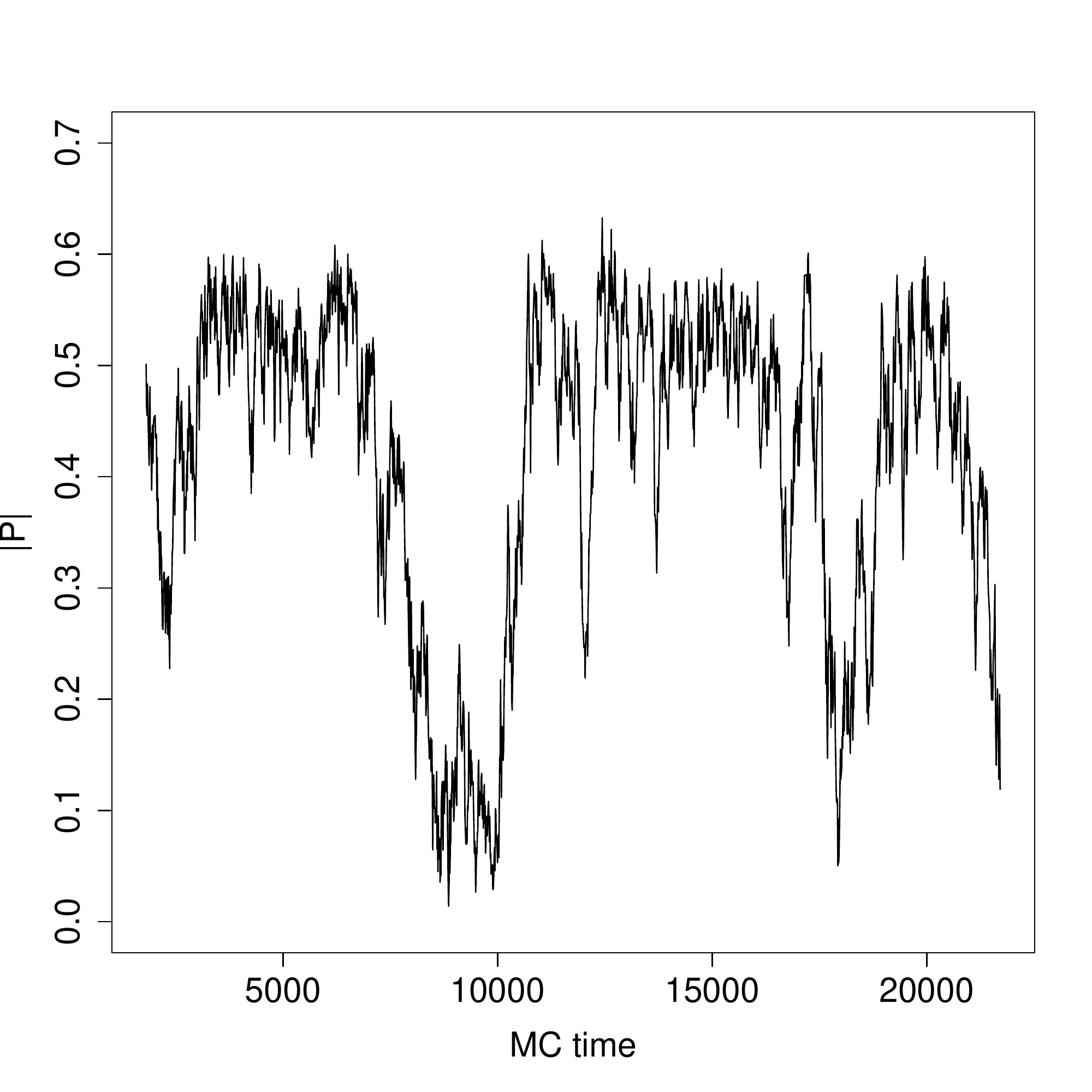}}}
\rotatebox{0}{
	\scalebox{0.25}{		\includegraphics{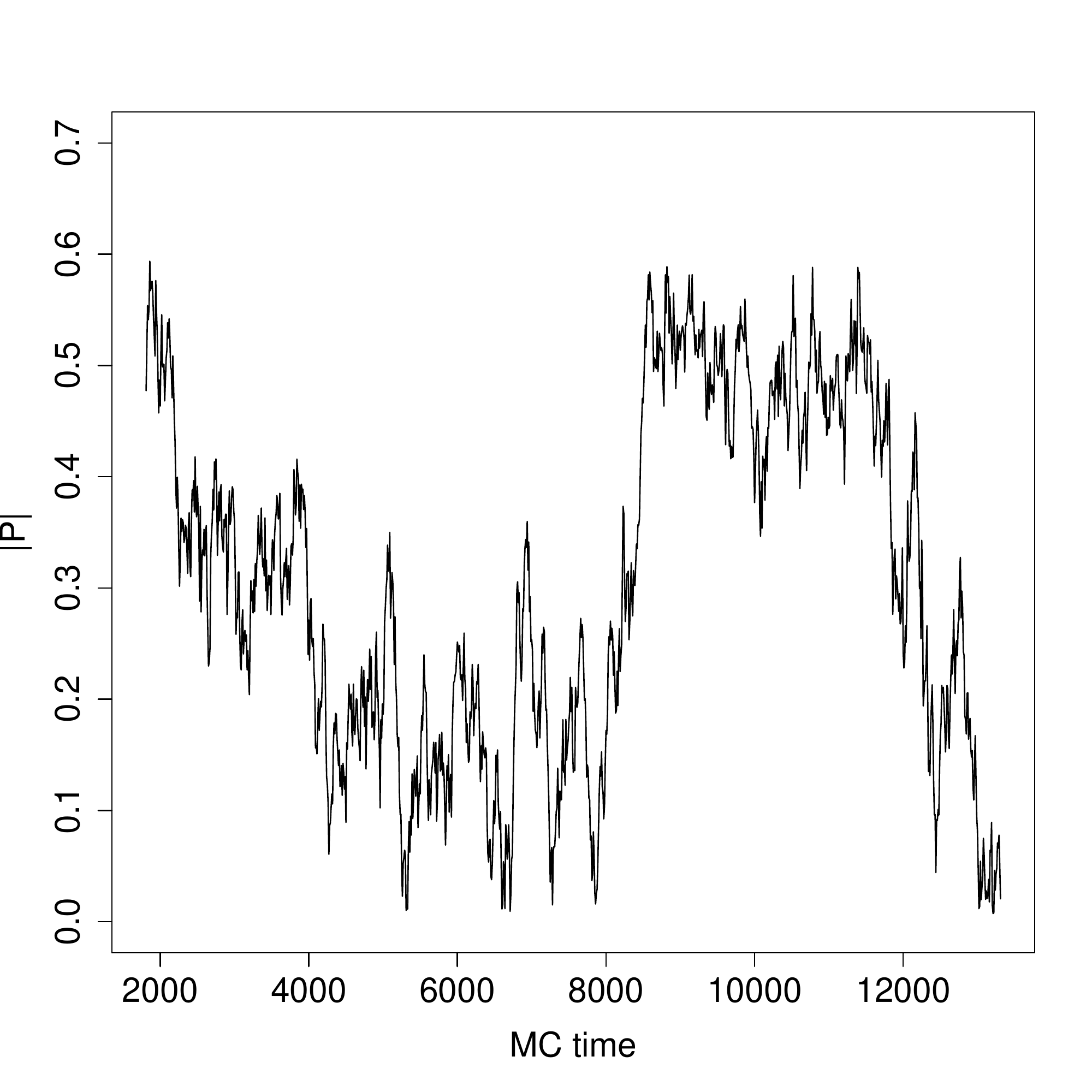}}}
			\rotatebox{0}{
			\scalebox{0.25}{		\includegraphics{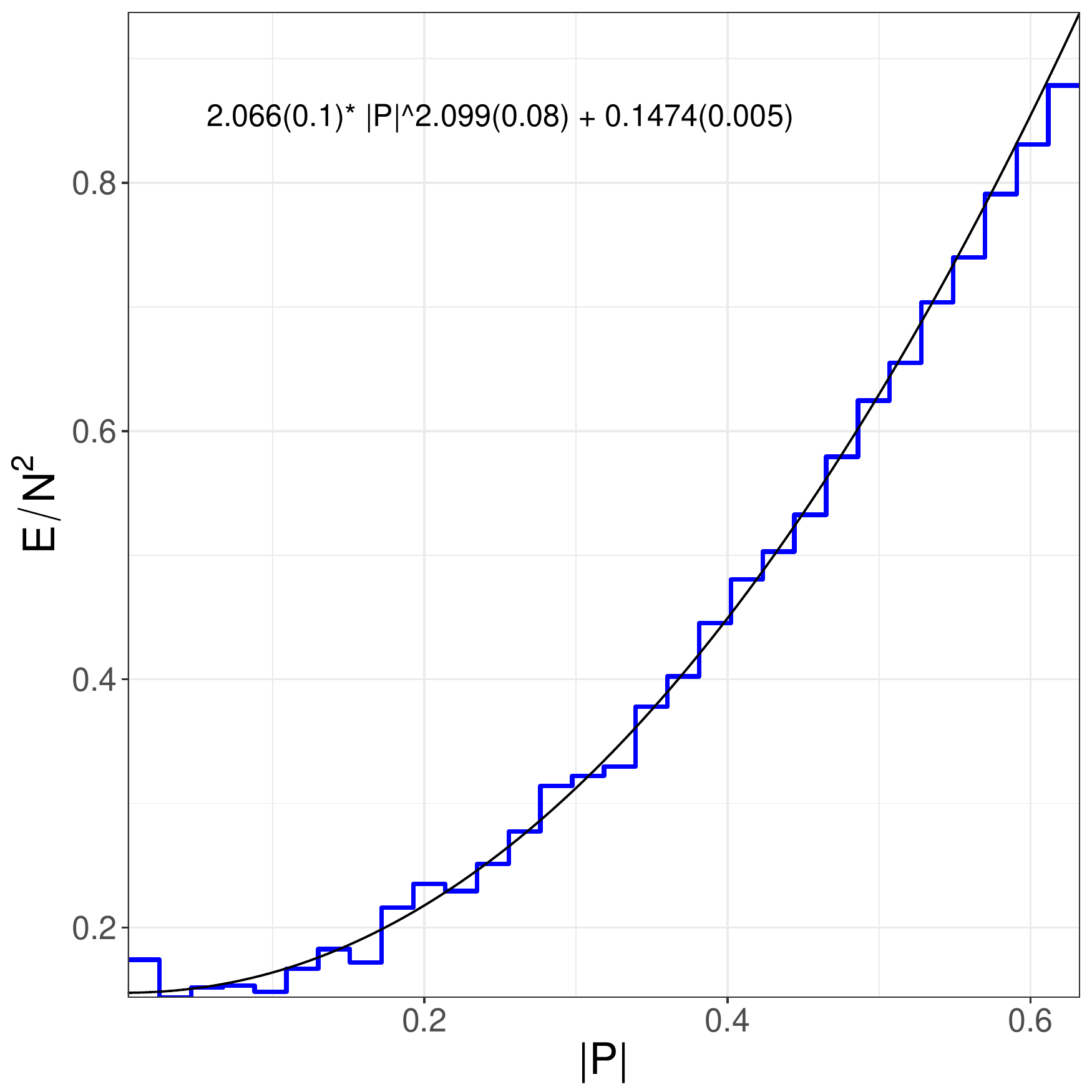}}}
		\rotatebox{0}{
			\scalebox{0.25}{
\includegraphics{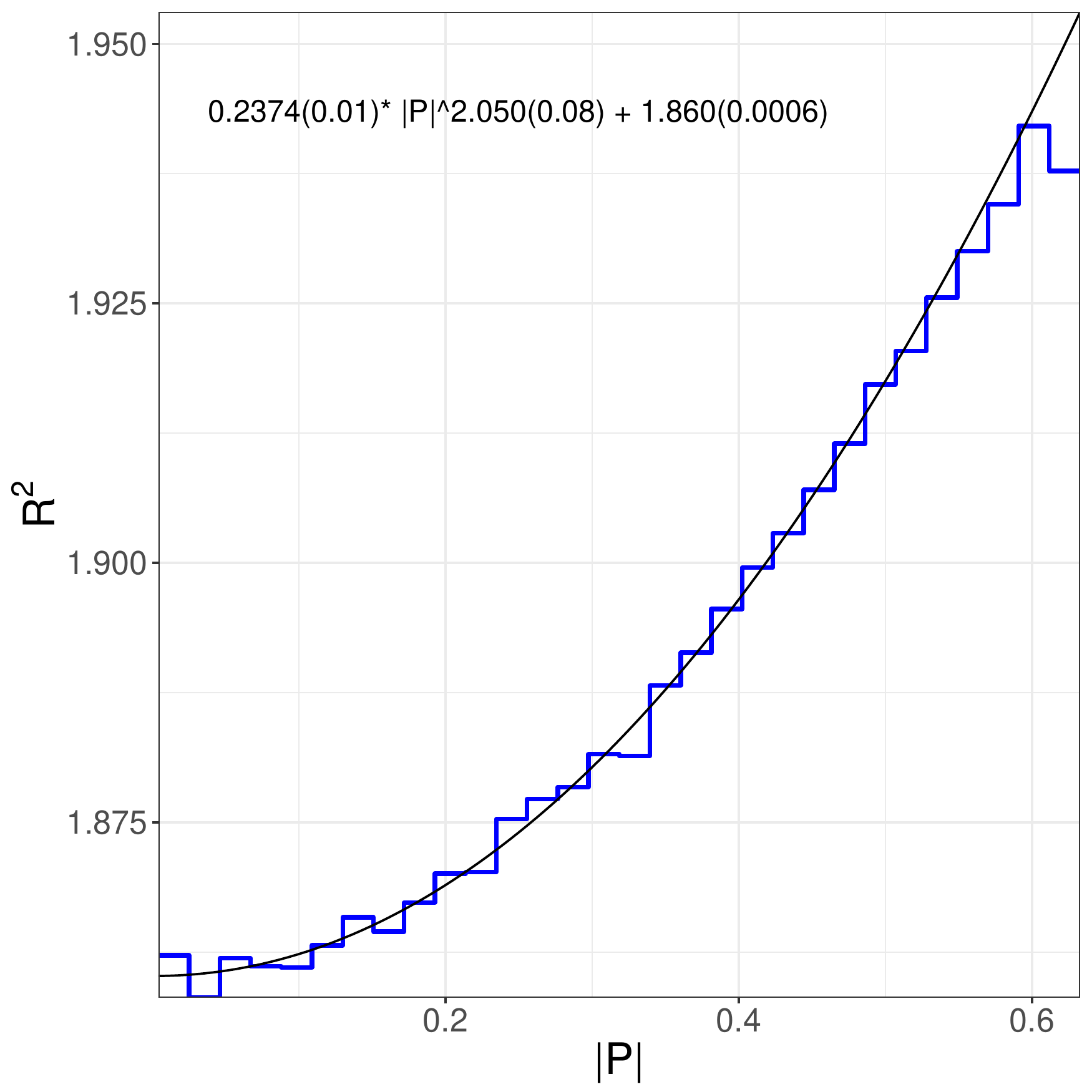}}}
	\caption{Full BMN model, flux $\mu=3.0$, matrix size $N=32$, lattice size $L=12$ at temperature $T=0.756$. [Top] From left to right, histogram of $|P|$ close to the critical temperature, and Monte Carlo history for the Polyakov loop at the same parameters for two typical runs with different sequences of random numbers.  A two-state signal is observed.
	[Bottom] Binned $E$ vs $|P|$ and binned $R^2$ vs $|P|$. Larger values of $|P|$ (more deconfined) corresponds to larger values of $E$ and $R^2$, as expected.
		}\label{fig:PDM3}
\end{figure}

In this region, the transition between the trivial-confined and trivial-deconfined phases can easily be identified.
We observe convergence to the perturbative limit as $\mu$ becomes large, as shown in Fig.~\ref{fig:tc_mu}. The tunneling to a nontrivial background close to $T_c$ is not relevant and we are able to determine $T_c$ precisely from a two state signal in the Polyakov loop. This indicates a first order phase transition, as shown in Fig.~\ref{fig:PDM3}. By monitoring ${\rm Tr}X_I^2$ and ${\rm Tr}X_1[X_2,X_3]$, we can confirm that the two peaks are both consistent with a trivial background. Due to the stability of the results, no constraint part has to be added to the action.
The phase transition takes place at a rather high temperature, where the effect from the fermions is not so important.
Therefore, the transition temperatures become similar to that of the bosonic theory.

\begin{figure}[htbp]
	\begin{center}
		\scalebox{0.5}{
			\includegraphics[trim={4cm 0cm 4cm 0},clip]{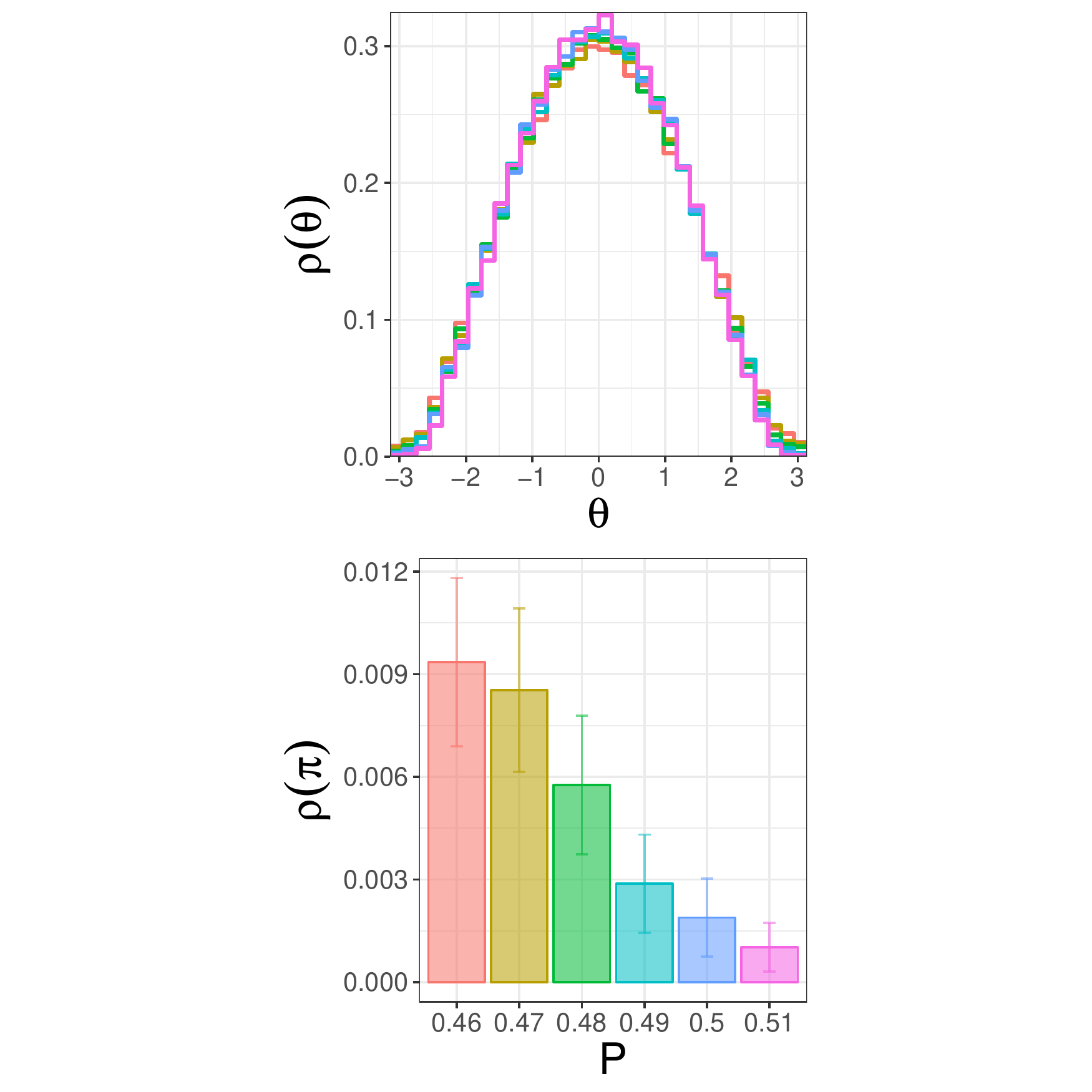}}
		~~~~~~~~~~~~~~
		\scalebox{0.5}{
			\includegraphics[trim={4cm 0cm 4cm 0},clip]{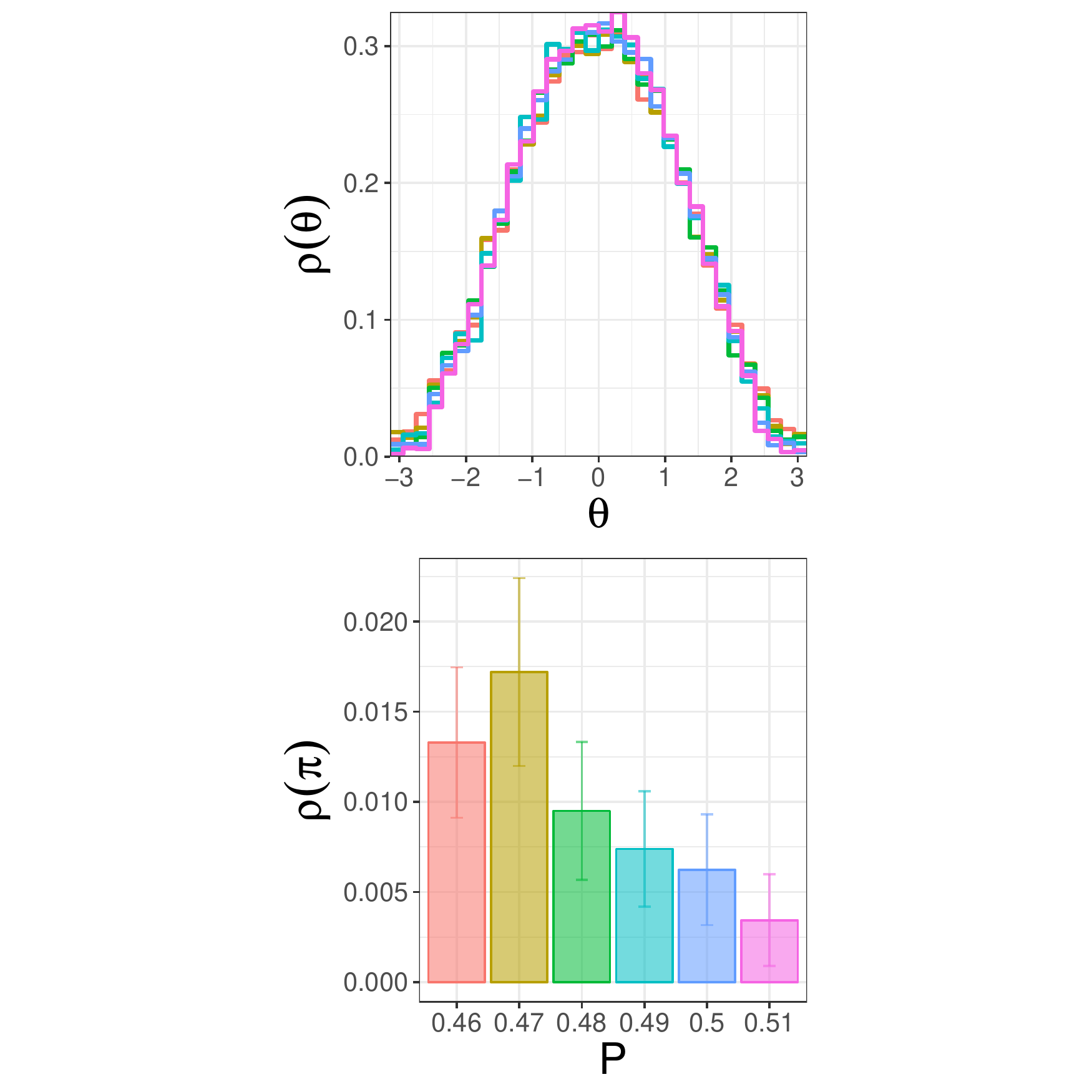}}
	\end{center}
	\caption{Full BMN model:
		[Top] The histogram of the Polyakov line phases for $[P-\Delta P,P+\Delta P]$, where $\Delta P=0.005$ and $P=0.46, 0.47, \cdots,0.51$.
		[Bottom] $\rho(\pi)$ or more precisely, the height of the right-most bin in the histogram.
		Top and bottom rows use the same color for the same value of $P$.
		[Left] $T=0.756$ (close to $T_1$) for $\mu=3.0$, $N=32$, $L=12$. The Myers term was not constrained in this simulation.
[Right] $T=1.18$ (close to $T_1$) for $\mu=5.0$, $N=32$, $L=24$.
	}\label{fig:phase_bins_N32L12M3M5}
\end{figure}

\begin{figure}[htbp]
	\begin{center}
		\scalebox{0.4}{
			\includegraphics[trim={0 0 0 0},scale=1.0, clip]{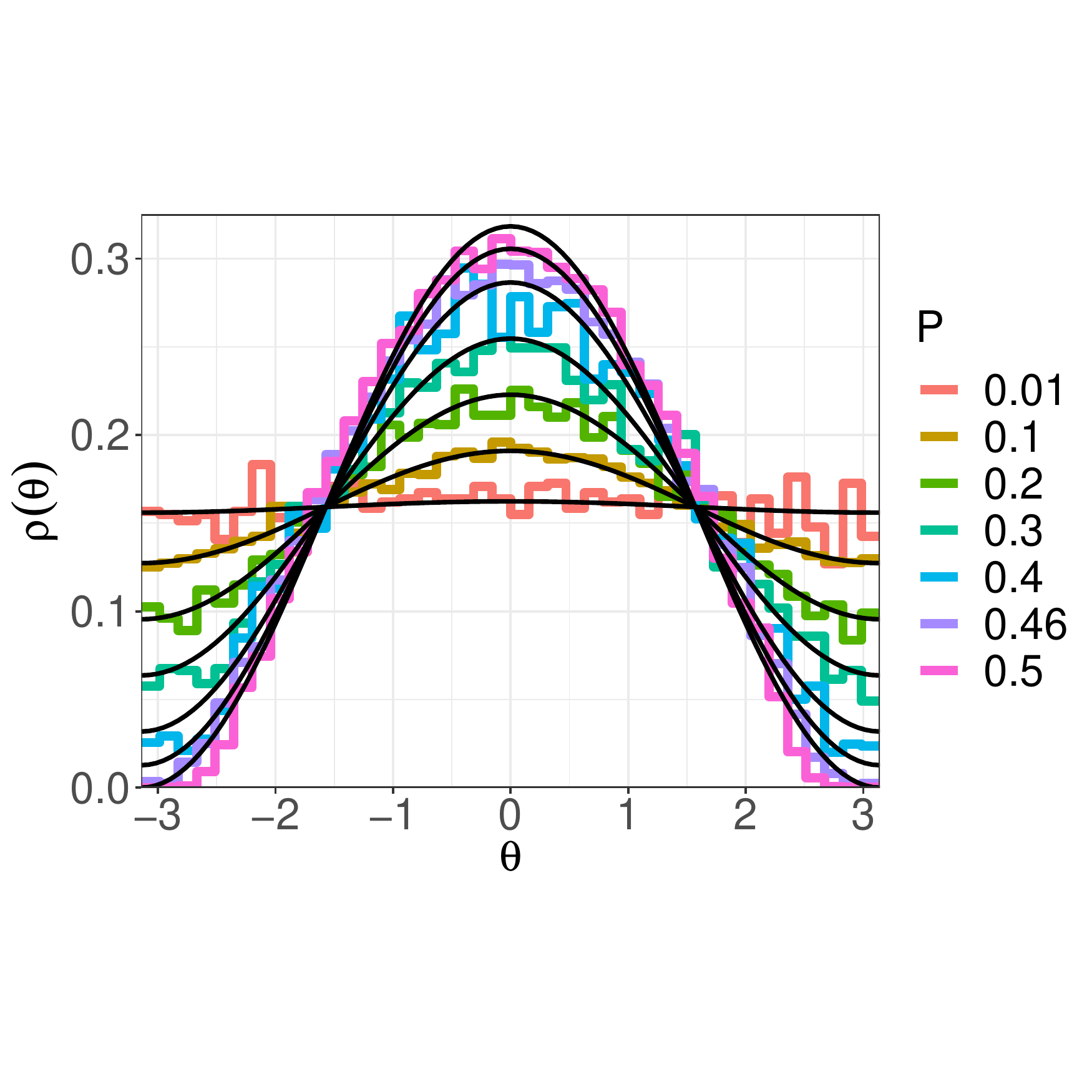}}
		\scalebox{0.4}{
			\includegraphics[trim={0 0cm 0 0cm},clip]{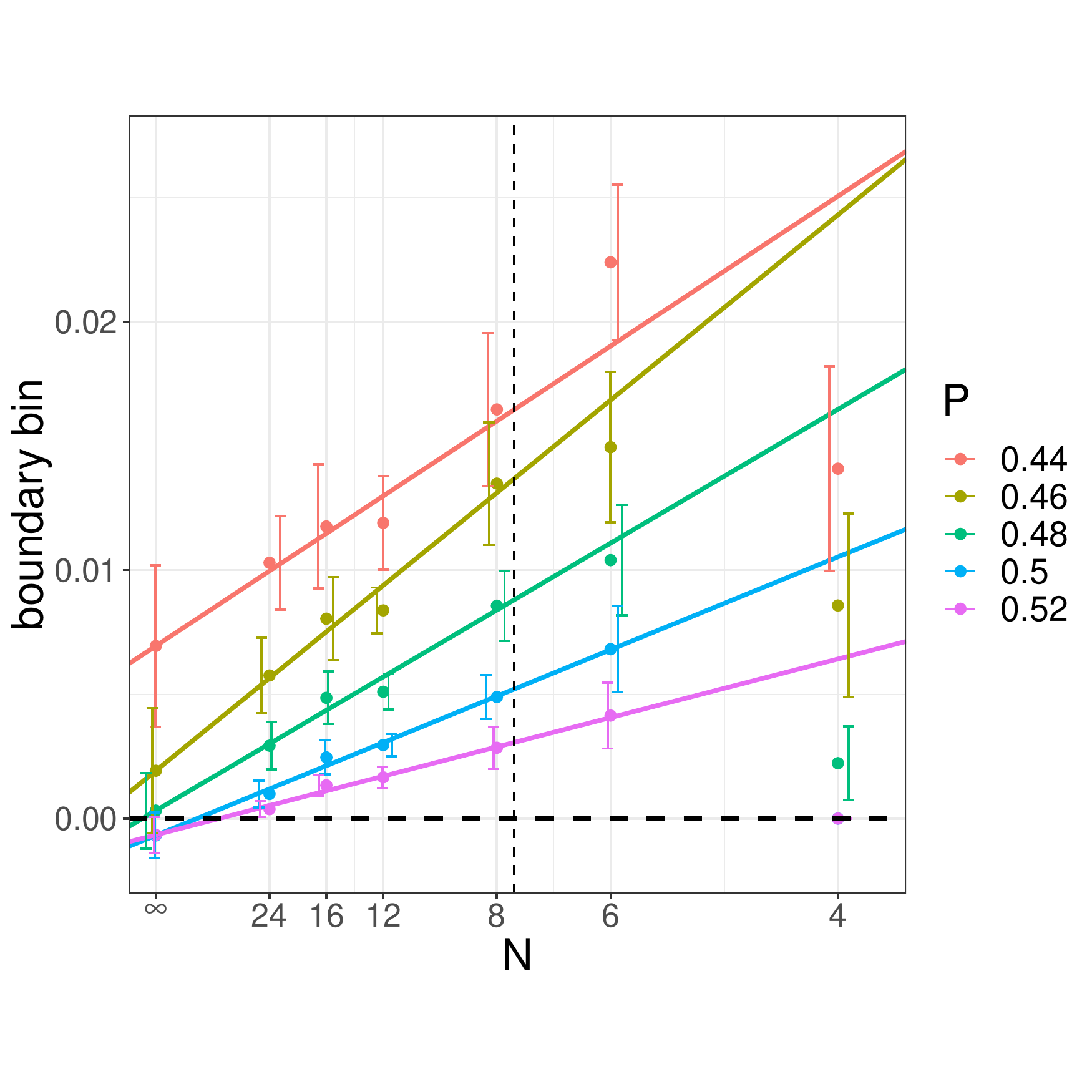}}
	\end{center}
	\caption{Full BMN model:
	[Left] Histogram of the Polyakov line phases in the interval $[P-\Delta P,P+\Delta P]$, for $\Delta P=0.01$ and various values of $P$, for $\mu=2.0$, $T=0.543$ (close to $T_1$), matrix size $N=32$, and lattice size $L=12$.
	 The large $\mu$ behavior $\rho(\theta)=\frac{1+2P\cos\theta}{2\pi}$ is shown in black. A clear deviation is observed close to $P=0.5$. It appears that the GWW-point is between $P=0.4$ and $P=0.46$.
		The Myers term was not constrained in this simulation.
[Right] Large $N$ extrapolation of the size of the boundary bin for $L=12$ and $\Delta P=0.01$. The GWW transition is seen to take place around $P=0.46$. Only data points left of the dashed vertical line were included in the linear fit.
	}\label{fig:mu2distortion}
\end{figure}

We can also check more detailed features of the deconfinement transition and confirm the picture that we have already investigated in the bosonic BFSS model \cite{Bergner:2019rca}: for each fixed $T$ and $\mu$ in the first-order-transition region, the energy and $\langle{\rm Tr}X^2\rangle$ depend on $P$
as $a+bP^2$. For large $\mu$, the distribution of the phases of the Polyakov line is consistent with $\rho(\theta)=\frac{1+2P\cos\theta}{2\pi}$,
and the GWW transition takes place at around $P=\frac{1}{2}$; see Fig.~\ref{fig:phase_bins_N32L12M3M5}. However, already at $\mu=2.0$, we can see a subtle but clear deviation from the large $\mu$ behavior: the GWW transition takes place slightly below $P=1/2$, and the distribution of the phases of the Polyakov line becomes distorted, as we can see from Fig.~\ref{fig:mu2distortion}.

\subsubsection{Convergence to dual gravity prediction at $0.8\le\mu< 2.0$}
\begin{figure}[htbp]
	\centering
	\begin{subfigure}{.25\textwidth}
		\rotatebox{0}{
			\scalebox{0.3}{
				\includegraphics{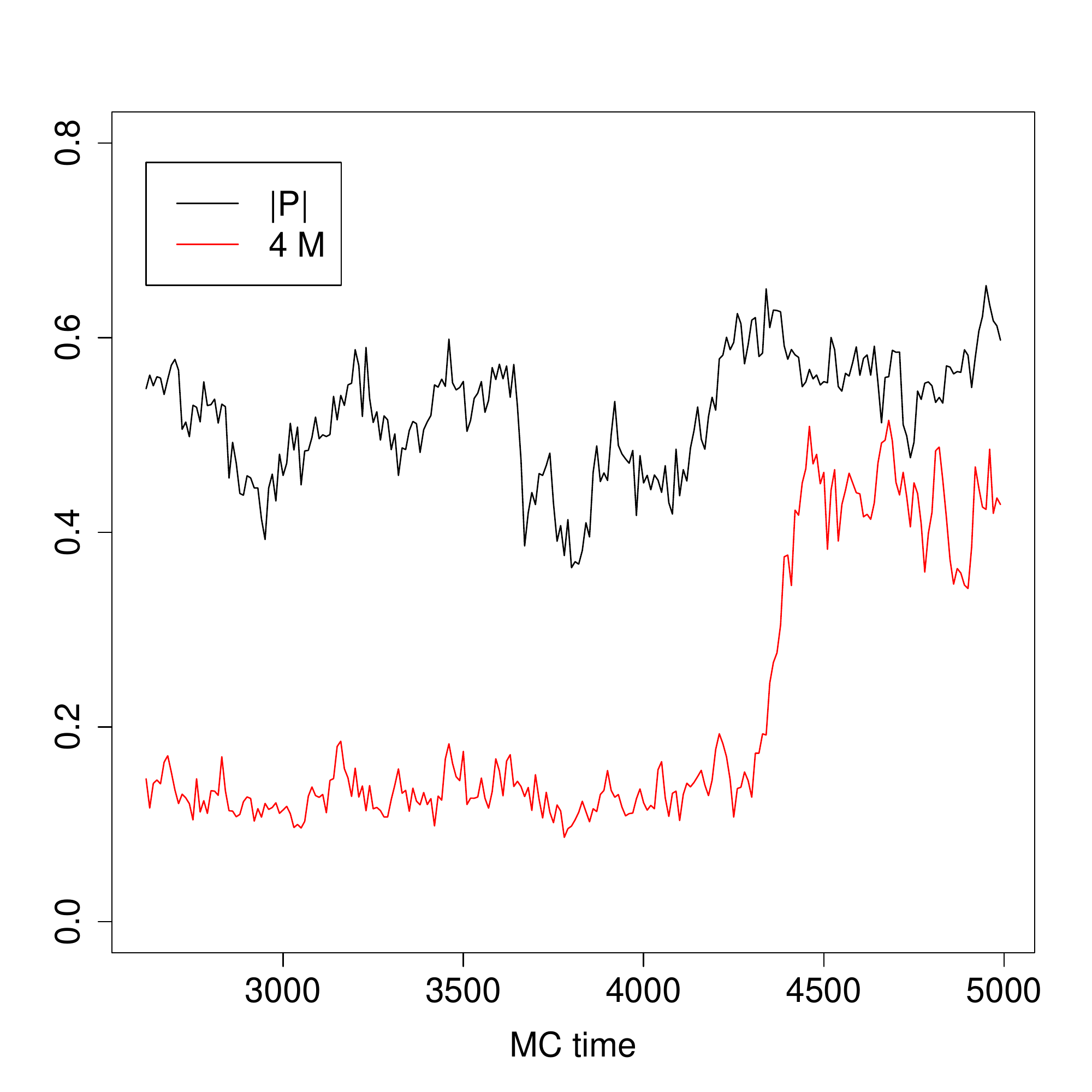}}}
		\caption{\mbox{} }
	\end{subfigure}
	\caption{Full BMN model: Monte Carlo history of $|P|$ and $M$ for $N=24$, $L=24$, $\mu = 1.5$, $T=0.43$. The system is initially in the deconfined trivial background and tunnels to a fuzzy-sphere background around trajectory 4500. }\label{fig:M15-instability}
\end{figure}

\begin{figure}[htbp]
	\begin{center}
		\scalebox{0.3}{
			\includegraphics[trim={0 0cm 0 0},clip]{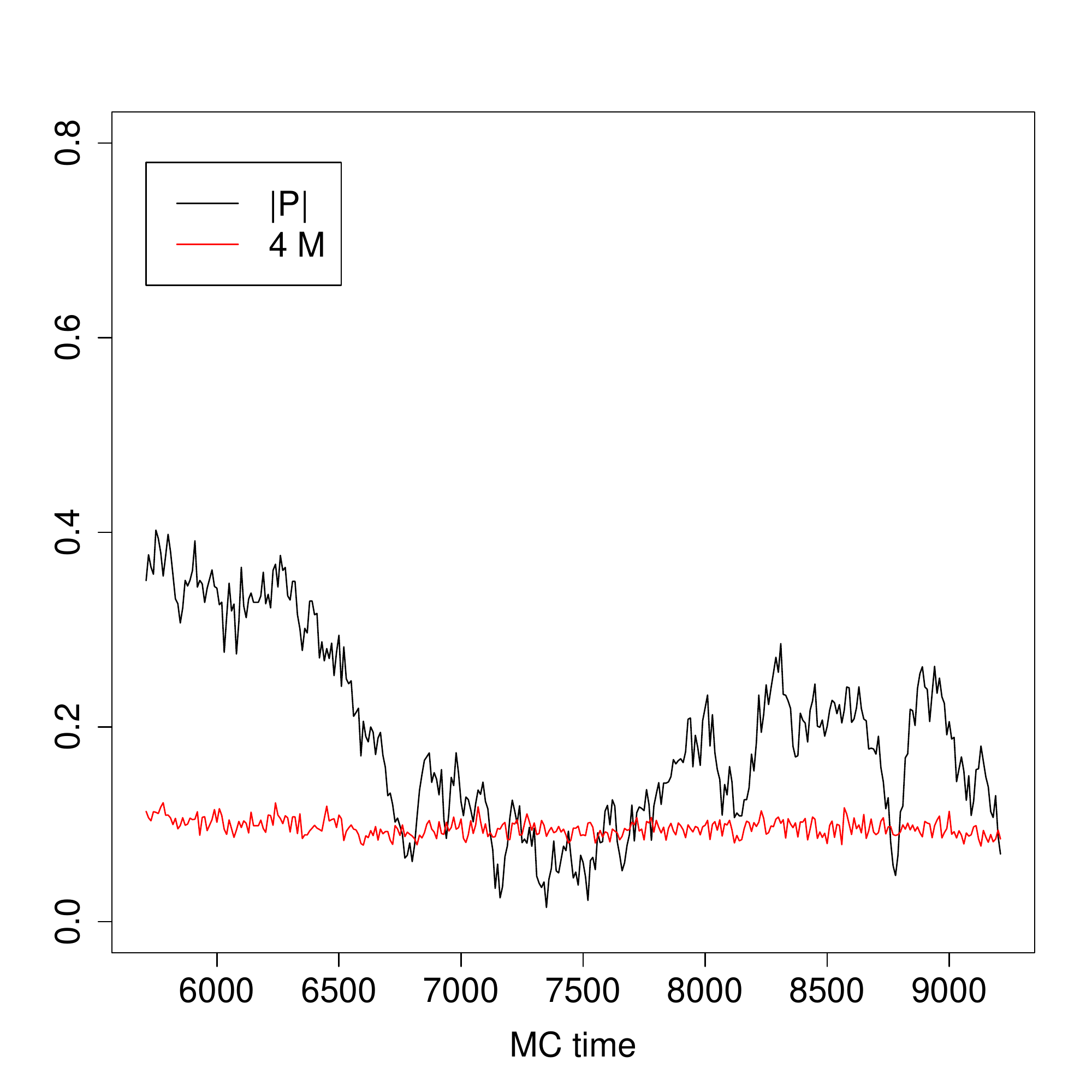}}
		\scalebox{0.3}{
			\includegraphics[trim={0 0cm 0 0},clip]{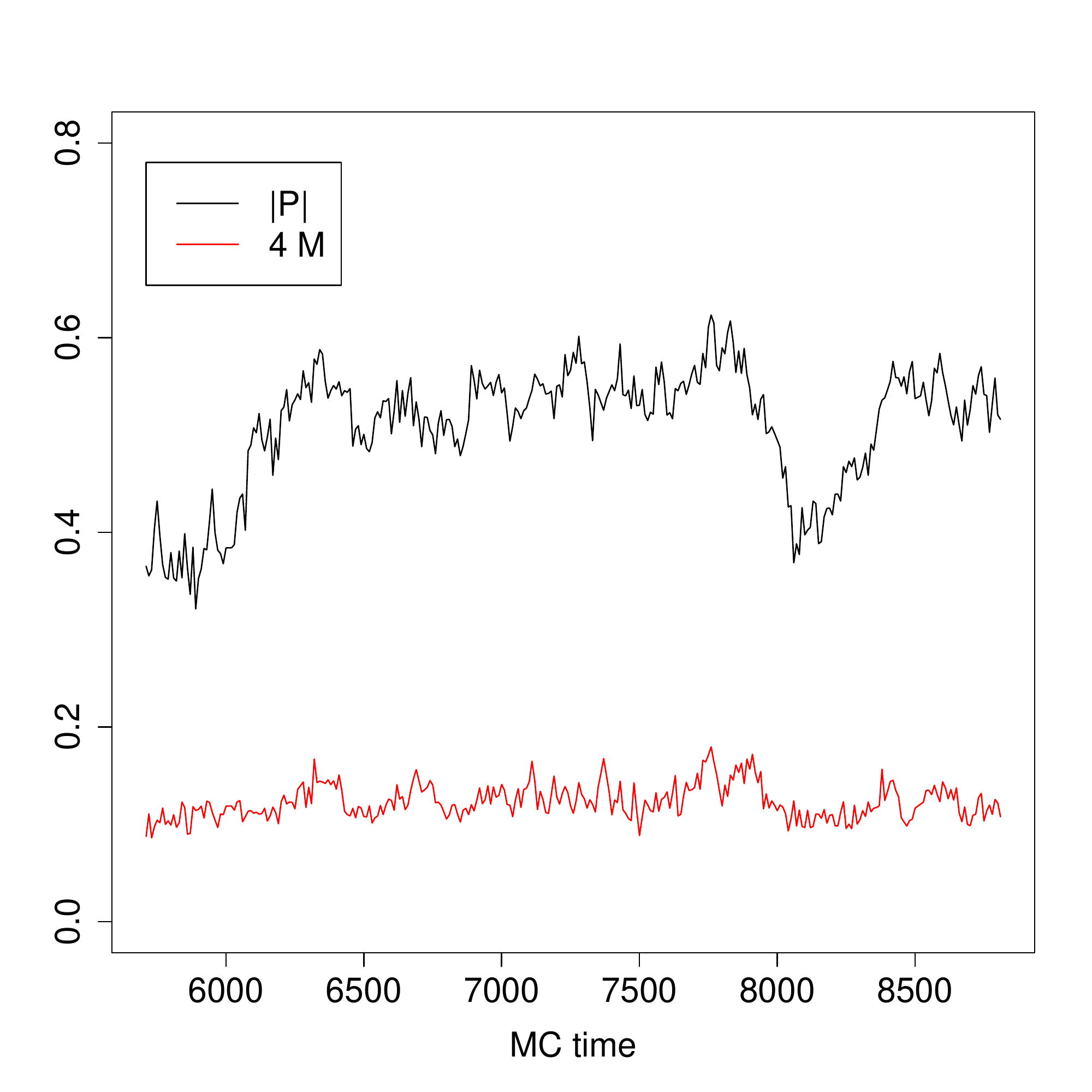}}
	\end{center}
	\caption{
		Full BMN model: $\mu=1.6$, $N=32$, $L=24$, $T=0.452$. Starting from the same initial configuration, a confined and deconfined stream is obtained from two different random number generator seeds. For these runs, no constraint on the Myers term is imposed and the simulation remained in the trivial background.
	}\label{BMN_mu1.6}
\end{figure}

\begin{figure}[htbp]
	\begin{center}
		\scalebox{0.27}{
		\includegraphics[trim={0.9cm 0cm 0cm 0},clip]{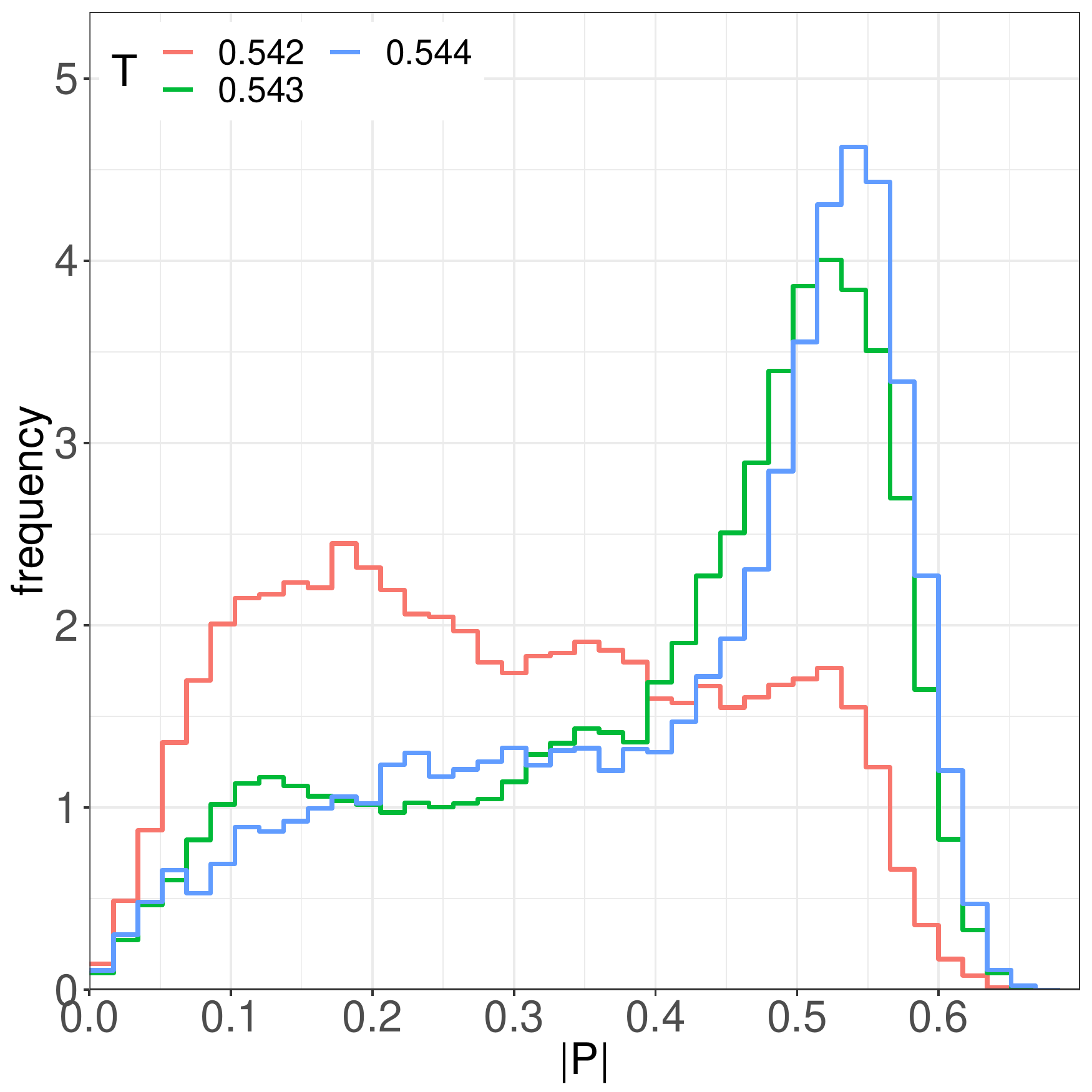}}
				\scalebox{0.25}{
\includegraphics{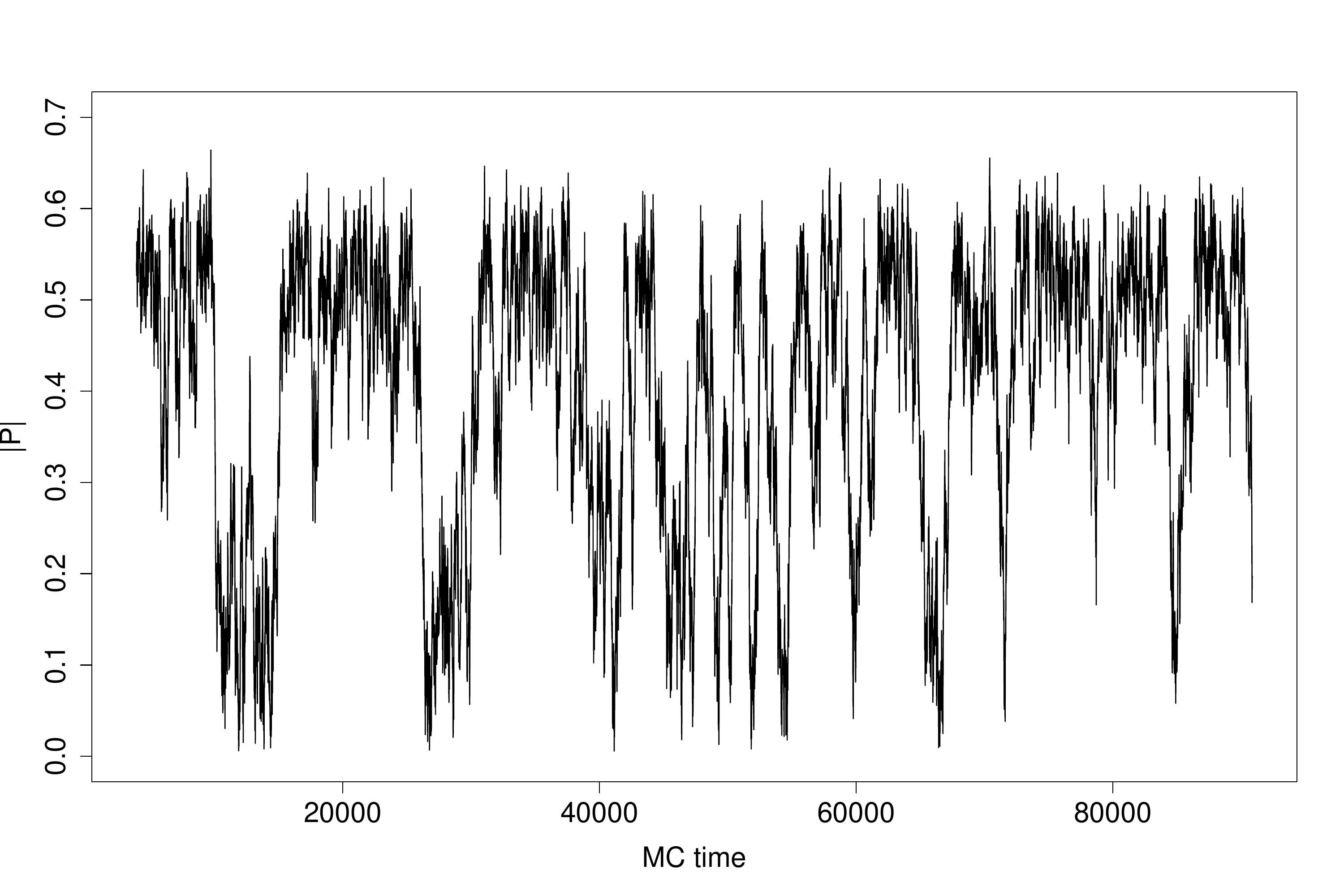}}
		\scalebox{0.27}{			\includegraphics[trim={0.9cm 0cm 0cm 0},clip]{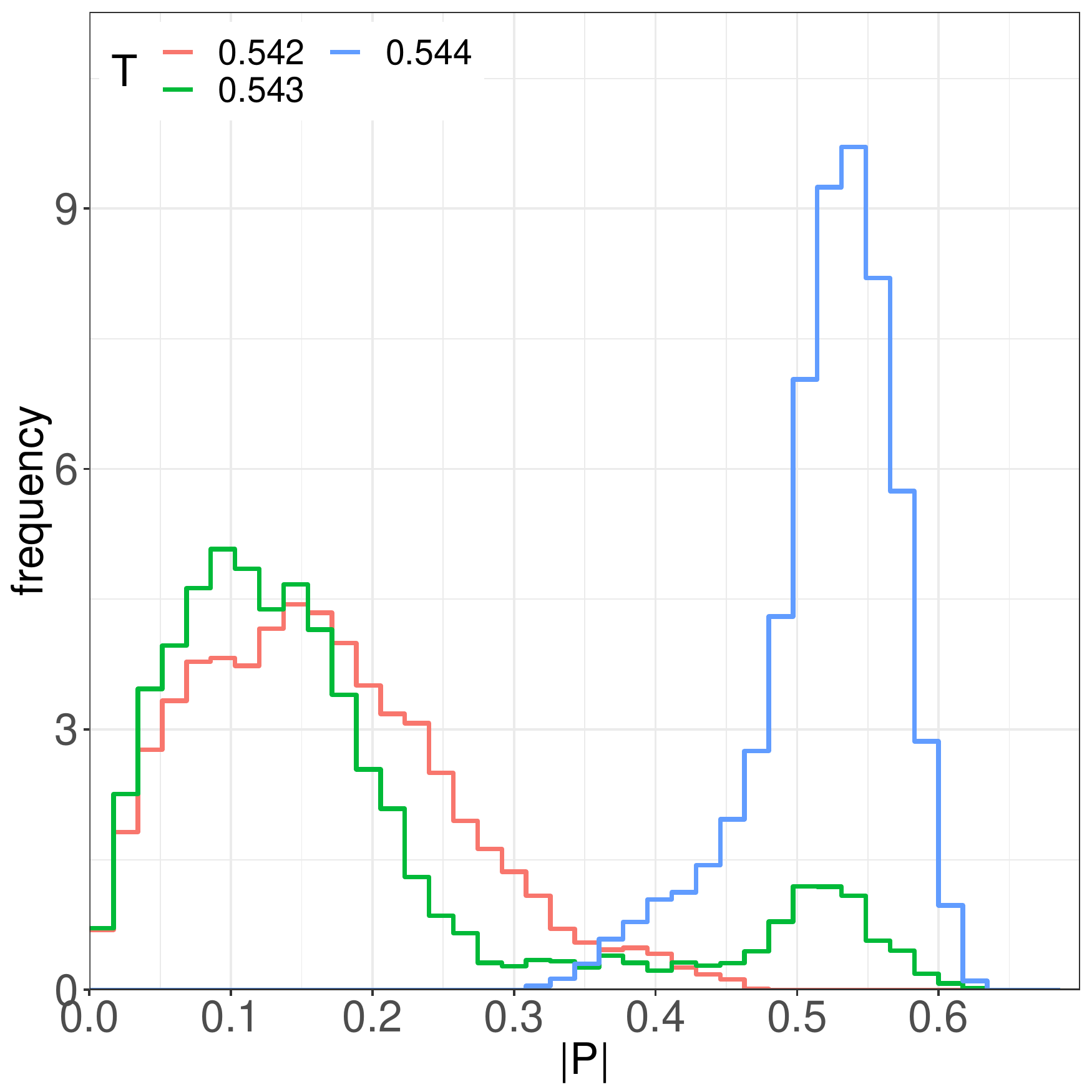}}
		\scalebox{0.25}{		\includegraphics{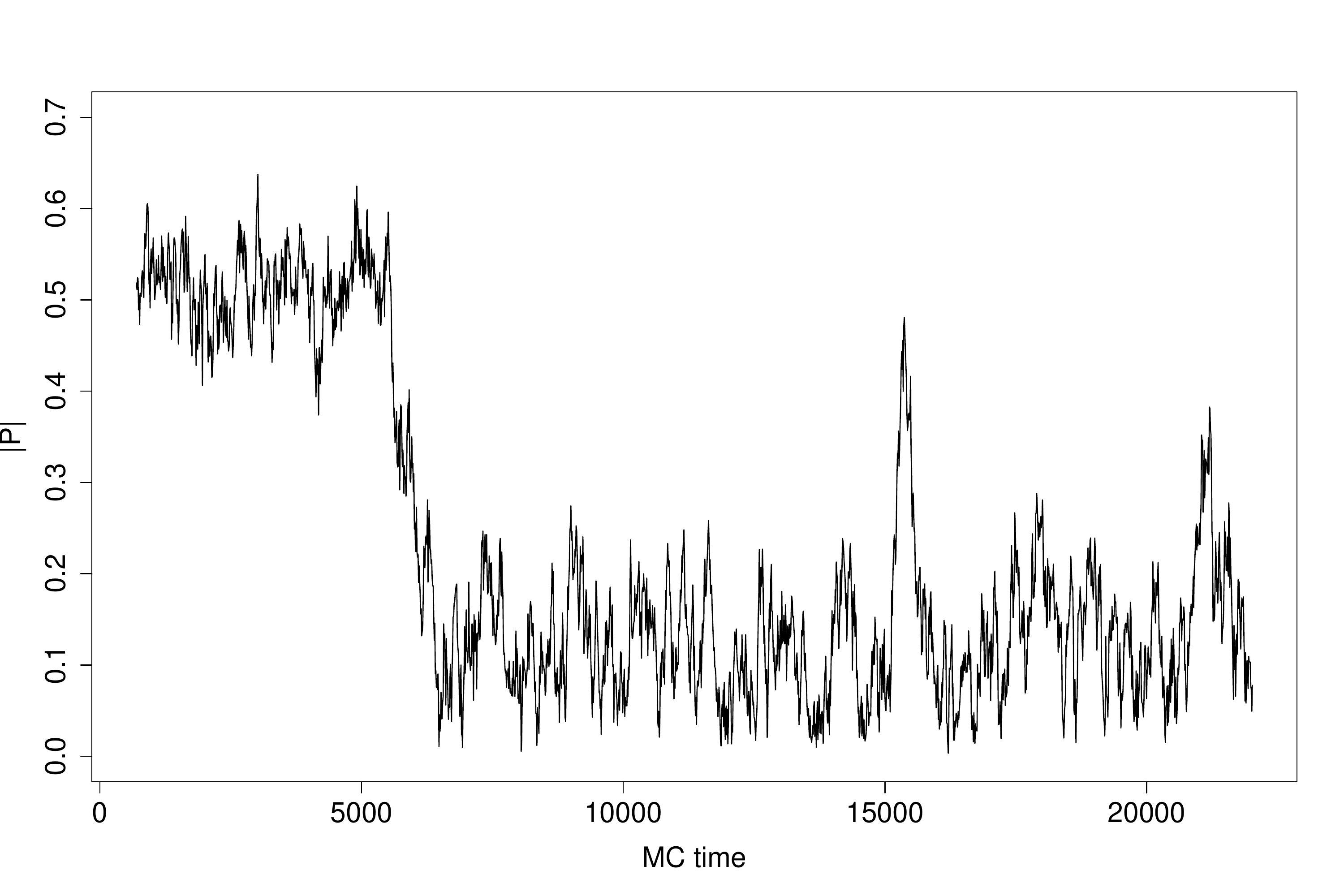}}
	\end{center}
	\caption{
Full BMN model: the histogram (left) and typical Monte Carlo history (for $T=0.543$) (right) of $P$ close to the critical temperature for $\mu=2.0$, for $N=24$, $L=12$ (top) and $N=32$, $L=12$ (bottom).
	}\label{fig:Pol_mu=2_neat_Tc}
\end{figure}

\begin{figure}[htbp]
	\begin{center}
		\scalebox{0.27}{
			\includegraphics[trim={0.9cm 0cm 0cm 0},clip]{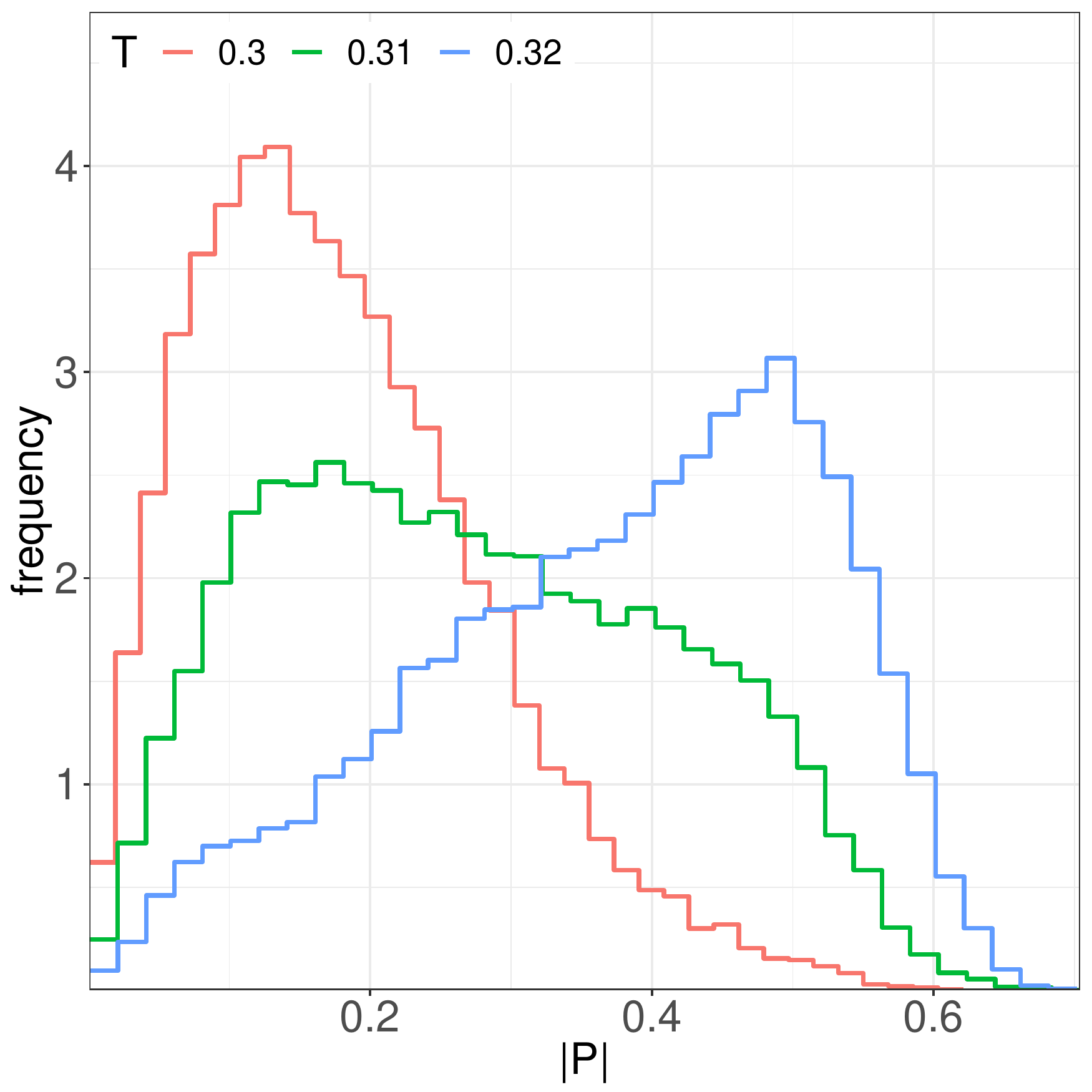}}
		\scalebox{0.27}{
			\includegraphics[trim={0.9cm 0cm 0cm 0},clip]{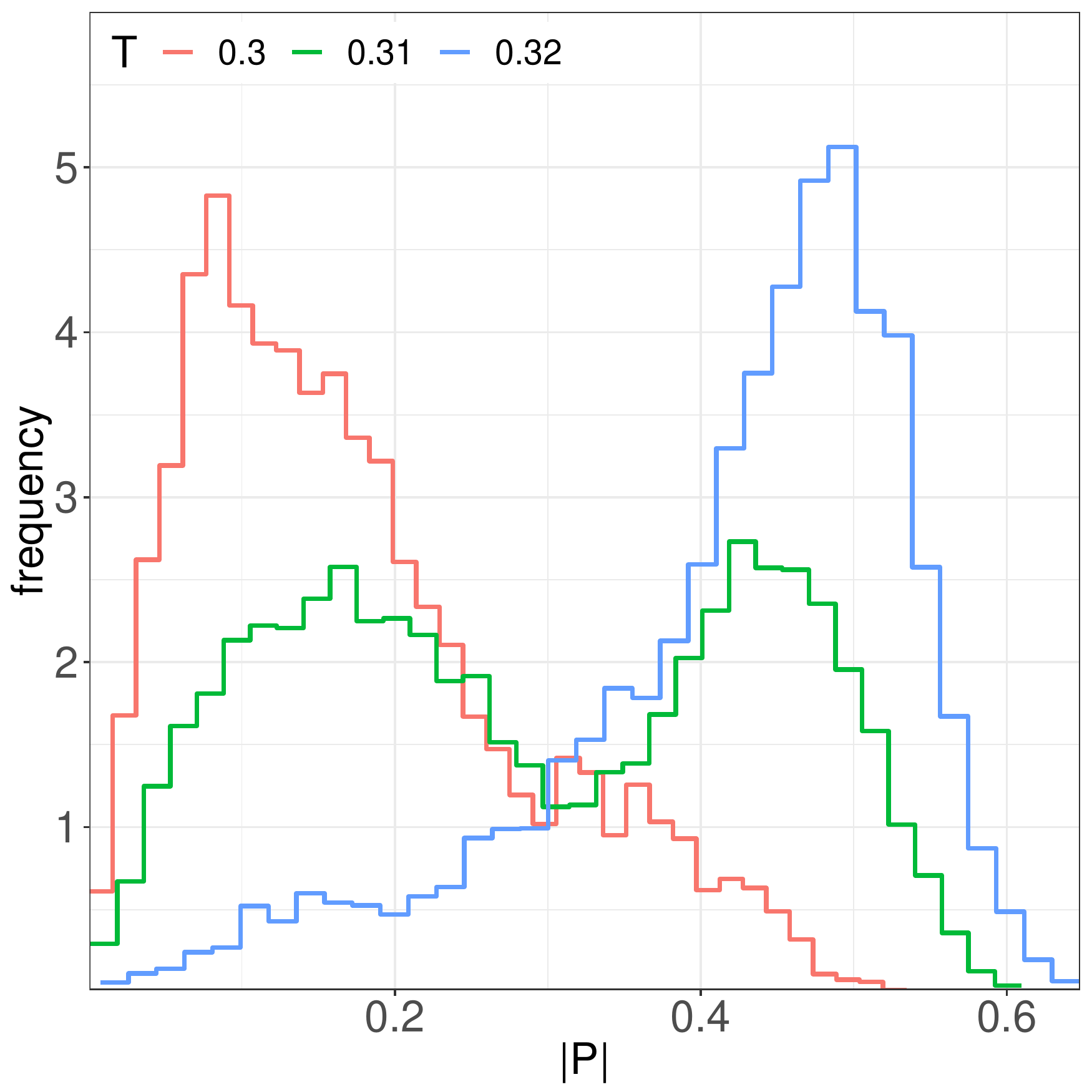}}
		\scalebox{0.27}{
			\includegraphics[trim={0.9cm 0cm 0cm 0},clip]{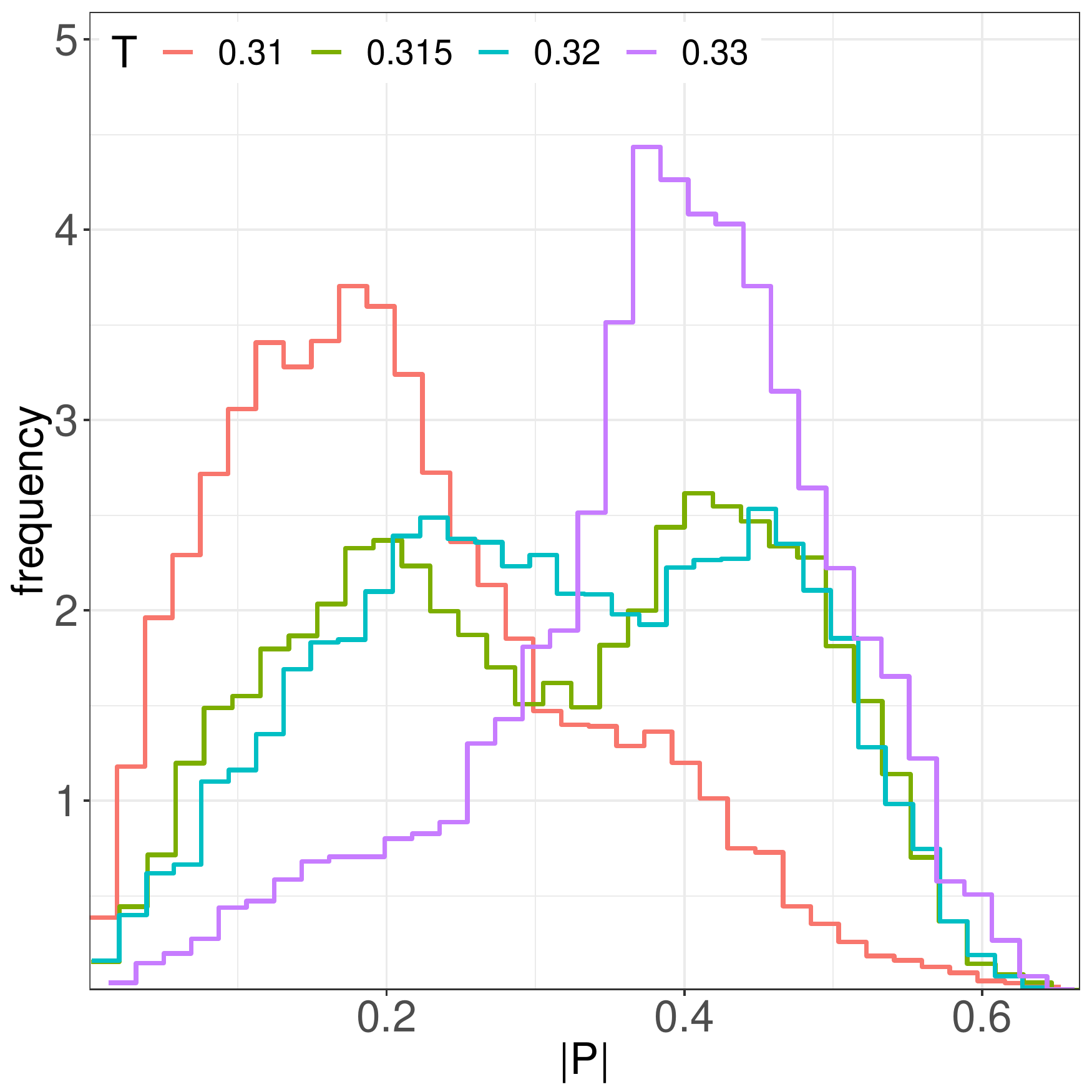}}
\end{center}
	\caption{
		Full BMN model: Histogram of $P$ close to the critical temperature for $\mu=1.0$.
		From left to right, $N=12$, $L=12$; $N=16$, $L=12$; $N=12$, $L=24$.
		A two-state signal is observed. A consistent signal with lower statistics has also been observed for $N=12$, $L=36$, confirming the trend of the critical temperature to slightly increase in the continuum limit.
	}\label{fig:Pol_mu=1_neat_Tc}
\end{figure}

\begin{figure}[htbp]
	\begin{center}
		\scalebox{0.25}{
			\includegraphics[trim={0.9cm 0cm 0cm 0},clip]{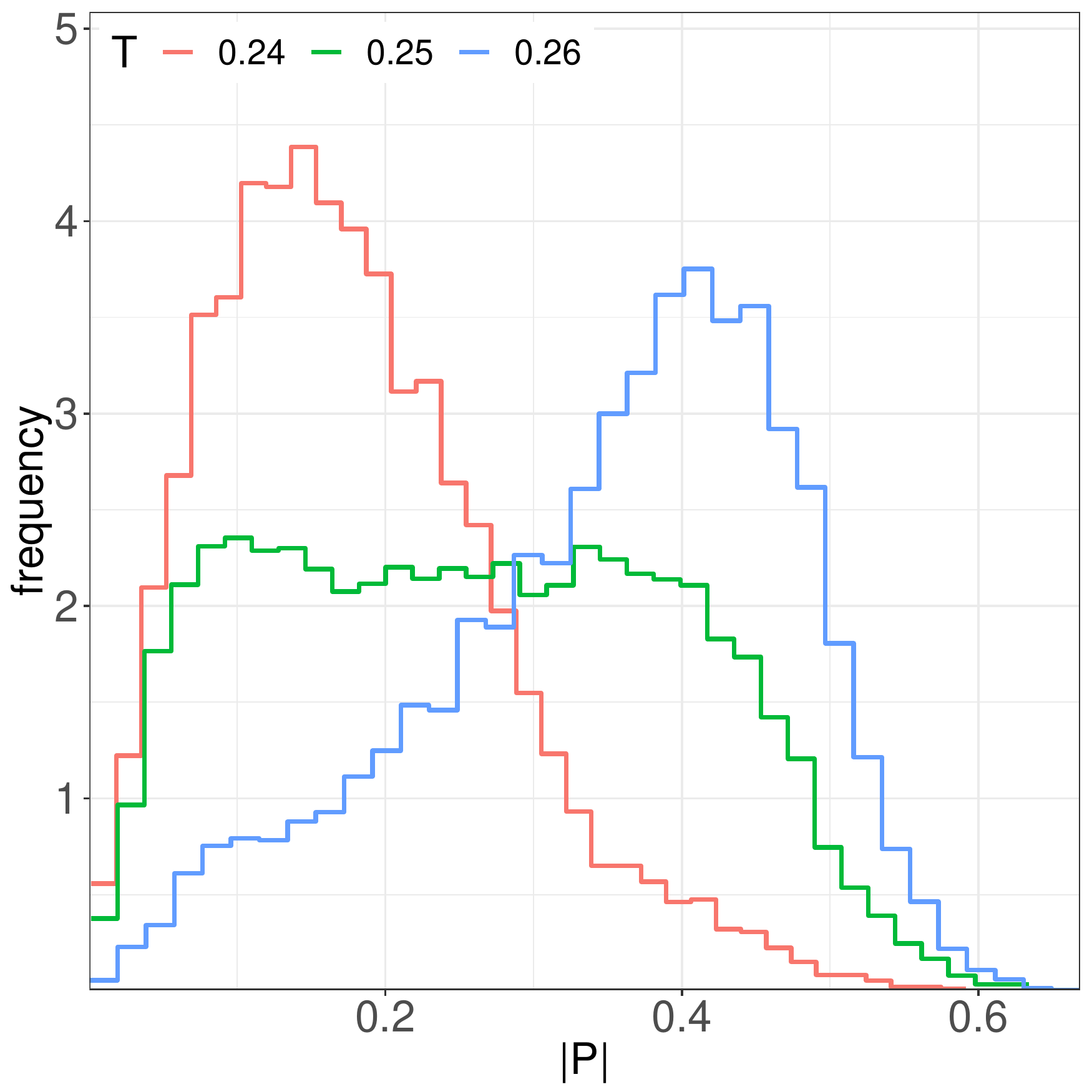}}
		\scalebox{0.25}{
			\includegraphics[trim={0.9cm 0cm 0cm 0},clip]{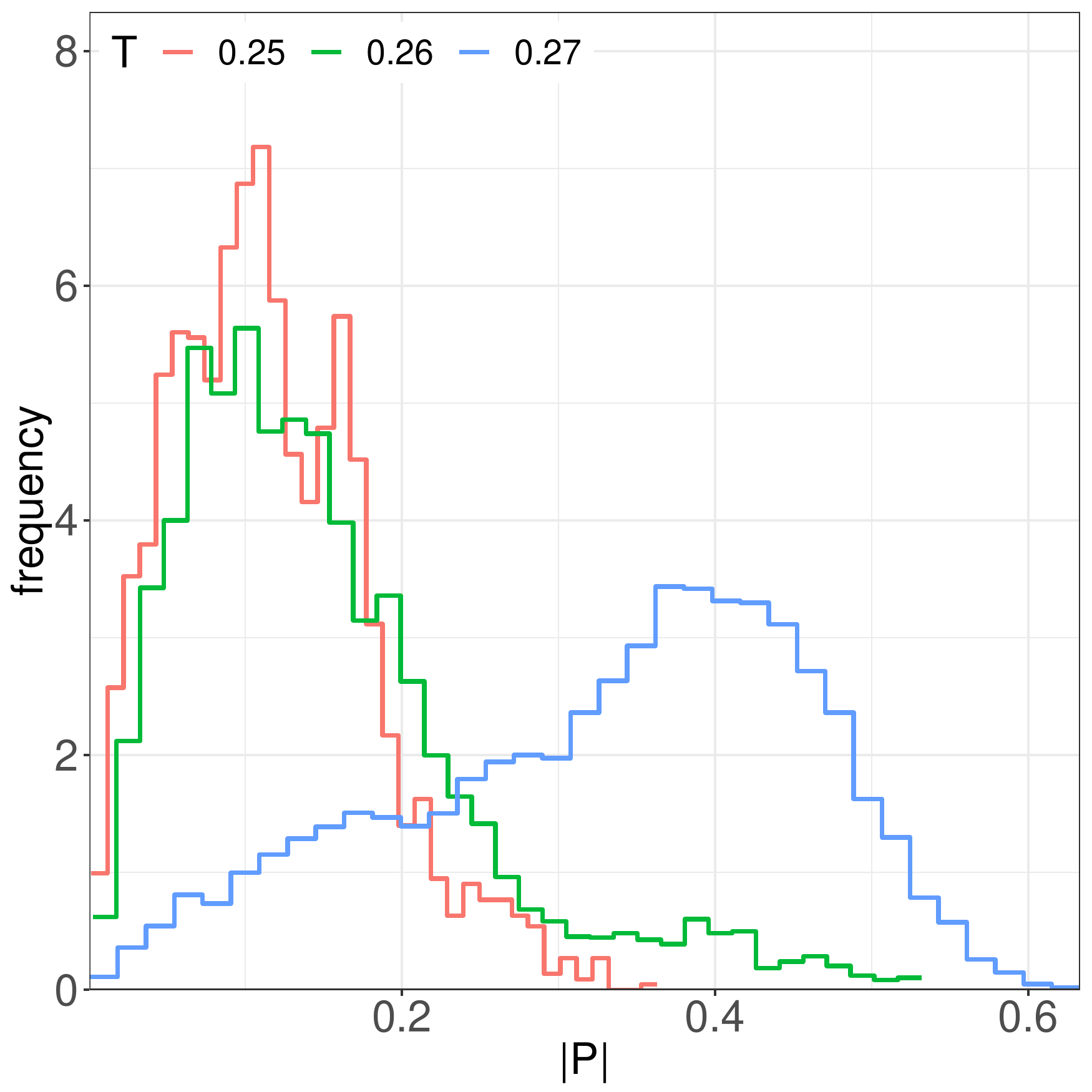}}
\end{center}
	\caption{
		Full BMN model: Histogram of $P$ close to the critical temperature for $\mu=0.8$, $N=12$, $L=12$ (left); $N=12$, $L=24$ (right). A blurred two-state signal is observed.
		A consistent signal with lower statistics has been observed also for $N=8$, $L=36$; $N=8$, $L=48$, confirming the trend of the critical temperature to slightly increase in the continuum limit.
}\label{fig:jointbins_Polyak_GN12S12M08D9}
\end{figure}

As we can see in Fig.~\ref{fig:tc_mu},
the deconfinement transition line shifts from the perturbative to the dual gravity prediction as decreasing $\mu$ towards $\mu=1.0$
and stays compatible with the latter down to $\mu=0.8$. The probability of tunneling to fuzzy-sphere backgrounds increases at smaller $\mu$.
Fig.~\ref{fig:M15-instability} shows an example of such a transition from trivial background to fuzzy-sphere background in the deconfined phase. Note that the Polyakov loop stays almost constant at this transition.
Already below $\mu=1.6$, we have to add a constraint for the Myers term to the action to stay in the trivial background. Without a constraint, the deconfinement transition also induces a transition in the Myers term. We expect this effect to be suppressed in the limit of large $N$. By using the constraint, we are able to confirm the trivial-confinement/trivial-deconfinement transition. As shown in Fig.~\ref{BMN_mu1.6}, Fig.~\ref{fig:Pol_mu=2_neat_Tc}, Fig.~\ref{fig:Pol_mu=1_neat_Tc}, and Fig.~\ref{fig:jointbins_Polyak_GN12S12M08D9}, we observe two-state signals indicating the first-order transition.

\begin{figure}[htbp]
	\begin{center}
		\rotatebox{0}{
			\scalebox{0.4}{
				\includegraphics[trim={0 0cm 0 0},clip]{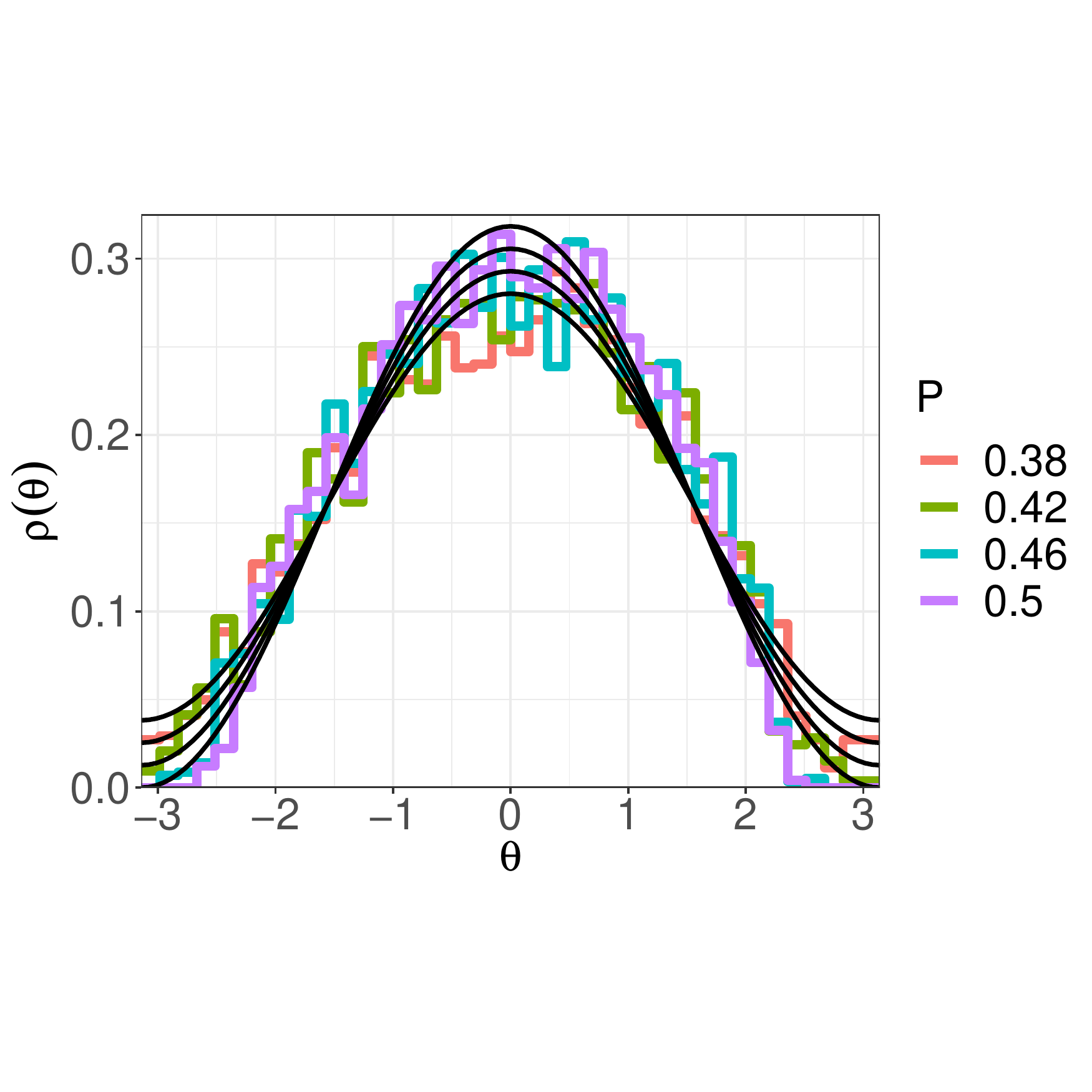}}}
		\rotatebox{0}{
			\scalebox{0.4}{
				\includegraphics[trim={0 0cm 0 0},clip]{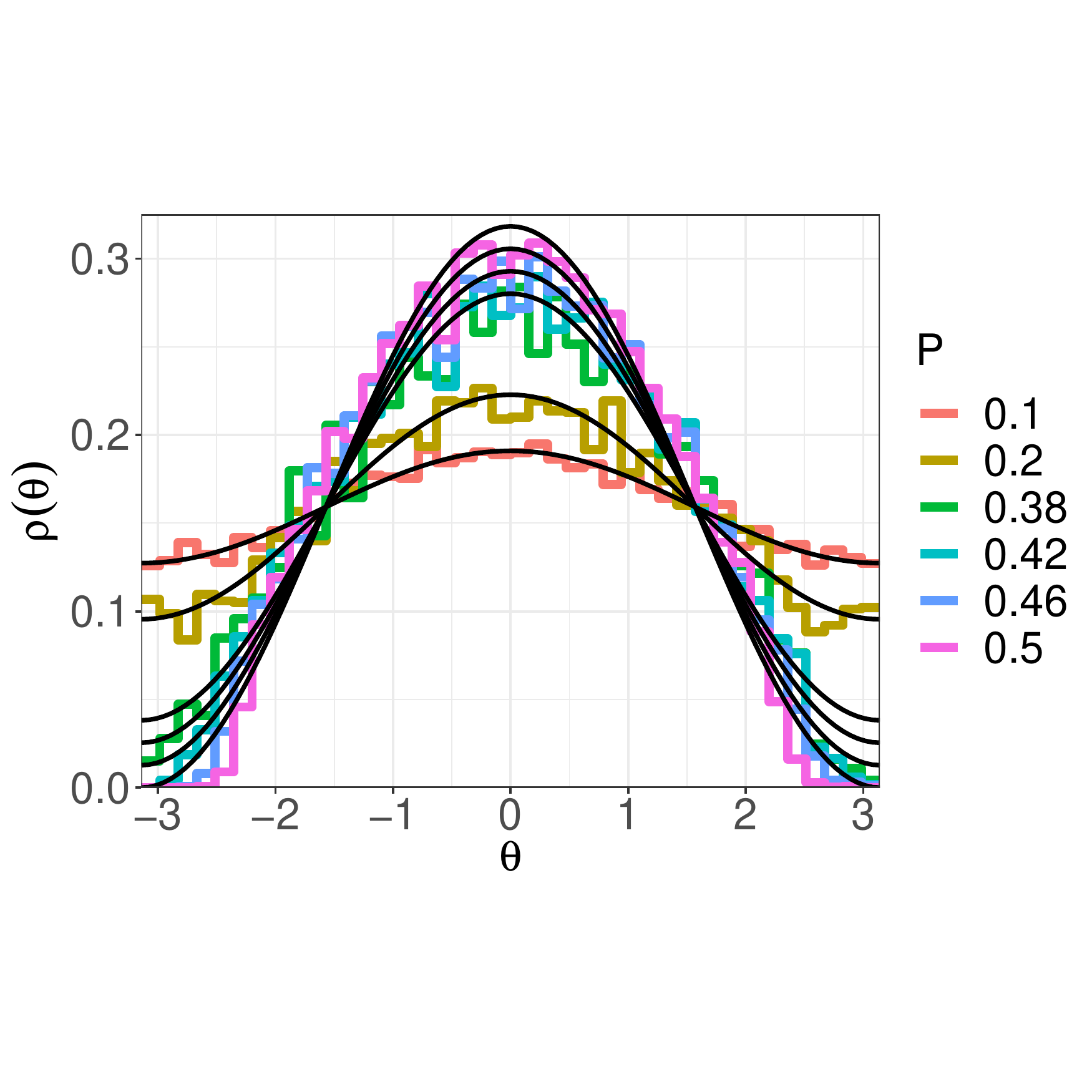}}}
	\end{center}
	\caption{
		Full BMN, the histogram of the Polyakov line phases for $[P-\Delta P,P+\Delta P]$, where $\Delta P=0.02$,  [Left] $\mu=1.5$, $N=24$, $L=24$, $T=0.429$ (close to $T_c$) and [Right] $\mu=1.6$, $N=32$, $L=24$, $T=0.45$ (close to $T_c$) along with the large $\mu$ behavior $\rho(\theta)=\frac{1+2P\cos\theta}{2\pi}$ in black. A clear deviation is observed close to $P=0.5$.
The Myers term was constrained in the simulation for $\mu=1.5$.
	}\label{fig:phase_bins_N24N32L12T0543M2}
\end{figure}

\begin{figure}[htbp]
	\begin{center}
		\scalebox{0.37}{
			\includegraphics[trim={0 0cm 0 0},clip]{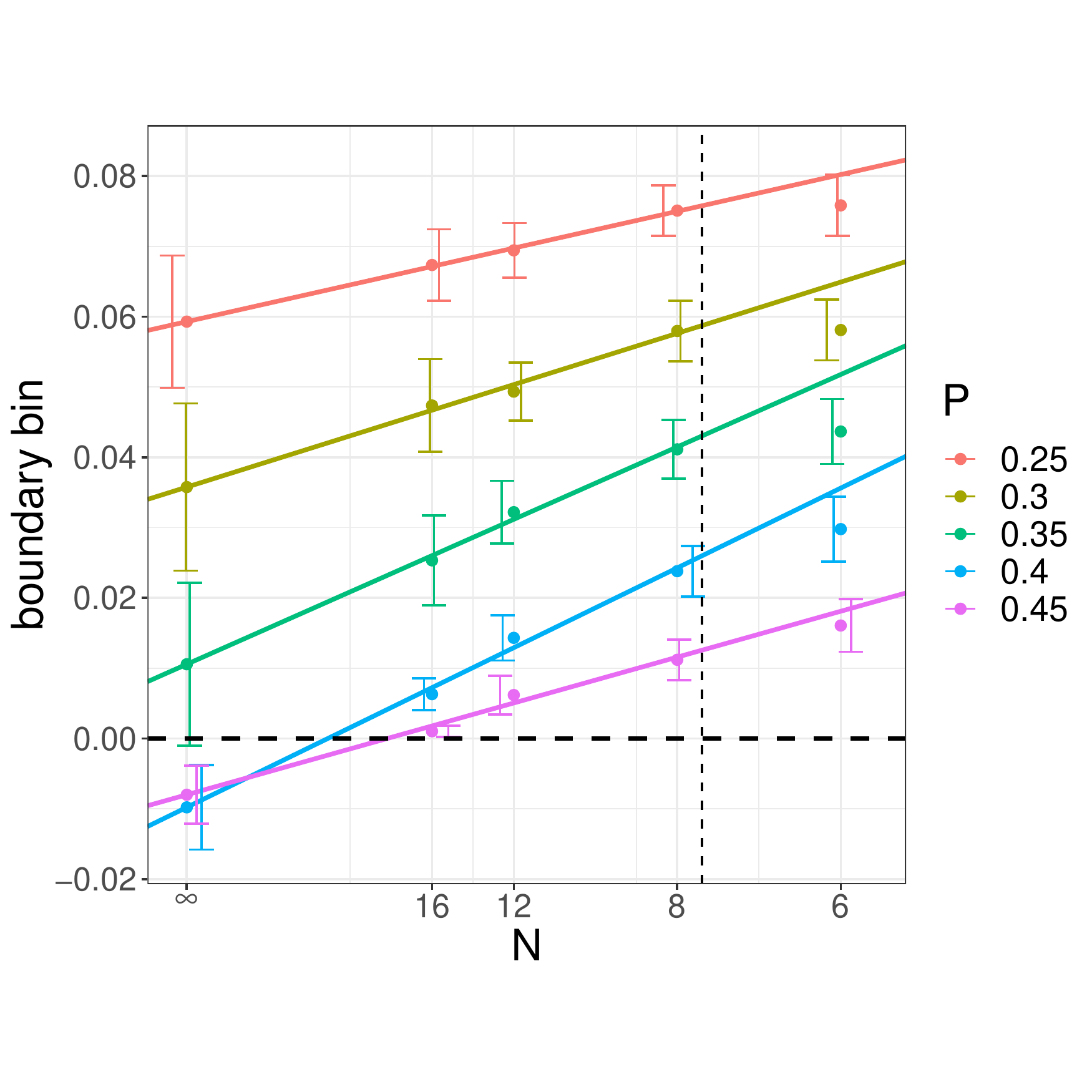}} ~~~~
		\scalebox{0.37}{
			\includegraphics[trim={0 0cm 0 0},clip]{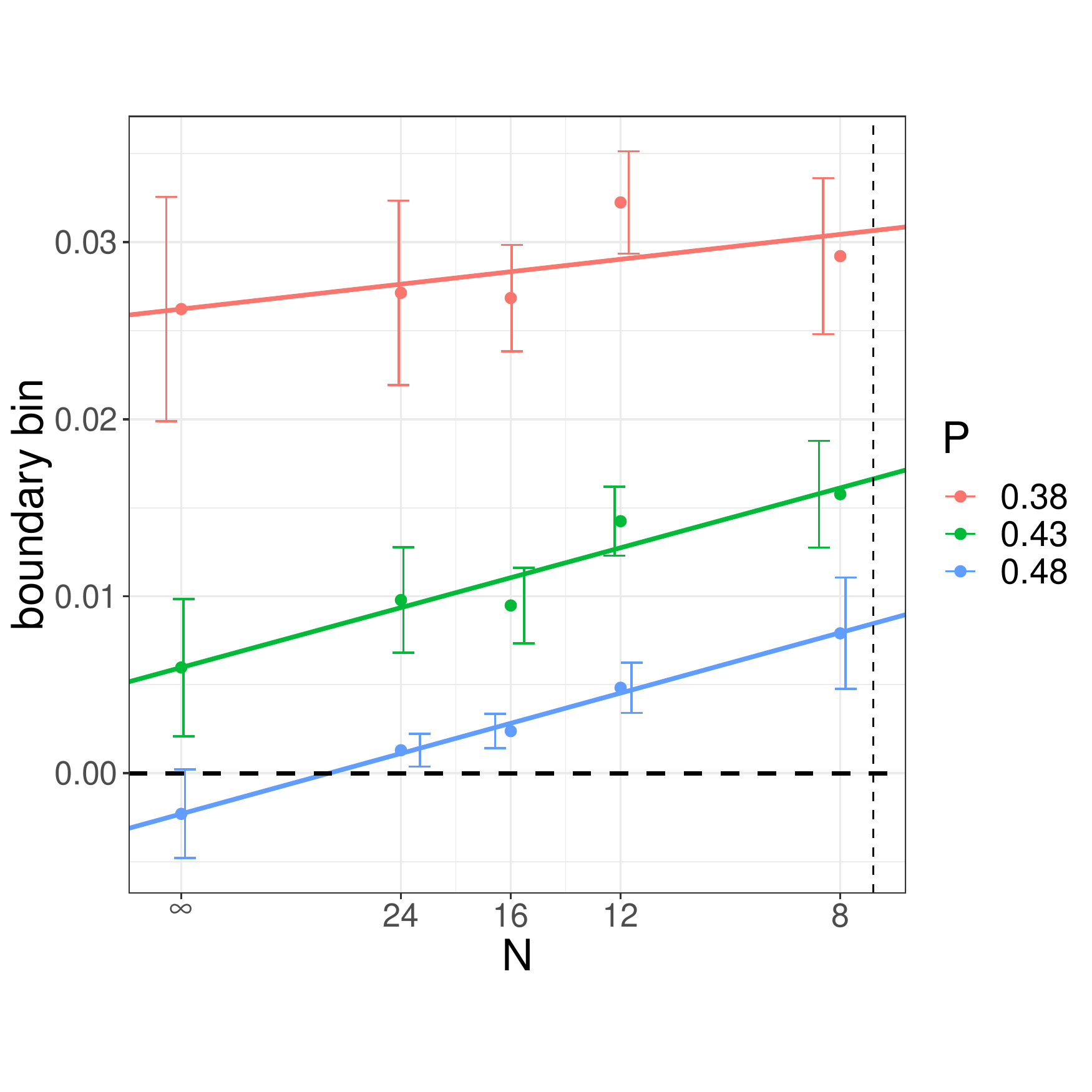}}
\end{center}
	\caption{
		Full BMN, the large-$N$ extrapolation of boundary bin, near $T_1$, for
		$T=0.31$, $\mu=1.0$, $L=12$ (left) and
		$T=0.429$, $\mu=1.5$, $L=24$ (right),
		as well as $\Delta P=0.025$.
The Myers term is constrained in these simulations.
	}\label{Pgap_large_N_M<2}
\end{figure}

In the considered region of $\mu$ we are able to investigate in more detail the nature of the phase transition.
As $\mu$ becomes smaller, the shape of the distribution of Polyakov line phases changes.
In Fig.~\ref{fig:phase_bins_N24N32L12T0543M2}, we plot the distribution of phases for different values of $P$,
for $\mu=1.5$ and $\mu=1.6$ close to the critical temperature $T_c$. A clear deviation from the large $\mu$ behavior, now more pronounced compared to $\mu=2.0$, is observed.
We can see that the gap in the phase distribution of the Polyakov line closes around $P=0.42$.
In Fig.~\ref{Pgap_large_N_M<2}, we show an extrapolation of $\rho(\theta=\pm\pi)$ to the large-$N$ limit for $\mu=1.0$ and $1.5$. The extrapolated value is consistent with zero for decreasing $P$ as $\mu$ decreases.
This is clearly different from the distribution at the large-$\mu$ region, that develops a gap at $P=\frac{1}{2}$, and continues the trend already observed at $\mu=2.0$.

\begin{figure}
	\centering
	\scalebox{0.27}{
		\includegraphics[trim={0.9cm 0cm 0 0},clip]{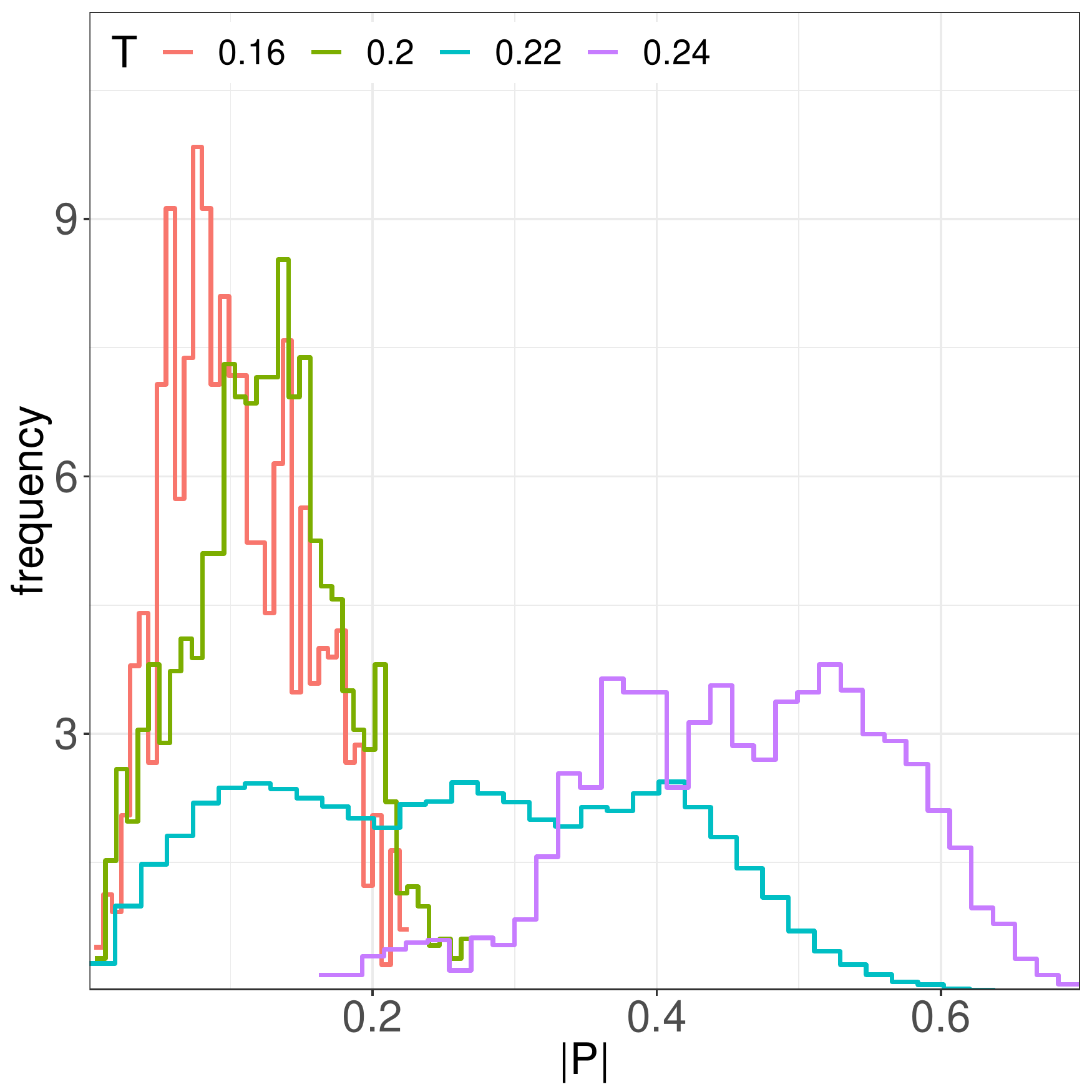}}
		\scalebox{0.27}{
		\includegraphics[trim={0.9cm 0cm 0 0},clip]{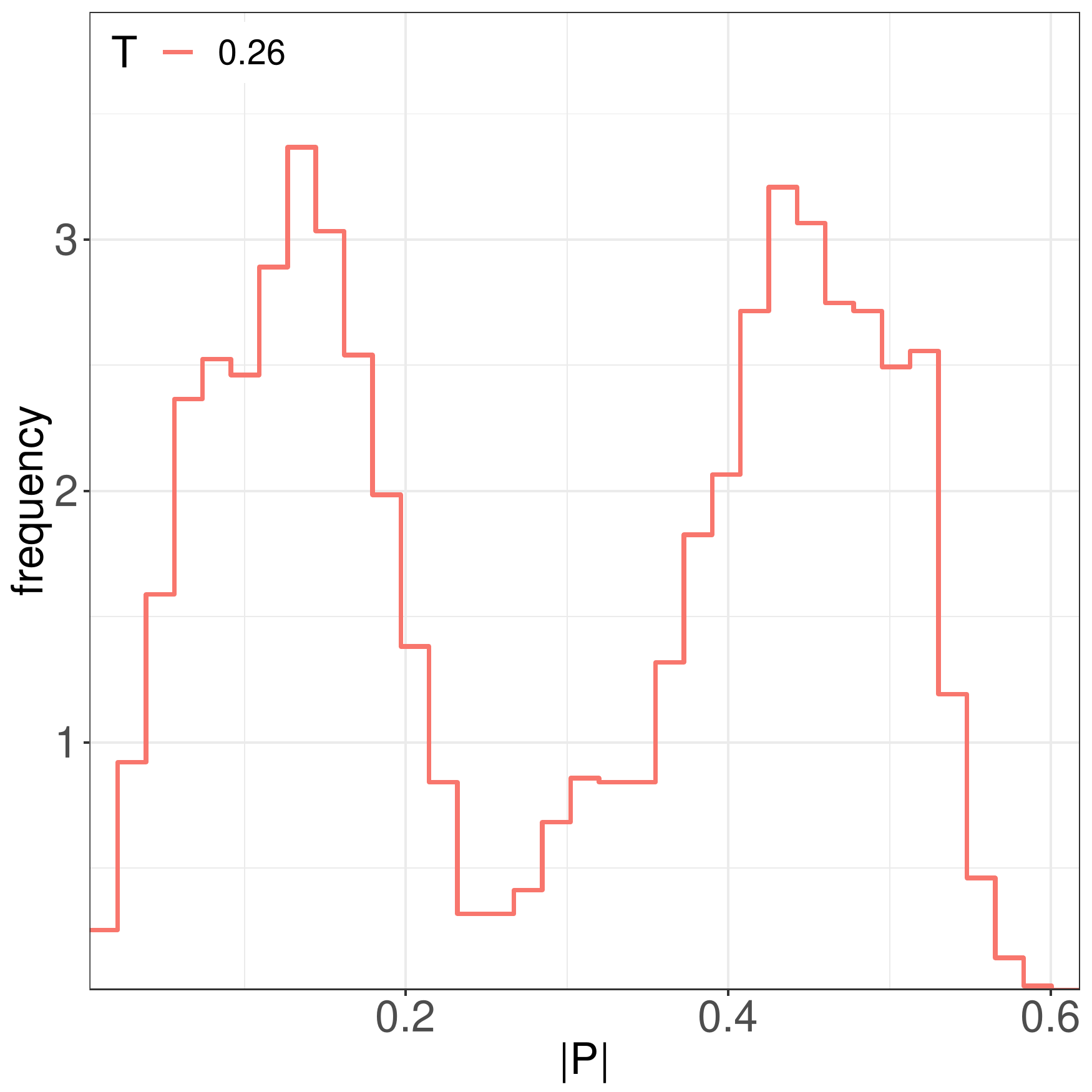}}	\\
\scalebox{0.23}{\includegraphics[trim={0 0cm 0 0},clip]{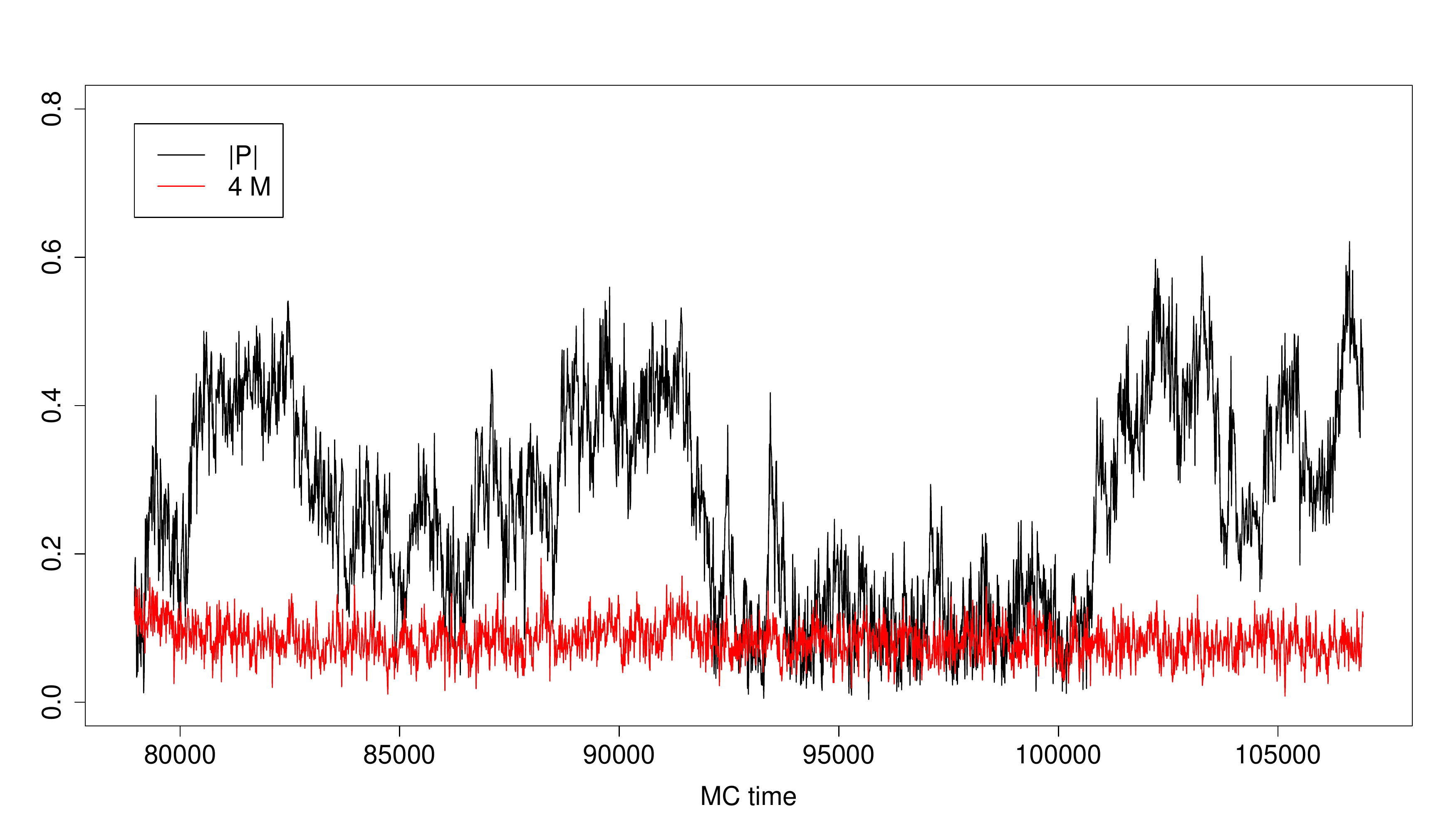}}
		\scalebox{0.28}{\includegraphics[trim={0 0cm 0 0},clip]{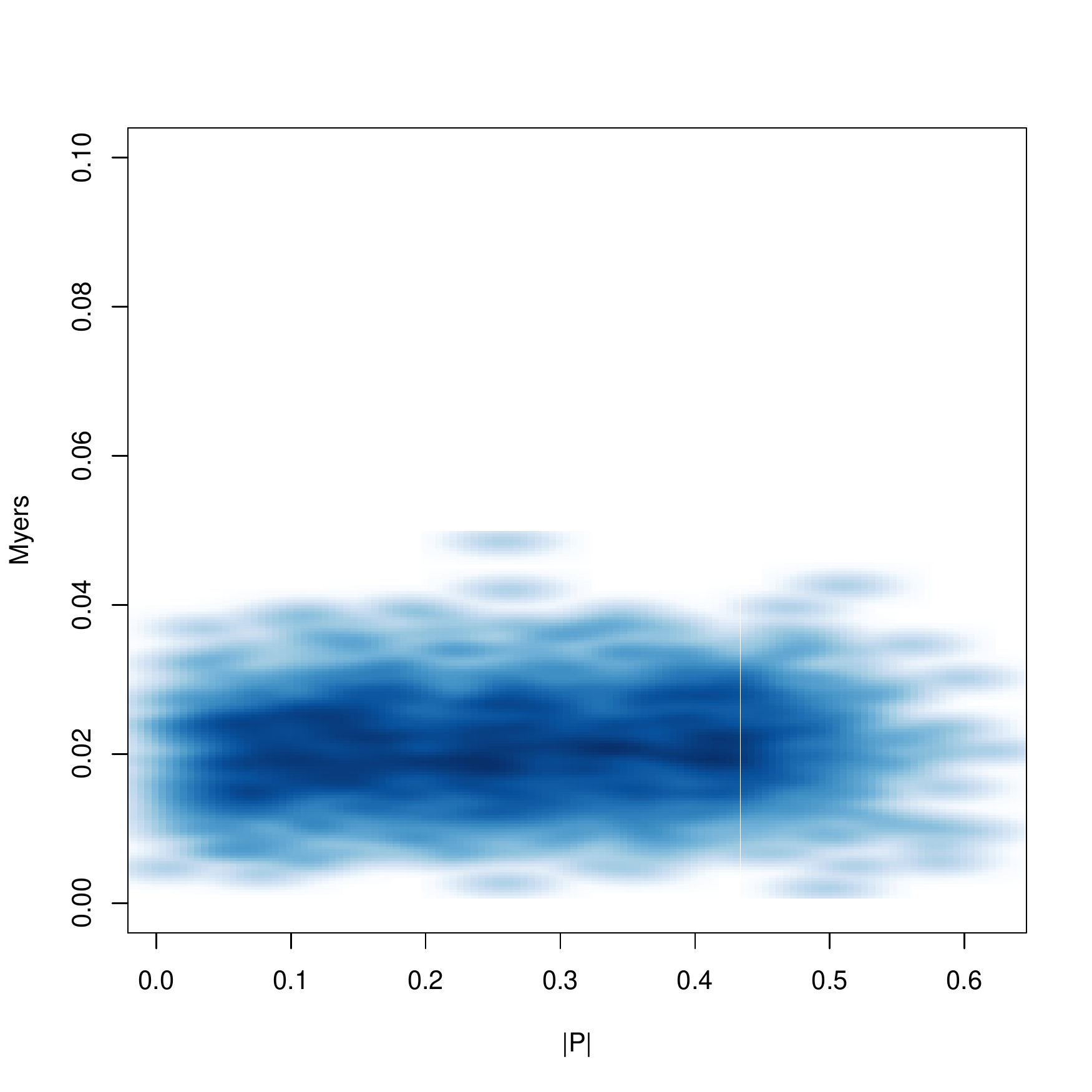}}
	\caption{ Full BMN, the histograms of the Polyakov distribution, and relative Monte Carlo histories for small $\mu$. [Top-Left] $N=12, L=24$ at $\mu=0.3$. A possible two-state signal is observed. At $T=0.24$ we observe runaway behavior of $R^2$, as well as in the deconfined sector at $T=0.22$. [Top-Right] $N=16, L=24$ at $\mu=0.5$. For $\mu=0.5$, the histogram was obtained by adding two independent streams which were initially prepared in a confined / deconfined state and remained there throughout a sufficiently long Monte Carlo evolution time due to the relatively large $N=16$.
	[Bottom-Left] MC history of $N=12, L=24$ at $\mu=0.3, T=0.22$ from the up-left histogram. The signal suggests repeated tunneling between confined and deconfined states.
	Note that, at such a small value of $\mu$, the distinction between trivial background and fuzzy-sphere background is not clear. [Bottom-Right] The relation between the Myers term and the Polyakov loop for the up-left histogram.
	No significant correlation is observed.
Note that the Myers term was constrained to be below $M=0.02$, but this cutoff was hit frequently, and often $M$ became larger than the cutoff 0.02. (This is possible because the coefficient for the constraint term is large but finite.)
	}\label{fig:phase_bins_smallmu}
\end{figure}

\subsubsection{Small $\mu\leq 0.8$ regime and convergence to BFSS limit}\label{sec:small-mu-region}
The point $\mu=0.8$ is the first point starting to depart from the gravity line and for even smaller $\mu$ our estimate for the critical temperatures shown in Fig.~\ref{fig:tc_mu} is always larger than $T^{\rm (gravity)}_{\rm c}$.
However, this apparent deviation from the gravity prediction may be simply a finite-$N$ effect:
we used at most $N=16$ to study the small-$\mu$, small-$T$ region. It is not unreasonable to compare to $\mu=0$ as a first approximation.
Then, a rough estimate of the critical temperature from the gravity side is $T_c\sim T_2\sim N^{-5/9}\gtrsim 0.251$ for $N=12$ and $0.214$ for $N=16$, up to an order one multiplicative factor. As we have seen in Sec.~\ref{sec:trivial-conf-trivial-deconf}, $T_1$ should also be similar.
Therefore, the critical temperatures observed in our simulations are in the right ballpark. In fact, if we observed the values much closer to the gravity line shown in Fig.~\ref{fig:tc_mu}, this could even be seen as evidence against the duality conjecture.
As $N$ increases, $T_2$ and $T_c$ should decrease while $T_1$ should increase, and the hysteresis should become more pronounced (Fig.~\ref{fig:Pol-1st-order-scenario}). Thus, the numerical results should converge to the gravity line even at small $\mu$.

Possible caveats of our analysis have been explained earlier. It is difficult to stabilize the trivial background, especially in the deconfined state.\footnote{
It may not be an issue because the trivial background and fuzzy-sphere backgrounds are indistinguishable at $\mu\lesssim 0.018$ for $N=16$, as explained in Sec.~\ref{sec:backgrounds_definition}. However we never reached that small $\mu$ values.
} In addition, the run-away of the scalar fields becomes more important towards the BFSS limit because the mass term $\sim\mu^2{\rm Tr}X^2$ is turned off. On the other hand, in this low-temperature region, we observe an increased metastability which counteracts these effects. The trivial-confined phase gets stabilized, and simulations without the constraint are possible in some parameter regions.

The metastability in the small-$\mu$, low-$T$ region is indicated by a strong dependence on the initial conditions of the simulations which tend to stay in either a confined or deconfined state when started with the respective initial configuration.
The observed metastability indicates a large separation of the temperatures $T_1$ and $T_2$ (see right panel of Fig.~\ref{fig:Pol-1st-order-scenario}).
As we will see in Sec.~\ref{sec:BFSSLimit}, the trivial-confined phase turns out to be rather stable at low temperatures even in the BFSS limit ($\mu=0$), while we could not see the two-peak signal at $\mu<0.3$ (i.e., at $\mu<0.3$ and $T\lesssim 0.3$ we could not see the deconfined phase).

At very small $\mu$, the difference in the Myers term for different fuzzy-sphere backgrounds is significantly reduced (see Sec.~\ref{sec:backgrounds_definition}).
Fuzzy-spheres can be buried in quantum fluctuations, and we expect that there is not a strong dependence of the deconfinement transition on the chosen background configuration.
Therefore, the constraint on the Myers term should not affect the result in the small-$\mu$ region.
Indeed we can, for example, observe very similar signals for metastable confined and deconfined state from the Polyakov loop distribution at $\mu=0.3$, $N=16$, $L=36$, $T=0.27$ independent of the constraint.
The metastability extends over a  large temperature range indicating the difference between  $T_1$ and $T_2$.
Fig.~\ref{fig:phase_bins_smallmu} presents evidence for these assertions. For $\mu=0.5$, a clear two-state signal is observed by joining a confined and deconfined stream with the same simulation parameters into a single histogram. Both streams remain in their respective state for a sufficiently long time to conclude that there is a strong hysteresis. For $\mu=0.3$, repeated tunneling between a confined and deconfined state has been observed. In the deconfined parts of the Monte Carlo history, we observe runaway behavior in $R^2$. For our choices of $N$ and $L$, $\mu=0.4$ was the smallest mass parameter for which we could observe a stable deconfined phase at temperatures close to the transition temperature.  Due to the choice $N=12$, as opposed to $N=16$ for $\mu=0.5$, the two-state signal is not clear in the histogram, but we were able to see repeated tunneling due to the increased tunneling probability and simulation speedup at smaller $N$.

\begin{figure}
	\centering
	\scalebox{0.2}{
		\includegraphics[trim={0 0cm 0 0},clip]{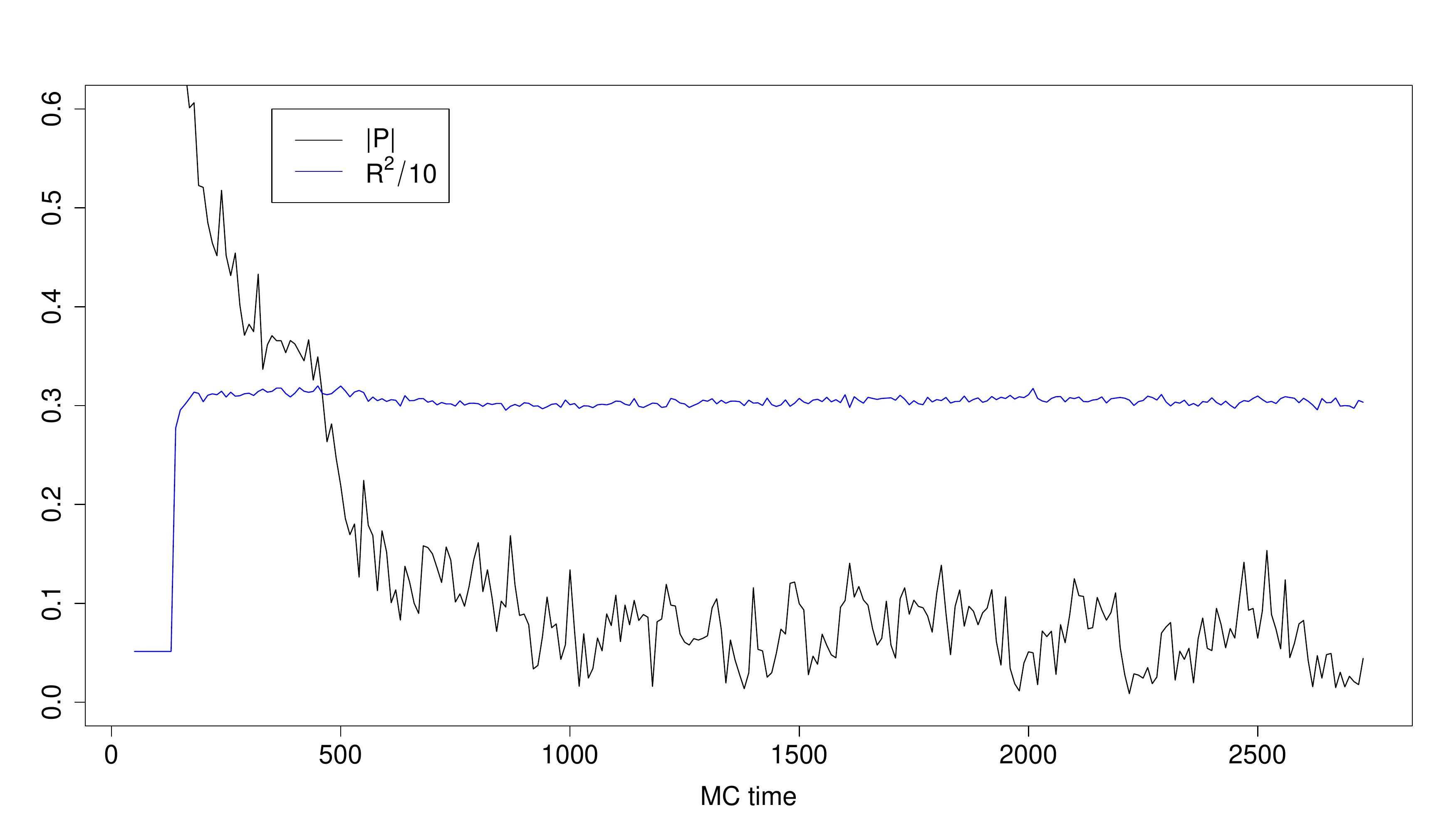}}
	\scalebox{0.2}{
		\includegraphics[trim={0 0cm 0 0},clip]{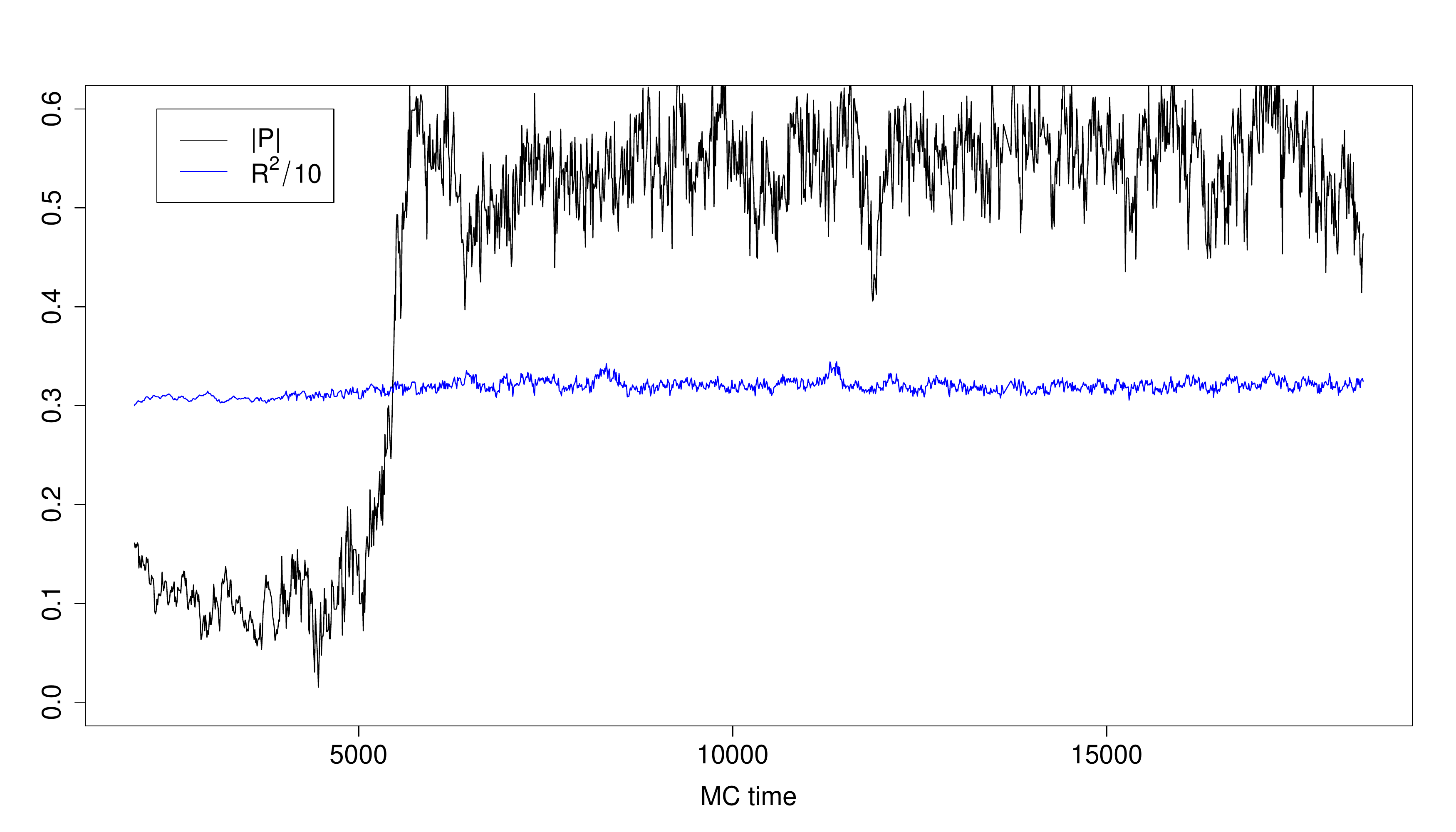}}
	 \caption{ Full BMN model: [Left] Monte Carlo history for $N=16,  L=24,  \mu=0.6, T=0.22$. The confined phase ($|P|\simeq 0$) is observed.
	 [Right] For $N=16,  L=24,  \mu=0.6, T=0.29$, starting with a confined configuration. A transition to the deconfined phase is observed. The temperature window is of order $\Delta T\approx 0.07$ at these parameters.}
	\label{Fig:mu06T1T2}
\end{figure}

Let us give some more details on how we estimate $T_1$, $T_c$, and $T_2$ from results as those shown in Fig.~\ref{Fig:mu06T1T2} . In this example the parameters are $N=16, L=24$ at $\mu=0.6$. It is challenging to observe a precise two state signal at these parameters and requires large statistics.  Another way to give an estimate for $T_c$ in a first-order scenario is by locating $T_1$ and $T_2$ since $T_2<T_c<T_1$ (see Fig.~\ref{fig:Pol-1st-order-scenario}). We then face the problem of locating $T_2$ which is indicated by the formation of a gap in the Polyakov eigenvalue distribution and $T_1$ which is the highest possible temperature with a metastable confined state. Both determinations require large statistics to be trustable, but one may be able to roughly estimate the difference $\Delta T=T_1-T_2$ at the current level of statistics. Hence we do the following to locate $T_2$: we start by considering deconfined initial configurations and gradually lower the temperature until we see the simulation converge to a confined state. When this transition is accompanied by a gap in the eigenvalue distribution of the Polyakov loop, it provides an estimate of $T_2$. Following this procedure, see Fig.~\ref{Fig:mu06T1T2}, we have located the temperature $T_2$ to be approximately $T_2\approx 0.22$. Even though a precise determination would require larger $N$ and much more statistics, we emphasize that  this is nevertheless a valid estimate.

To locate $T_1$, we took the opposite direction. A confined initial configuration was used from smaller temperatures and the temperature was gradually raised  to check if the configuration remains confined or not. In the right panel of Fig.~\ref{Fig:mu06T1T2}, we observe a transition to deconfined state for $T_1\approx 0.29$ for the same parameters $N,L,\mu$. This will be roughly the maximum confined temperature for this configuration. Then we see that the temperature window $\Delta T$ can be rather large even at these parameters. Towards lower $\mu$, the situation starts becoming even more difficult to analyze, but the same pattern is observed.

\begin{figure}
	\centering
	\scalebox{0.2}{
		\includegraphics[trim={0 0cm 0 0},clip]{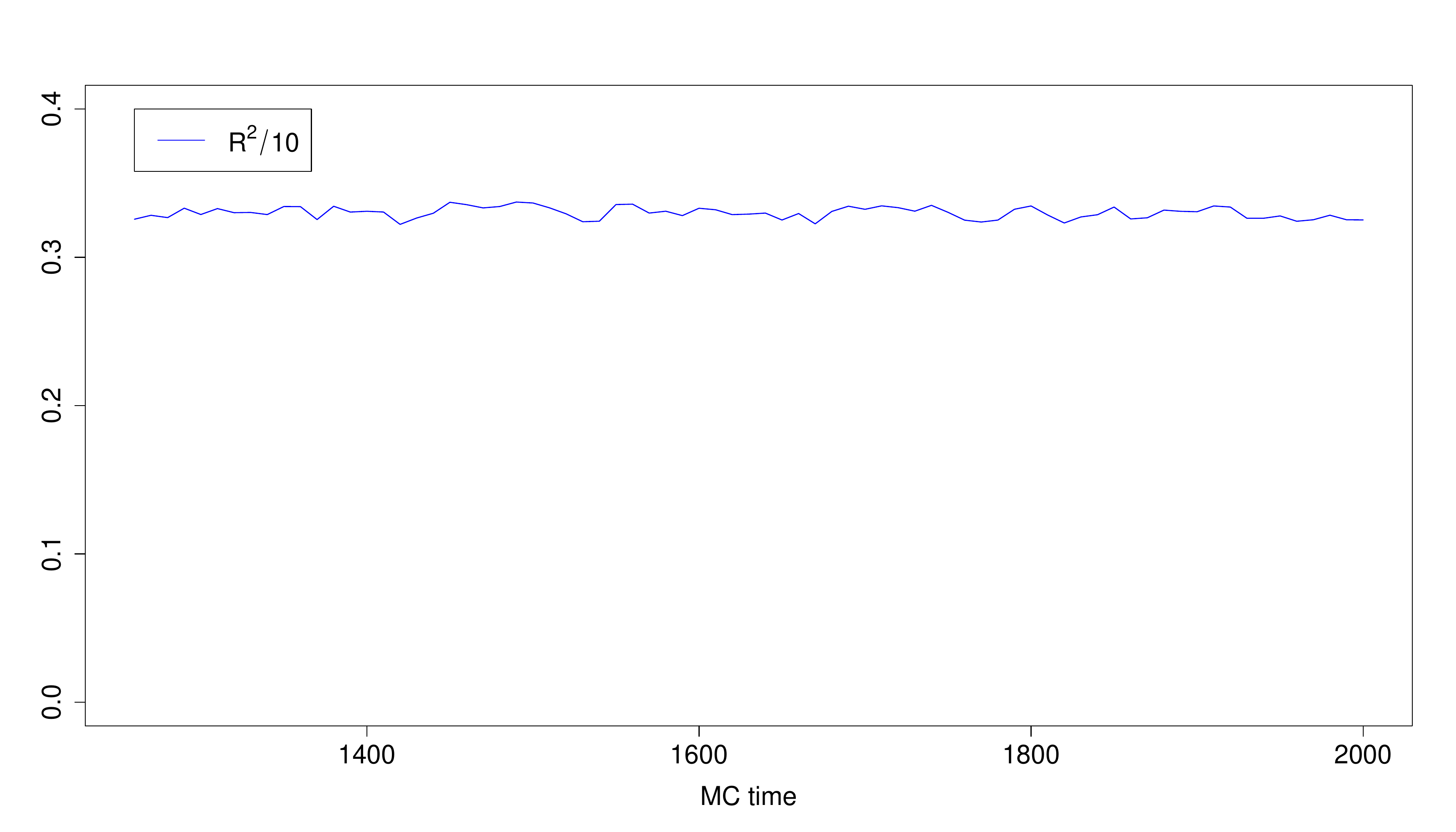}}
		\scalebox{0.2}{
		\includegraphics[trim={0 0cm 0 0},clip]{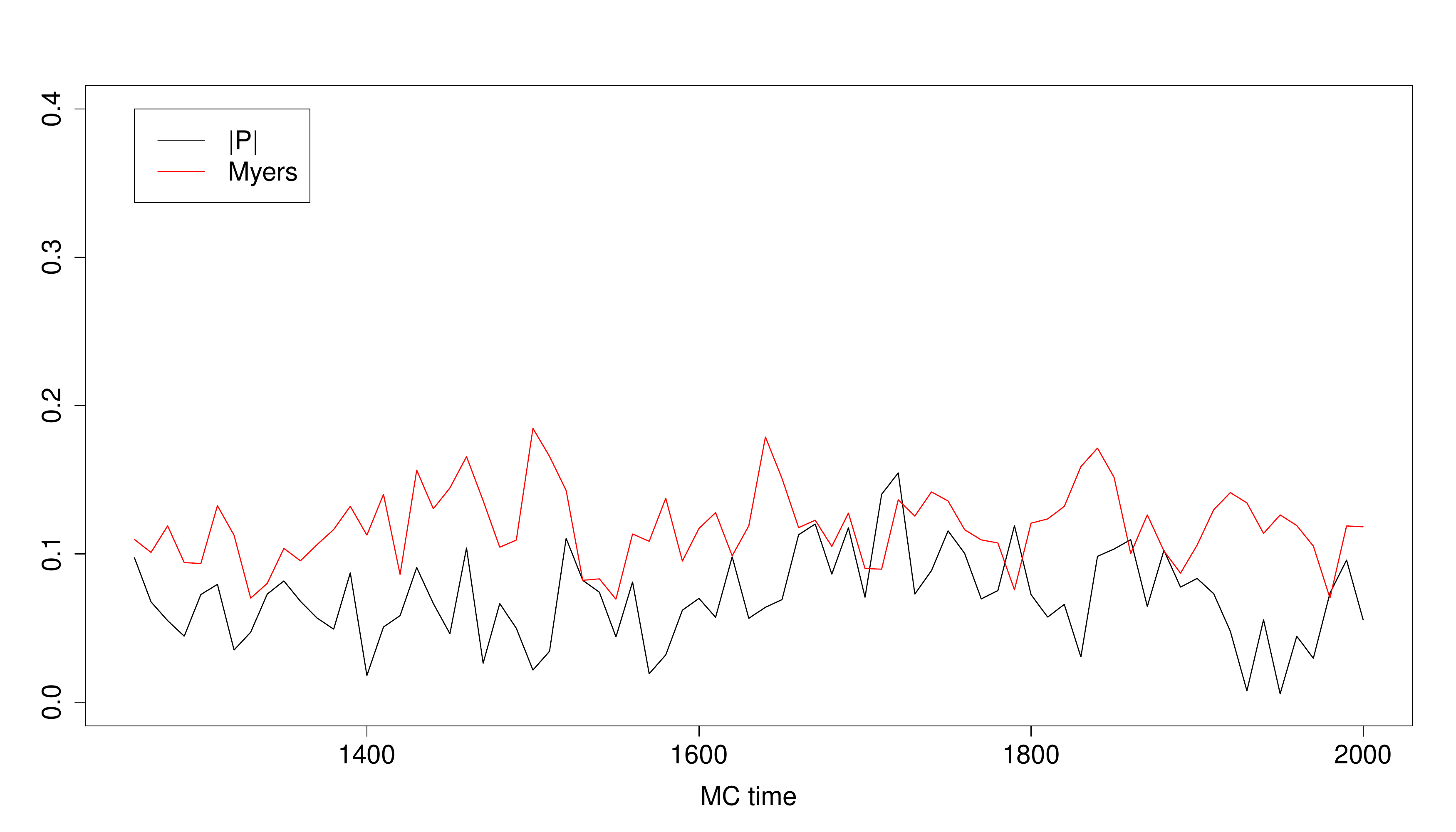}}	\\
	\scalebox{0.2}{
		\includegraphics[trim={0 0cm 0 0},clip]{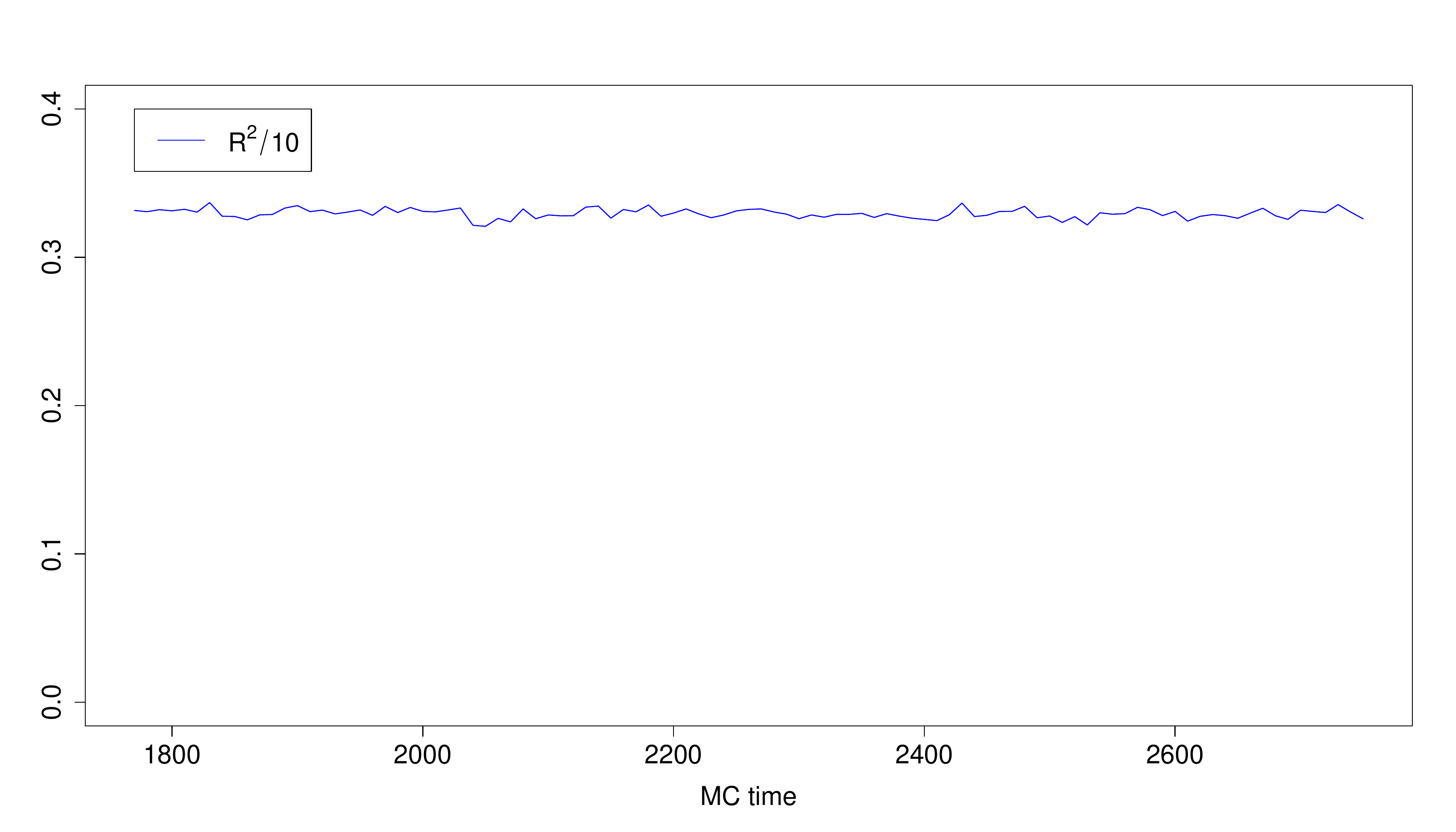}}
		\scalebox{0.2}{
		\includegraphics[trim={0 0cm 0 0},clip]{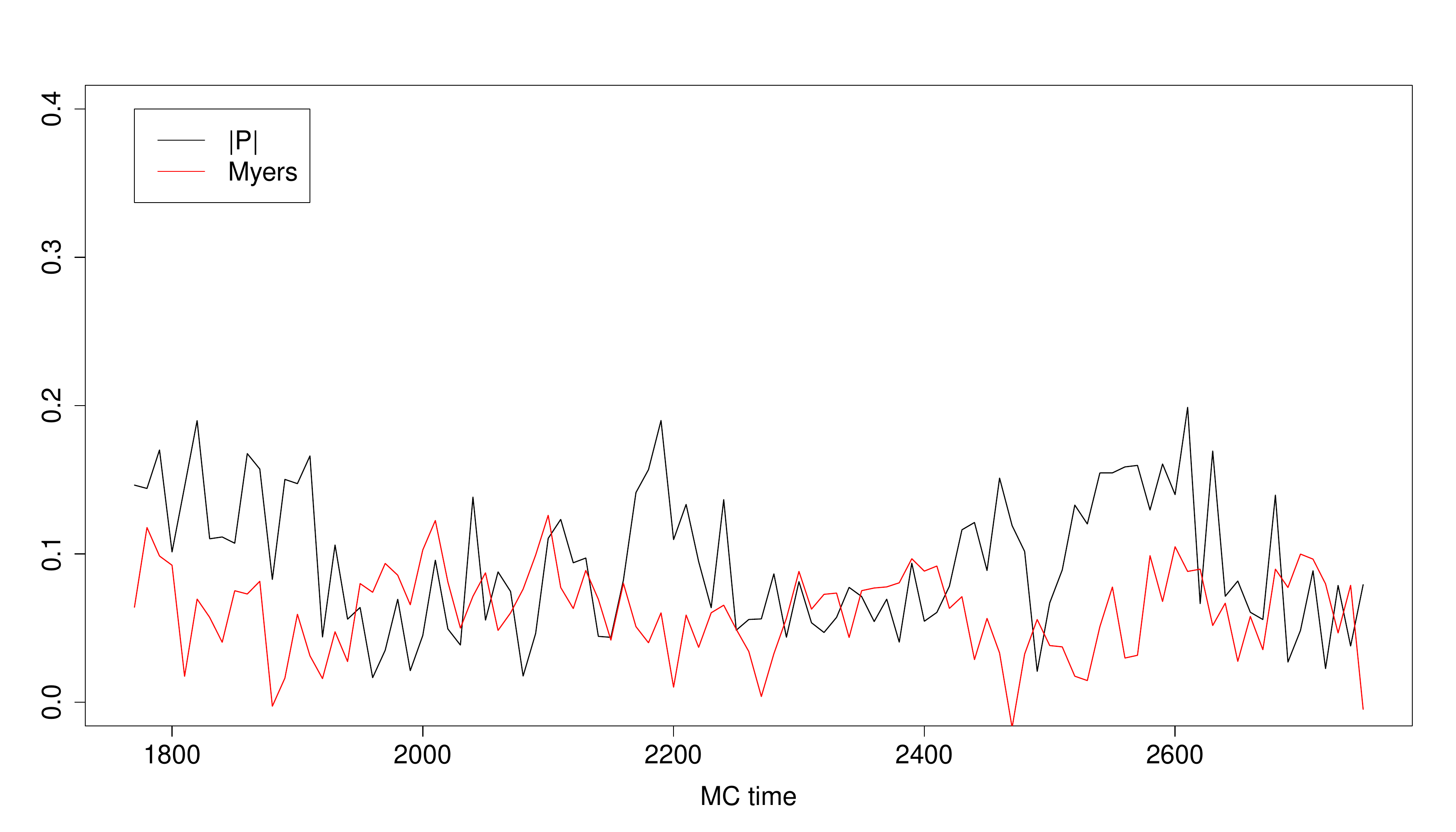}}	\\
\scalebox{0.2}{\includegraphics[trim={0 0cm 0 0},clip]{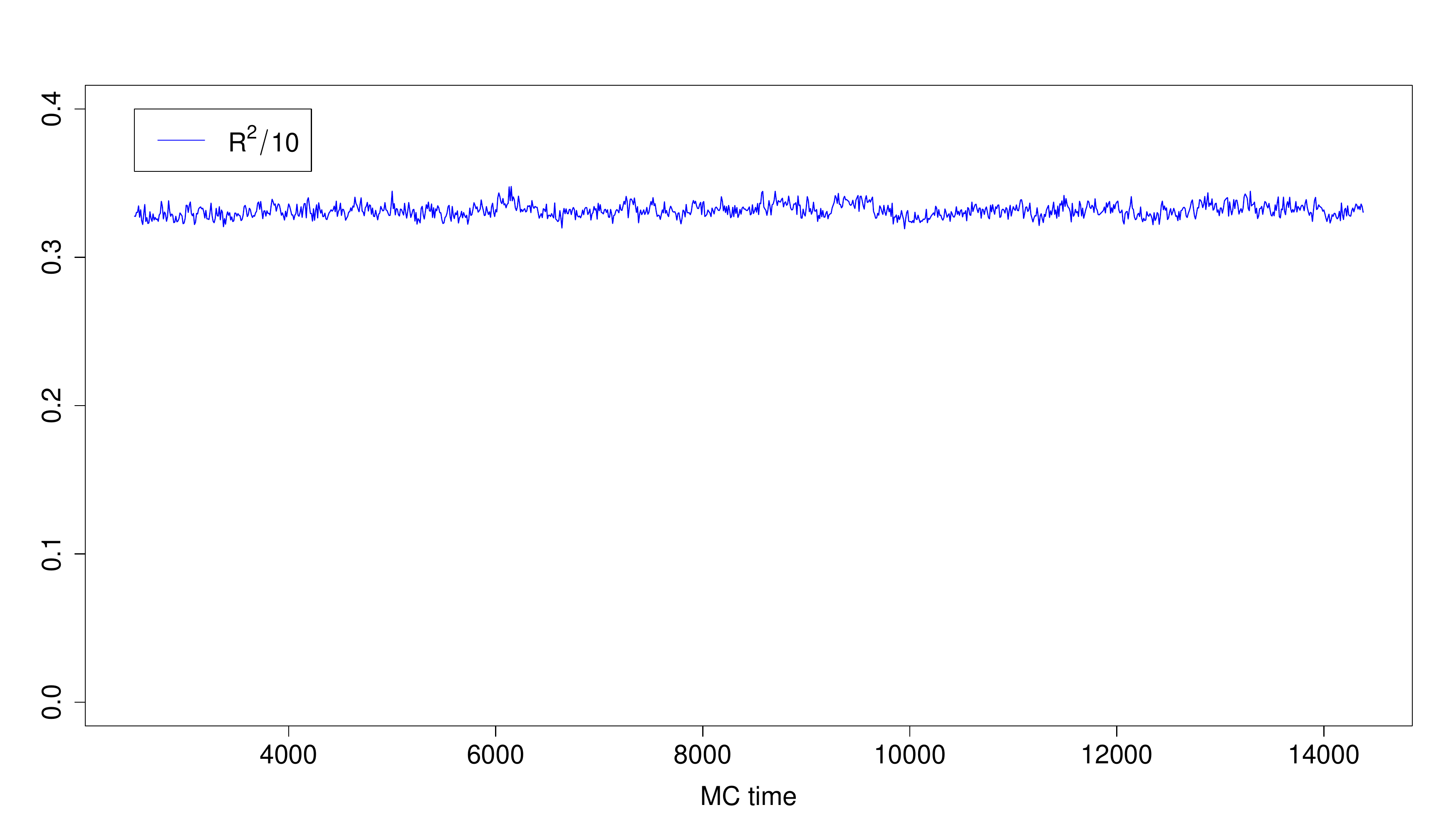}}
	\scalebox{0.2}{
		\includegraphics[trim={0 0cm 0 0},clip]{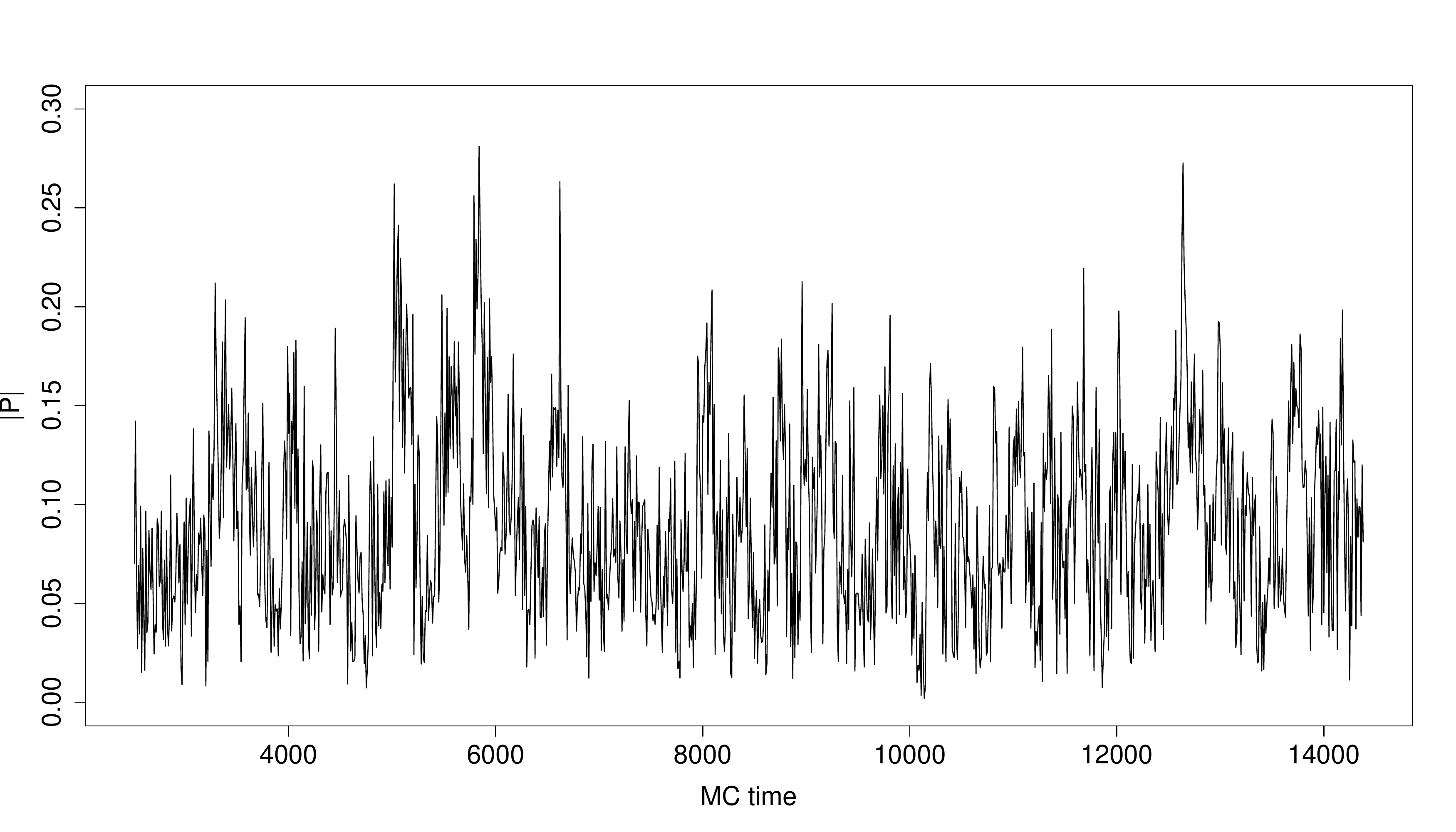}}
	\caption{Monte Carlo histories showing confined, stable states for $N=12, L=48$ and temperature $T=0.2$. From first row to third we have $\mu=0.2, 0.1,$ and $\mu=0$ respectively.
	No constraint was imposed for the simulation.
	}
	\label{fig:BFSSlimit_confined}
\end{figure}

\subsubsection{$\mu=0$: BFSS limit}\label{sec:BFSSLimit}
In this subsection, we review the information extracted from our simulations on the BFSS limit.
As expected, the run-away of scalar fields becomes more relevant as we approach $\mu=0$, especially at smaller $N$ and $L$ \cite{Anagnostopoulos:2007fw,Catterall:2008yz,Berkowitz:2016jlq}.
An interesting new observation is that the signal for metastability in a confined state even persists in the BFSS limit, see Fig.~\ref{fig:BFSSlimit_confined}.
Such a metastable state can exist because the tendency for the scalar field to diverge is reduced in the confined phase.
On the other hand, we did not observe the deconfined phase in such a low-temperature region.
As we have already seen in Sec.~\ref{sec:small-mu-region}, this deviation from the gravity expectation can be due to finite $N$ corrections. The expectation is that $T_2$ and $T_c$ decrease with increasing $N$, as $T_c, T_2\sim N^{-5/9}$, which is still sizable in our simulations (e.g., $12^{-5/9}\simeq0.251$, $16^{-5/9}\simeq0.214$). Therefore it is possible that the simulations at finite $N$ and small $T$ see only the confined phase, even though the deconfined phase exists all the way down to $T_2=T_c=0$ in the large $N$ limit.
$T_1$, on the other hand, increases with increasing $N$, which means larger $N$ stabilizes the confined metastable state up to higher temperatures. Nevertheless, this is not in contradiction with $T_c=0$ since the deconfined  state has lower free energy.

\begin{figure}
	\centering
	\scalebox{0.2}{
		\includegraphics[trim={0 0cm 0 0},clip]{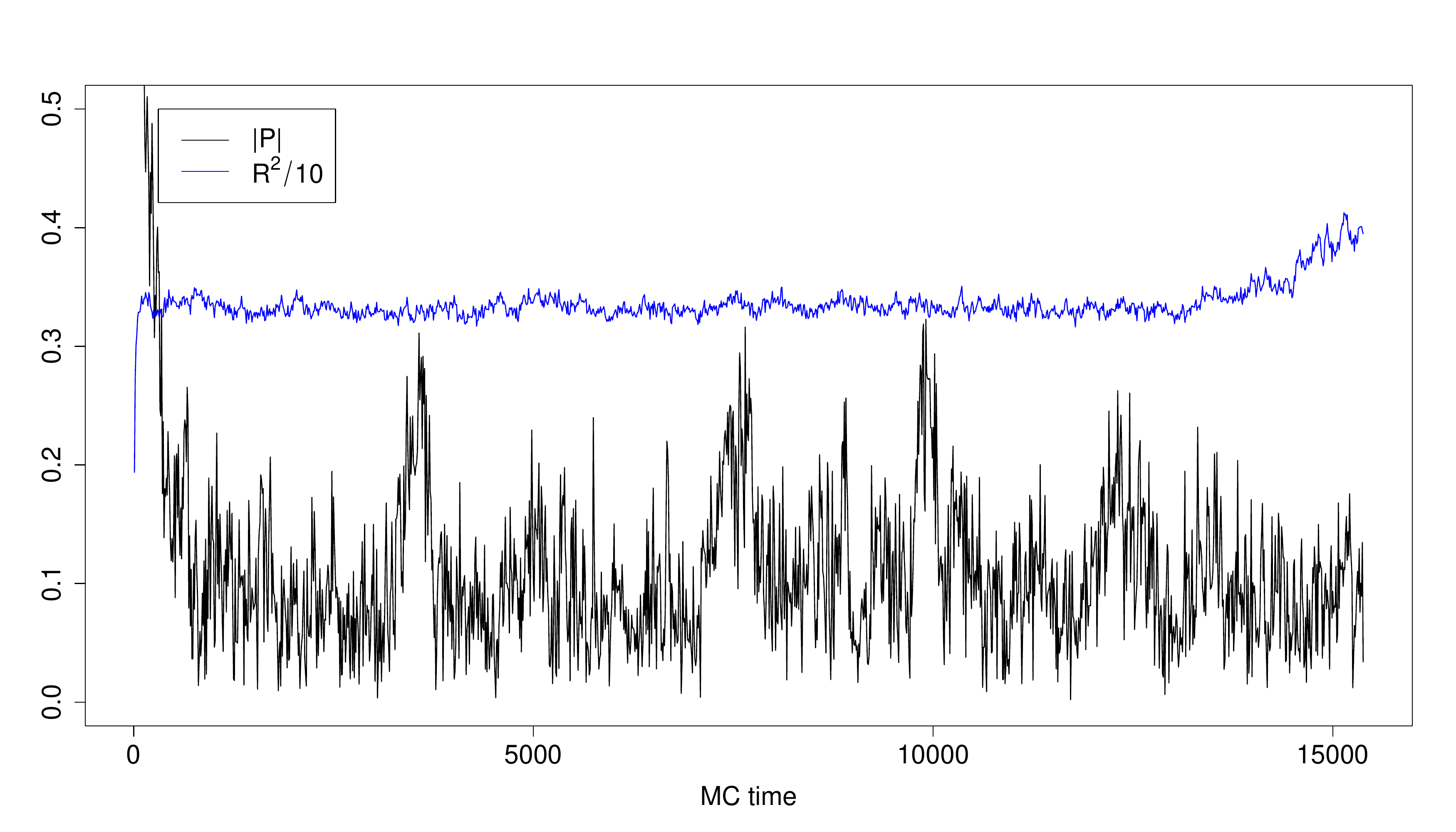}}
	\scalebox{0.2}{
		\includegraphics[trim={0 0cm 0 0},clip]{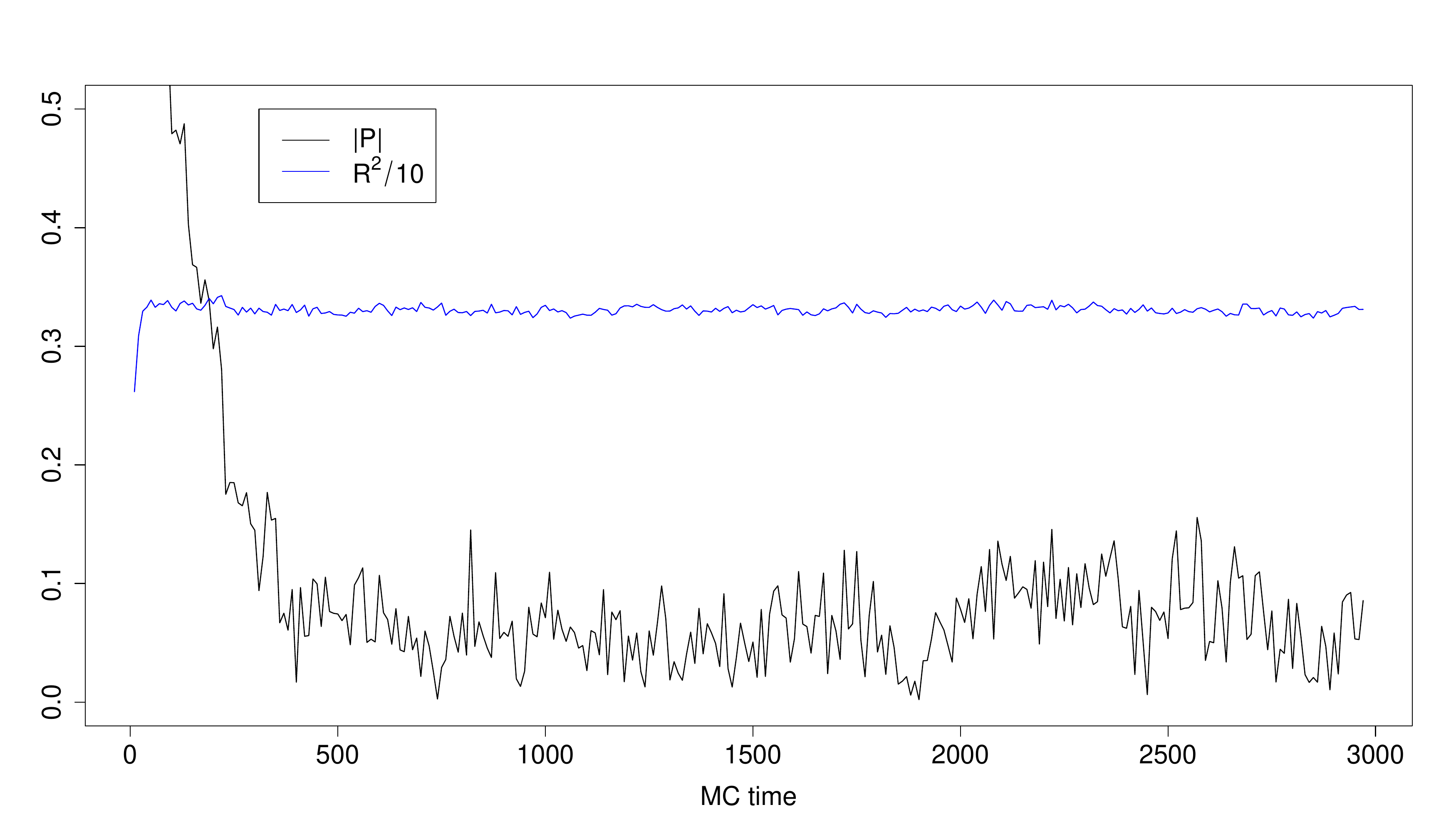}}
	\caption{Monte Carlo histories from cold starts ($X_1=X_2=\cdots=X_9=0$) for $\mu=0$, $T=0.2$, $L=48$, $N=10$ (left) and $N=16$ (right).
	For $N=10$, the onset of the run-away behavior (i.e., the increase of $R^2$) can be seen at late time.}
	\label{fig:BFSScold_start}
\end{figure}

We performed a detailed investigation at $T=0.2$, where the confined phase is sufficiently stable.
To obtain the confined phase, the initial configuration has to be tuned appropriately.
We obtained confined configurations by taking the initial configuration to be a configuration from the BMN simulation at small $\mu$, and $X_I=0$ for all $I=1,2,\cdots,9$ (`cold start'). Due to the flat directions, the latter often does not converge to a confined state, but we were able to observe such a convergence for $N=16$, $L=48$ and $N=10$, $L=48$; see Fig.~\ref{fig:BFSScold_start}.
While the confined phase appears to stabilize the flat directions to a certain extent, instabilities may still appear after many Monte Carlo steps, as shown in the left part of Fig. \ref{fig:BFSScold_start}. We generally observe that the confined state is more stable at larger $N$ and/or larger $L$. We were not able to obtain a stable confined state for a sufficiently long time below $L=30$. However, for larger $L$, by slowly increasing the temperature, we were able to see a confined phase up to about $T=0.26$.

\begin{figure}
	\centering
	\scalebox{0.25}{
		\includegraphics[trim={0 0cm 0 0},clip]{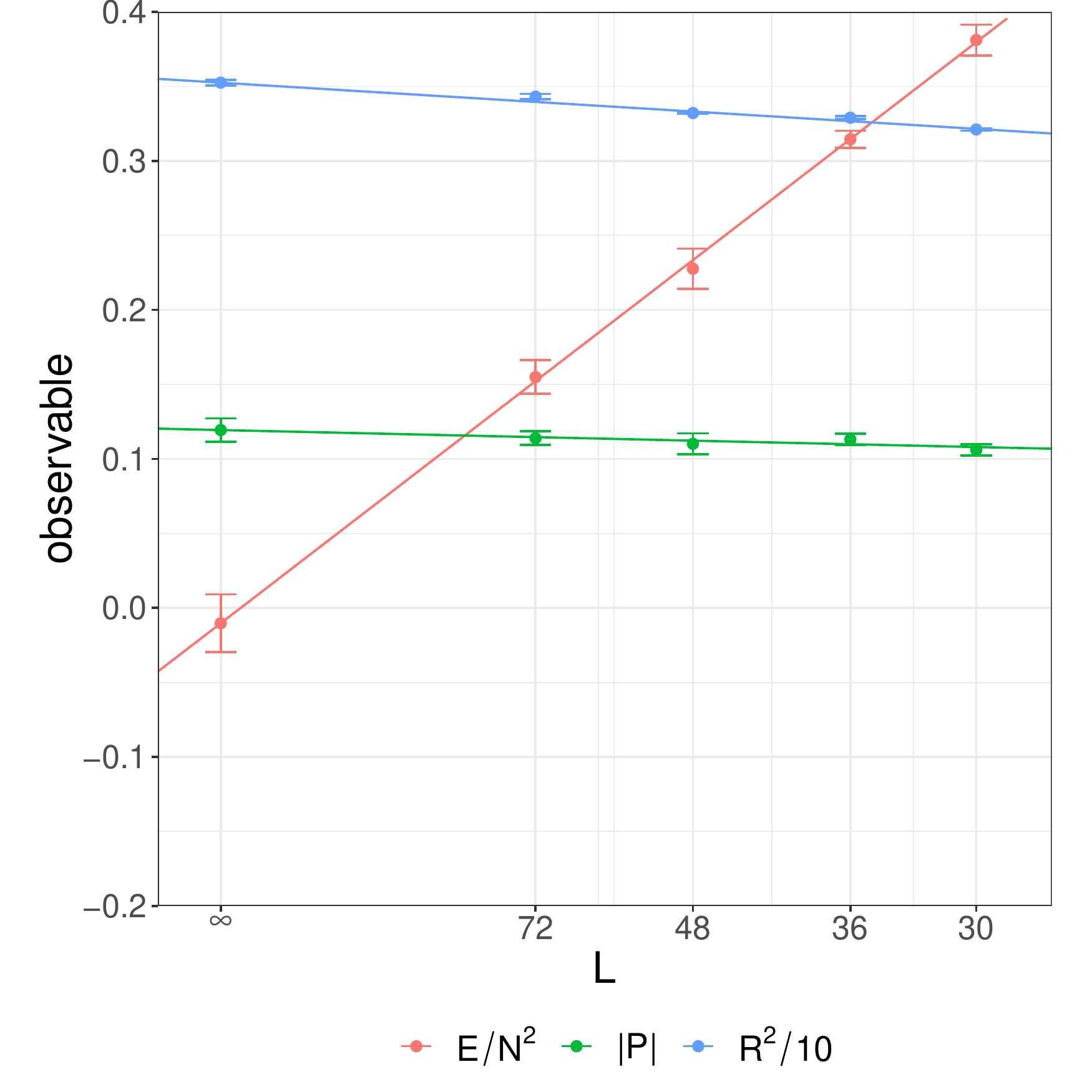}}
	\scalebox{0.25}{
		\includegraphics[trim={0 0cm 0 0},clip]{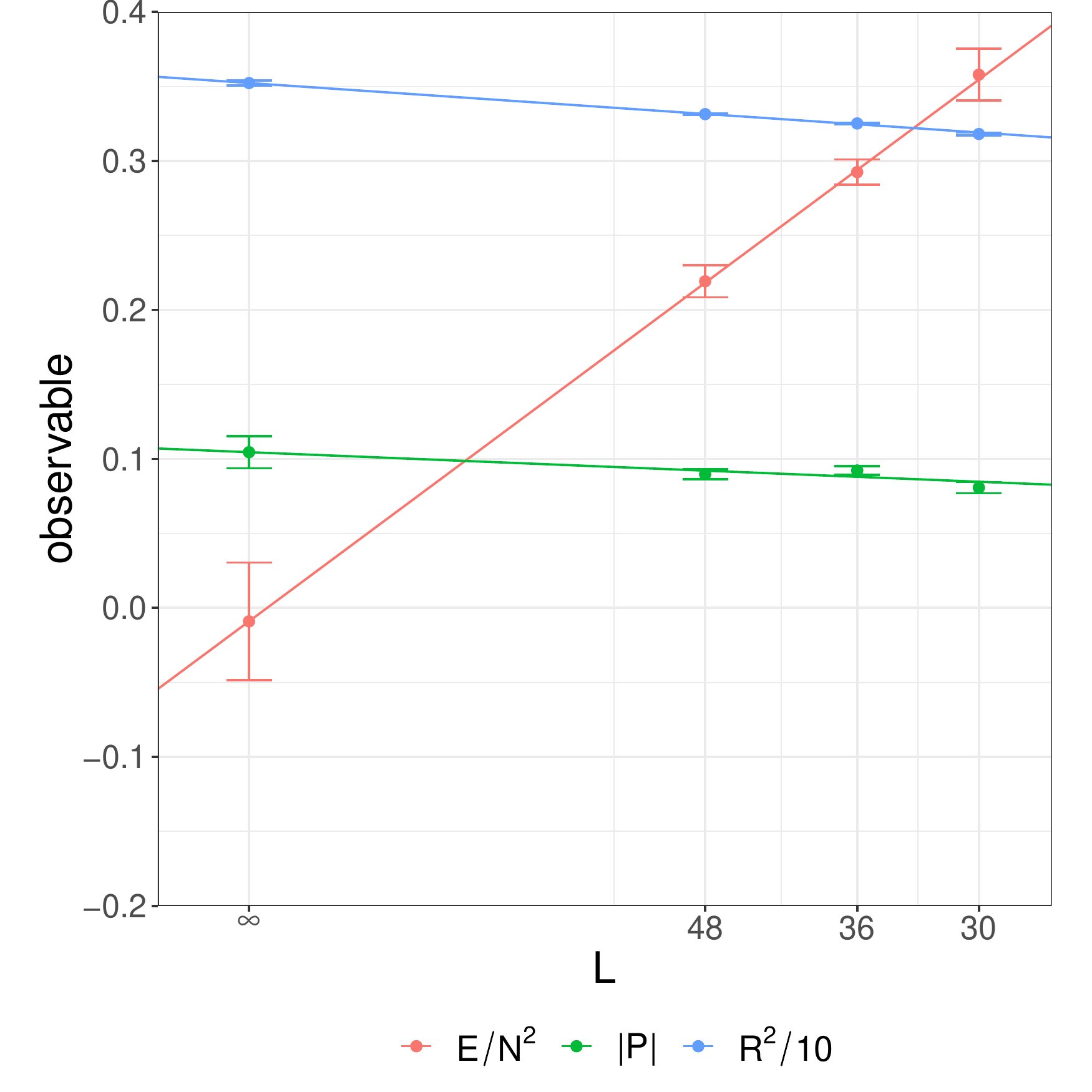}}
	\scalebox{0.25}{
		\includegraphics[trim={0 0cm 0 0},clip]{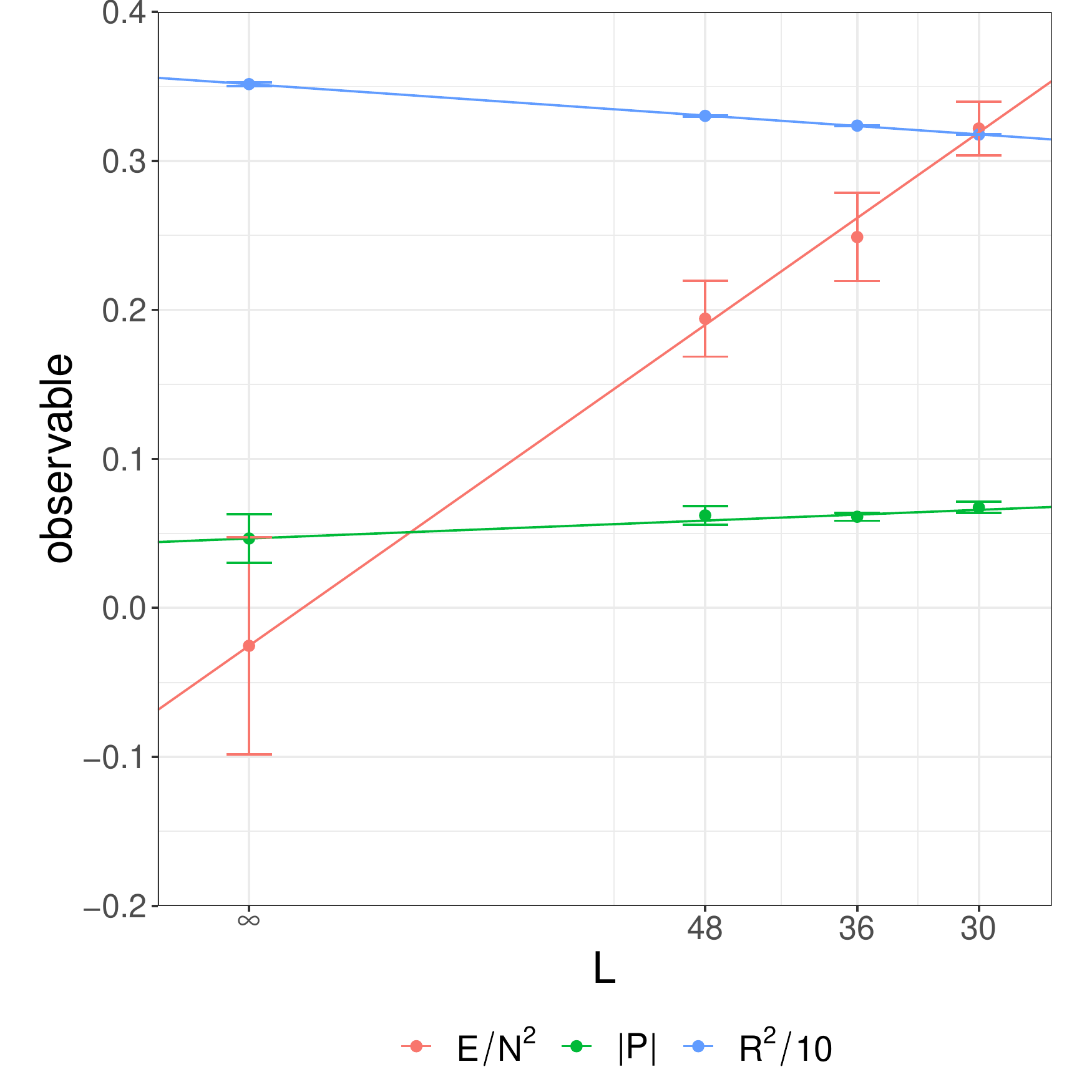}}
	\caption{Full BFSS (i.e., full BMN with $\mu=0$) at temperature $T=0.2$.
	From left to right: continuum extrapolation (lattice size $L\to\infty$) for matrix size $N=10,12,16$. The horizontal axis scales as $1/L$.
	We can see that the continuum extrapolation of $\frac{E}{N^2}$ is consistent with zero for all values of $N$. Therefore, $\frac{E}{N^2}$ is consistent with zero in the simultaneous continuum and large-$N$ limits.
	We note that for the simulation with parameters $L=30$, $N=16$, we discarded configurations with a too-large value of $R^2$ due to the run-away behavior. We show alternative plots and discuss the reasoning in Appendix~\ref{sec:appendix_L30N16}.
	}
	\label{fig:BFSSlargeL}
\end{figure}

\begin{figure}
	\centering
	\scalebox{0.25}{
		\includegraphics[trim={0 0cm 0 0},clip]{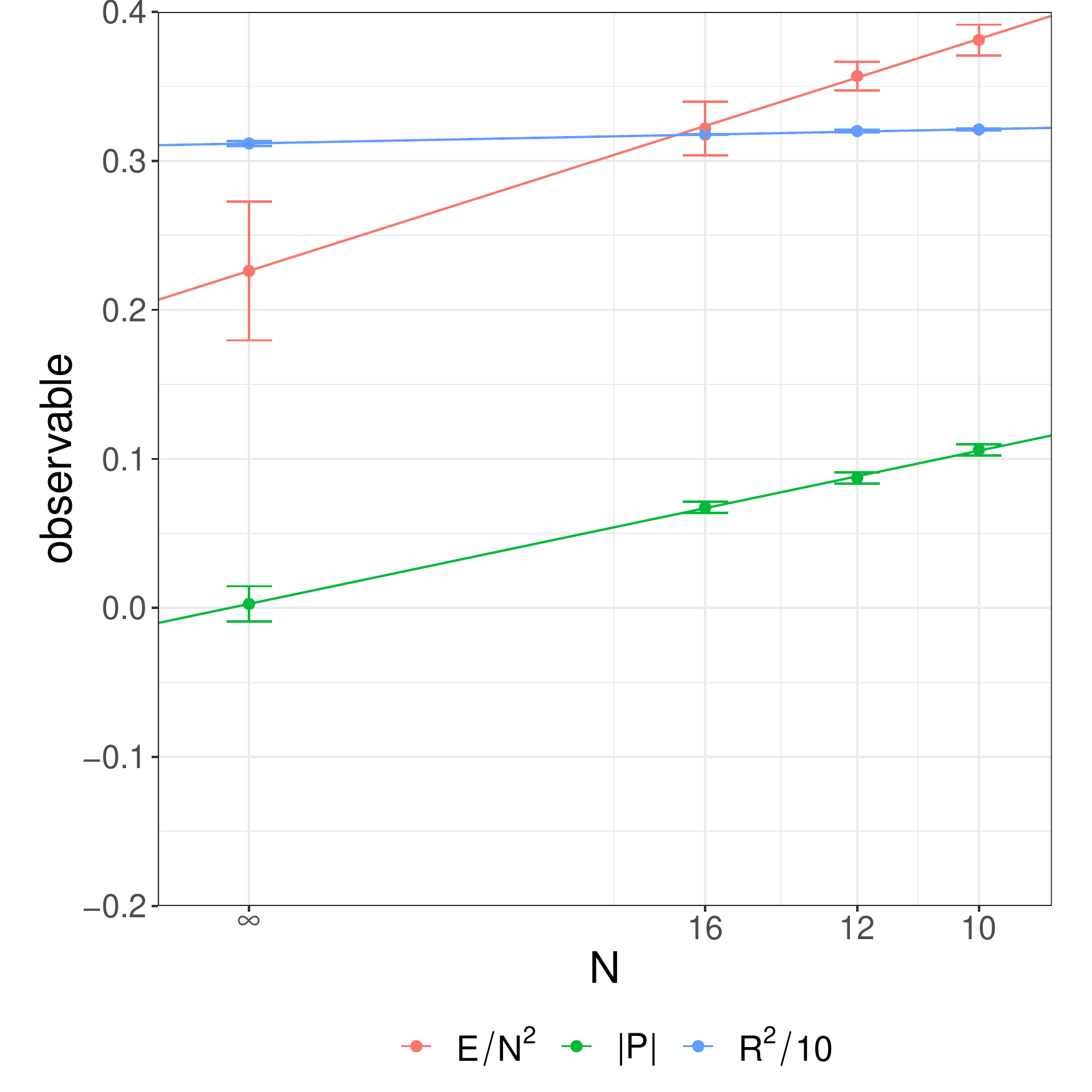}}
	\scalebox{0.25}{
		\includegraphics[trim={0 0cm 0 0},clip]{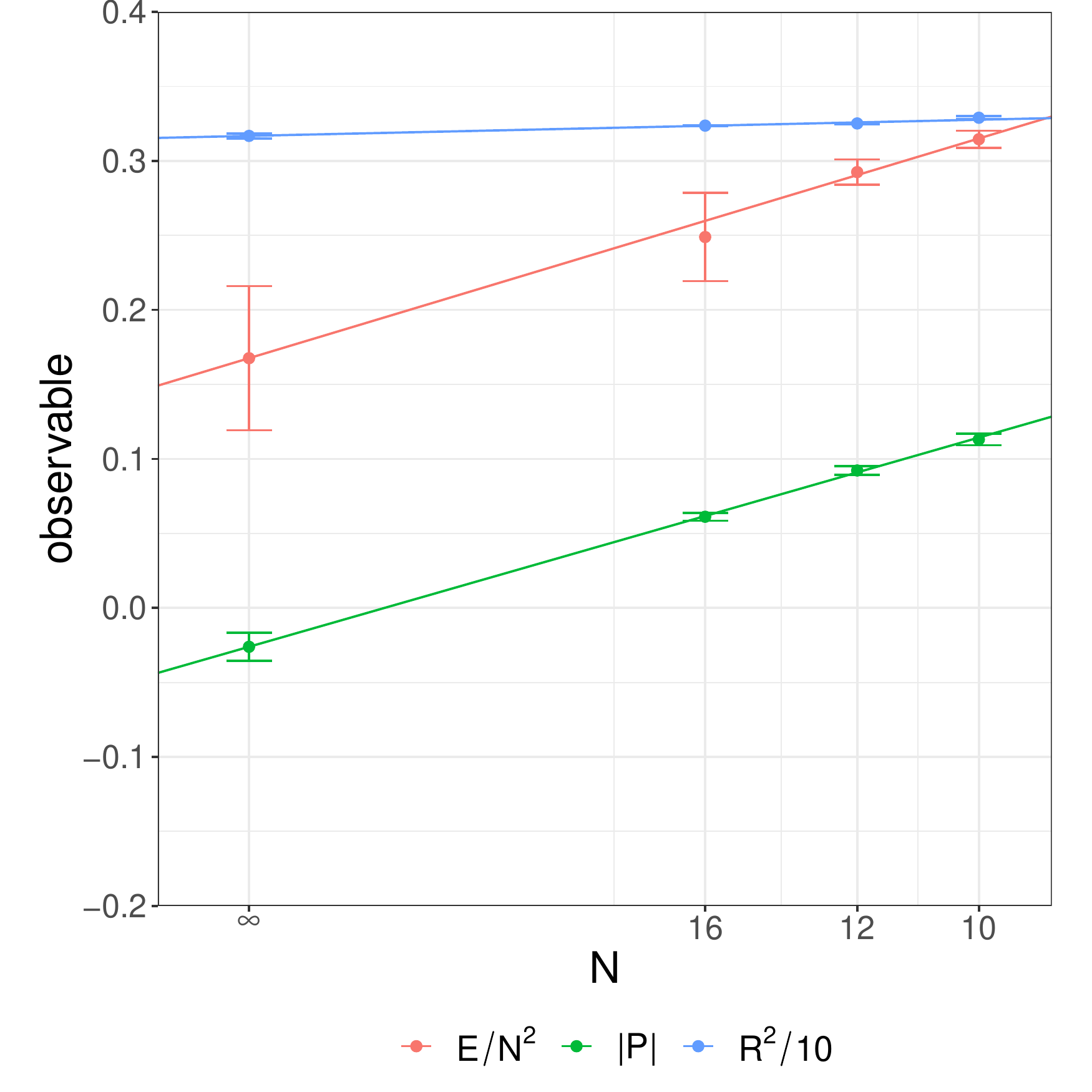}}
	\scalebox{0.25}{
		\includegraphics[trim={0 0cm 0 0},clip]{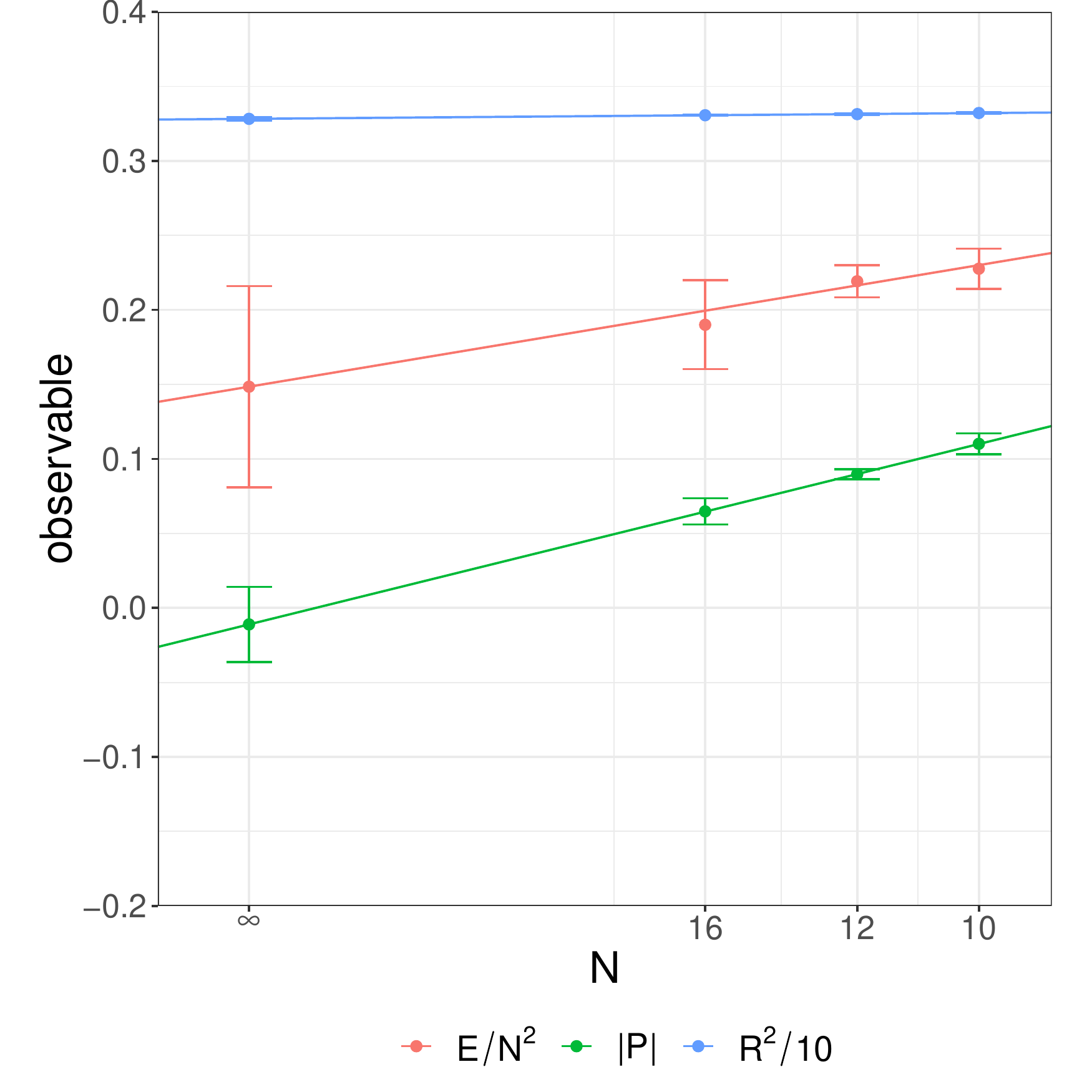}}
\caption[Caption for LOF]{Full BFSS (i.e., full BMN with $\mu=0$) at temperature $T=0.2$.
	From left to right: large $N$ extrapolation for lattice size $L=30,36$ and $48$. The horizontal axis scales{\footnotemark} as $1/N$. We can see that the large-$N$ extrapolation of the Polyakov loop is very close to zero for all values of $L$. As a result, the Polyakov loop is consistent with zero in the simultaneous continuum and large-$N$ limits. We again refer to Appendix~\ref{sec:appendix_L30N16} for the datapoint with $L=30$, $N=16$.}
		\label{fig:BFSSlargeN}
\end{figure}

To make sure that we are observing the confined phase, we performed continuum extrapolations (linear in $1/L$) for $N=10, 12$ and $16$, as well as large $N$ extrapolations (linear in $1/N$) for $L=30,36,48$, at $T=0.2$; see Figs.~\ref{fig:BFSSlargeL}, \ref{fig:BFSSlargeN}.
(To obtain these data points, we only used the parts of the Monte Carlo histories before the instabilities due to the flat directions set in.)
We observe that the Polyakov loop $P$ is consistent with zero in the large $N$ limit, and thus consistent with a confined phase. Similarly, the energy is consistent with zero in the continuum limit independently of $N$.\footnote{
Due to supersymmetry, the zero-point energy is zero.}
Note that, in the deconfined phase, we expect larger values of the energy and the Polyakov loop that are clearly distinguishable from zero.
The energy of the deconfined phase predicted by the gravity dual (the black zero-brane in type IIA superstring theory) is $\frac{E}{N^2}\simeq 7.41T^{14/5}\simeq 0.0818$ at $T=0.2$, up to string-theoretical corrections. As for the Polyakov loop in the deconfined phase, Ref.~\cite{Anagnostopoulos:2007fw} observed $P\simeq\exp\left(-\frac{0.15}{T}+0.072\right)$ at $0.475\le T\lesssim 1$. This fit agrees well with the values of $P$ at $T\ge 0.4$ obtained in Ref.~\cite{Berkowitz:2016jlq}. By assuming that this fit is valid at lower temperatures as well, we obtain $P\simeq 0.51$ at $T=0.2$.

\section{Conclusion and discussion}

In this paper, we studied the thermodynamic features of the BFSS matrix model and BMN matrix model. Specifically, we studied the confinement/deconfinement transition in the trivial background. 
We conjectured the phase structure based on dual gravity description~\cite{Itzhaki:1998dd,Costa:2014wya} and previous numerical studies. 
According to our conjecture, there are two kinds of confined phases: the completely-confined phase, which is the (local) minimum 
of the free energy, and the partially-confined phases (eleven-dimensional Schwarzschild black hole), which is the local maximum of the free energy separating the completely-confined phase (M-theory vacuum up to a small thermal fluctuation) and completely-deconfined phase (black zero-brane in type IIA string theory). 
We provided the evidence supporting this conjecture by performing lattice Monte Carlo simulations, modulo a few subtleties associated with the finite-$N$ effects. 
It is natural to expect that the confined phase of the BFSS matrix model can describe M-theory, and hence, we might have finally reached the M-theory region.

\footnotetext{As opposed to the deconfined phase, it is not clear a priori whether $\frac{E}{N^2}$ and $R^2$ should have $1/N$ or $1/N^2$ corrections. Both options provide a reasonable fit to our data and we chose the $1/N$ option in figure \ref{fig:BFSSlargeL}.}

The relationship between the flux parameter $\mu$ and critical temperature $T_c$ is shown in Fig.~\ref{fig:tc_mu}. 
Naively, it may appear that the matrix model and gravity disagree at small $\mu$. However, this is not the case; the gravity line is from the strict large-$N$ limit, while the matrix model results are obtained at finite $N$, and the apparent disagreement is consistent with the gravity analysis and finite-$N$ correction. This apparent deviation is actually good news!

Because such a confined phase was not observed in the past, the consistency has to be discussed carefully. 
All studies but Ref.~\cite{Hanada:2013rga} focused on the high-temperature region (for example, Ref.~\cite{Berkowitz:2016jlq} studied $T\ge 0.4$, $N\le 32$, $L\le 32$), by gradually lowering the temperature, and hence, it could be a problem if the confined phase had been observed. It is straightforward to go to larger $N$, say $N=64$ or $128$, at $0.4\lesssim T$, and we might be able to see the confined phase there if the sufficiently fine lattice is used and the initial configuration is carefully tuned. It would be a good consistency check. 
Ref.~\cite{Hanada:2013rga} studied lower temperature ($T\sim 0.1$) and much smaller $N$ ($N=3,4,5$). 
They introduced the cutoff for ${\rm Tr}X_I^2$ and studied the unstable phase, whose energy was consistent with the deconfined phase. 
This phase may be stabilized at sufficiently large $N$ and become the stable deconfined phase. 

One surprise, which we did not expect when we started this project, was that the confined phase exists up to a rather high temperature, even in the BFSS limit. Concerning Monte Carlo simulations, it is very good news: at higher temperatures, the temporal circle is smaller, and hence, the simulation cost is smaller. 
Therefore, the detailed study of M-theory via Monte Carlo simulation of the BFSS or BMN matrix models might be doable with much smaller computational resources than previously expected. 
Nontrivial fuzzy-sphere backgrounds corresponding to M2-branes or M5-branes may also be tractable targets.  
To deepen the understanding of the duality in the M-theory region, it is important to perform calculations in the gravity side that can be compared to the simulation results in the matrix model side. Of equal importance is a similar confinement-to-deconfinement analysis for the fuzzy sphere backgrounds, since in this case the precise dependence of the relevant temperature with $N$ is not known. This is an interesting problem, not only from the matrix model side but also from the gravity analysis also, yet the difficulty to extract precise results is emanating from carefully manipulating multi-centered supergravity solutions \cite{LinTheSupergravityDual}. A better understanding of these solutions and their connection with analytic and Monte Carlo results, along the line of the trivial vacuum analysis, will lead to a better M-theory interpretation.   

Note that our analysis on the M-theory regime of the gravity side was only qualitative. In order to understand the duality better, it is important to improve the analysis.
It is natural to think that only a part of the matrix degrees of freedom is excited and form a black hole and graviton gas~\cite{Banks:1997hz,Banks:1997tn,Banks:1997cm,Hanada:2021ipb,Hanada:2016pwv}, but the rest of the degrees of freedom can still contribute to the emergence of the background spacetime~\cite{Hanada:2021ipb}. 
The number of degrees of freedom forming a black hole can be determined from the distribution of the Polyakov line phases~\cite{Hanada:2020uvt}. 
As a rough estimate, it would be reasonable to identify it with the entropy~\cite{Banks:1997hz,Banks:1997tn,Banks:1997cm,Horowitz:1997fr}. 

For extremely low temperatures, we expect to enter the M-theory regime \cite{Itzhaki:1998dd}. That is because the small temperature region is related to the small energy regime and the latter is expected to be connected with M-theory. Therefore, it might be possible that we can access the M-theory region by keeping $\mu$ fixed and taking the limit $T\ll1$. This would correspond to a BMN M-theory description with a finite deformation, and indeed for low temperatures, this is assumed to be described by the M5-brane of M-theory \cite{Berenstein:2002jq}. We can see hints for this feature since the clustering of the six matrices that construct the five-sphere leads to an average mean value bigger than the rest of the three matrices that construct the $SO(3)$ part. Indeed, recalling Fig.~\ref{fig:fluctuations} this phenomenon can be observed in simulations and we will report a more detailed analysis in a different project. We may also note that in the very low temperature regime  we cannot compare with results from the gravity side, since at this level there is no such precise analysis for the M5-brane.

The completely-confined phase is rather stable, and hence, can be studied straightforwardly. 
The study of the partially-confined phase can be tricky because it is unstable in the canonical ensemble. Still, it may be doable by constraining the value of the Polyakov loop appropriately. 
This phase, which is expected to be the dual of the eleven-dimensional Schwarzschild black hole, would offer us an ideal framework to study the black hole evaporation. It might be possible to study the region near the border between string theory and M-theory ($T\sim T_2$) in a similar manner by introducing appropriate constraints. 

Qualitatively, the BMN matrix model resembles four-dimensional maximal super Yang-Mills on three-sphere. For the latter, deconfinement is studied not just by using the thermal boundary partition function but also with the index~\cite{Choi:2018vbz}. The index is calculable analytically even at strong coupling, and impressive agreement with gravity has been observed~\cite{Copetti:2020dil}. 
Such an approach might be useful for the BMN matrix model as well, providing connections to the counting of the supersymmetric black hole in an analytically tractable manner. 

Rich phase diagrams are expected for other gauge theories as well.
Qualitative aspects of the phase diagrams of maximally supersymmetric Yang-Mills theories was discussed in Ref.~\cite{Itzhaki:1998dd} by utilizing dual gravity pictures and string dualities.
These theories may exhibit confinement, as we observed for the matrix model. 
Two- and three-dimensional theories can be studied on lattice without having the parameter fine-tuning problem, and simulations on small lattices are already tried; see Refs.~\cite{Catterall:2010fx,Catterall:2017lub,Kadoh:2017mcj} for the two-dimensional theory and Ref.~\cite{Catterall:2020nmn} for the three-dimensional theory. 
Lattice Monte Carlo simulation of these theories can be a powerful tool to reveal the nonperturbative aspects of string/M-theory and holography further. 

Last but not least, it is important to understand how the information of eleven-dimensional spacetime is encoded in the BFSS matrix model in which only ten of eleven dimensions can be seen manifestly; see e.g.~Ref.~\cite{Yoneya:2016aja}. We might be able to get some hints by studying the M-theory parameter region of  the BFSS matrix model and identifying the gravity dual precisely.

\acknowledgments
The authors would like to thank Oscar Dias, Yoshifumi Hyakutake, Jack Holden, Seok Kim, Andy O'Bannon, Juan Maldacena, Andreas Rabenstein, Jorge Santos, Steve Shenker, Lenny Susskind, Toby Wiseman and Tamiaki Yoneya for discussions. They thank the ECT* for its hospitality during the workshop Quantum Gravity meets Lattice QFT where this work was initiated.  
G.~B.\ acknowledges support from the Deutsche Forschungsgemeinschaft (DFG) Grant No.\ BE 5942/3-1.
N.~B.\ and S.~P.\ were supported by an International Junior Research Group grant of the Elite Network of Bavaria.
E.~R.\ is supported by Nippon Telegraph and Telephone Corporation (NTT) Research.
M.~H.\ was supported by the STFC Ernest Rutherford Grant ST/R003599/1.
H.~W.\ is supported in part by the JSPS KAKENHI Grant Number JP	21J13014.
The numerical simulations were performed on ATHENE, the HPC cluster of the Regensburg University Compute Centre, and QPACE 4. A.S. thanks the University of the Basque Country, Bilbao, for hospitality.

\bibliographystyle{JHEP}
\bibliography{matrix-model}
\newpage
\appendix
\section{Summary of simulations}\label{appendix:simulation_parameters}

In this appendix, we present a list of the simulations done to obtain the data underlying the findings of this paper. Very short Monte Carlo chains, especially at temperatures far away from the transition temperature, which did not enter the conclusions drawn in our work, are omitted here. None of the omitted data contradicts the presented findings.

In the column labeled ``constraint," we refer by ``$M: x_{\rm min}, x_{\rm max}, \gamma$" to a constraint on the Myers term and by ``$P: x_{\rm min}, x_{\rm max}, \gamma$" to a constraint on the Polyakov loop. Details on the constraints are given in Section~\ref{sec:simsetup}.

\renewcommand\theadalign{bc}
\renewcommand\theadfont{\bfseries}
\renewcommand\theadgape{\Gape[0pt]}
\renewcommand\cellgape{\Gape[0pt]}

{\scriptsize
\begin{tabular}{|p{0.02\linewidth}|p{0.02\linewidth}|p{0.02\linewidth}|p{0.21\linewidth}|p{0.18\linewidth}|p{0.35\linewidth}|}
\hline
 $\mu$ & $L$ & $N$ & $T$ & constraint & observation\\
 \hline \hline
 5.0  & 24 & 32 & 1.172, 1.176, 1.18, 1.182 & none & Stable confined phase at $T=1.172$, tunnelling at $1.176$, stable deconfined phase at $T\geq1.18$. ($\sim 5000-8000$ MC steps each)\\ \hline
  5.0  & 12 & 48 & 1.188, 1.189, 1.19 & none & Stable confined phase at $T=1.188$, tunnelling at $1.189$, stable deconfined phase at $T=1.19$. ($\sim 3000$ MC steps each)\\ \hline \hline
  3.0  & 12 & 32 & 0.752, 0.753, \ldots, 0.757 & none & See Fig. \ref{fig:PDM3} + stream mainly confined for $T=0.752, 0.753$. \\ \hline
    3.0  & 24 & 32 & 0.75, 0.752 & ~ & Repeated tunneling for $T=0.75$ ($\sim 12000$ MC steps), deconfined for $T=0.752$ ($\sim 8000$ Steps). \\ \hline \hline
        2.0  & 12 & 24 & 0.538, 0.539, \ldots, 0.544 & none & See Fig. \ref{fig:Pol_mu=2_neat_Tc} + more pronounced confinement at $T<0.542$. \\ \hline
        2.0  & 12 & 32 & 0.537, 0.538, \ldots, 0.544 & none & See Fig. \ref{fig:Pol_mu=2_neat_Tc} + more pronounced confinement at $T<0.542$. \\ \hline
        2.0  & 24 & 24 & 0.53, 0.535, 0.54 & none & Tunneling to confined phase at $T=0.53$, repeated tunneling at $T=0.535$, stable deconfined phase at $T=0.54$. . \\ \hline  \hline

         1.6  & 24 & 32 & 0.45, 0.451, \ldots, 0.454 & none & Mix of confined and deconfined signals without repeated tunneling and clear systematics, suggesting strong hysteresis in this temperature range. See Fig. \ref{BMN_mu1.6} for $T=0.452$. \\ \hline  \hline

         1.5  & 24 & 24 & 0.43, 0.432, \ldots, 0.44 & none & Initially confined simulations in the trivial background remain in this phase for some time and, if leaving this phase, tunnel first to a deconfined phase in the trivial background and shortly after to a deconfined fuzzy sphere phase, see Fig. \ref{fig:M15-instability}. Initially deconfined configurations in the trivial background all quickly tunnel to a fuzzy sphere background, see Fig. \ref{fig:M15-instability}. \\ \hline

1.5  & 24 & 8 & 0.429 & P: 0.34,~0.46,~5000 M: 0,~0.035, 30000 & Double constraint used to efficiently generate data for Fig. \ref{Pgap_large_N_M<2}.  \\ \hline
1.5  & 24 & 12 & 0.429 & P: 0.34,~0.46,~5000 M: 0,~0.035, 30000 & Double constraint used to efficiently generate data for Fig. \ref{Pgap_large_N_M<2}. \\ \hline
1.5  & 24 & 16 & 0.429 & P: 0.36,~0.48,~5000 M: 0,~0.035, 30000 & Double constraint used to efficiently generate data for Fig. \ref{Pgap_large_N_M<2}. \\ \hline
1.5  & 24 & 24 & 0.429 & P: 0.34,~0.46,~5000 M: 0,~0.035, 30000 & Double constraint used to efficiently generate data for Fig. \ref{Pgap_large_N_M<2}. Polyakov loop repeatedly moves between P-constraint boundaries, indicating that $T=0.429$ is very close to the critical temperature. \\ \hline
	 \hline
\end{tabular}
}

{\scriptsize
\begin{tabular}{|p{0.02\linewidth}|p{0.02\linewidth}|p{0.02\linewidth}|p{0.2\linewidth}|p{0.2\linewidth}|p{0.35\linewidth}|}
\hline
 $\mu$ & $L$ & $N$ & $T$ & constraint & observation\\
 \hline \hline

          1.0  & 12 & 12 & 0.29, 0.3, 0.31, 0.32 & M: 0,~0.03,~500000 & See Fig. \ref{fig:Pol_mu=1_neat_Tc} + clearly confined signal for $T=0.29$\\ \hline

         1.0  & 12 & 16 & 0.3, 0.31, 0.32 & M: 0,~0.03,~500000 & See Fig. \ref{fig:Pol_mu=1_neat_Tc} \\ \hline

          1.0  & 24 & 12 & 0.3, 0.31, 0.315, 0.32, 0.33 & M: 0,~0.03,~500000 & See Fig. \ref{fig:Pol_mu=1_neat_Tc}  + clearly confined signal for $T=0.3$\\ \hline

	  1.0  & 36 & 12 & 0.315, 0.32 & M: 0,~0.03,~500000 & Confined signal for $T=0.315$, repeated tunneling for $T=0.32$.\\ \hline

	  1.0  & 48 & 12 & 0.315 & M: 0,~0.03,~500000 & Confined signal for $T=0.315$.\\ \hline \hline

	   0.9  & 24 & 16 & 0.24, 0.25, 0.26, 0.27, 0.28, 0.29  & M: 0,~0.04,~30000 & Confined state at $T=0.27$.\\ \hline \hline

	 0.8 & 12 & 12 & 0.24, 0.25, 0.26 & M: 0,~0.025,~15000 & See Fig. \ref{fig:jointbins_Polyak_GN12S12M08D9}.  \\ \hline

	 0.8 & 24 & 12 & 0.25, 0.26, 0.27 & M: 0,~0.025,~15000 & See Fig. \ref{fig:jointbins_Polyak_GN12S12M08D9}.  \\ \hline

	 0.8 & 36 & 8 & 0.26 & M: 0,~0.025,~15000 & Some tunneling, mainly confined. Confirms increasing $T_c$ in continuum limit.  \\ \hline

	  0.8 & 48 & 8 & 0.26 & M: 0,~0.025,~15000 & Some tunneling, mainly confined. Confirms increasing $T_c$ in continuum limit. \\ \hline \hline

 0.6 & 24 & 16 & 0.2, 0.21, 0.22, 0.23, 0.24, 0.25, 0.26, 0.27, 0.28, 0.29, 0.3, 0.31 & M: 0,~0.04,~30000 & See Fig.~\ref{Fig:mu06T1T2}. \\ \hline \hline

 0.5 & 24 & 12 & 0.22, 0.25, 0.26 & M: 0,~0.02,~15000 & Blurred two-state signal at $T=0.25$ (repeated tunneling), confined at $T=0.24$, deconfined at $T=0.26$  \\ \hline
  0.5 & 24 & 16 & 0.26 & M: 0,~0.02,~15000 & Stable ($\sim$ 2000 MC steps) confined and deconfined signals at $T=0.26$ after starting with suitable initial conditions. Clear two-state signal once streams are combined, see Fig. \ref{fig:phase_bins_smallmu}. \\ \hline
  0.5 & 36 & 12 & 0.25, 0.4 & M: 0,~0.02,~15000 & Stable ($\sim$ 2000 MC steps) confined signal at $T=0.25$. Immediate tunneling from confined to deconfined phase with $P \approx 0.7$ for $T=0.4$. \\ \hline
  0.5 & 48 & 12 & 0.26 & M: 0,~0.02,~15000 & Stable ($\sim$ 1000 MC steps) confined signal at $T=0.26$ \\ \hline \hline

   0.3 & 24 & 12 & 0.16, 0.2, 0.22, 0.24 & M: 0,~0.02,~15000 & See Fig. \ref{fig:phase_bins_smallmu}  \\ \hline

      0.3 & 48 & 12 & 0.23 & M: 0,~0.02,~15000 & Confined signal.   \\ \hline \hline

      0.2 & 48 & 12 & 0.2 & M: 0,~0.02,~15000 & Confined signal.   \\ \hline

      0.2 & 48 & 12 & 0.2 & none & Very similar for both $P$ and $M$ to constrained simulation. \\ \hline \hline

      0.1 & 48 & 12 & 0.2, 0.24, \ldots, 0.36 & none & Starting from a confined configuration: stable confined signal for $T\leq 0.24$, immediate divergence of $R^2$ for $T \geq 0.28$. \\ \hline \hline

      0.01 & 48 & 12 & 0.2 & none & Starting from a confined configuration: stable confined signal. \\ \hline
 \hline
\end{tabular}
}

{\scriptsize
\begin{tabular}{|p{0.02\linewidth}|p{0.02\linewidth}|p{0.02\linewidth}|p{0.21\linewidth}|p{0.1\linewidth}|p{0.35\linewidth}|}
\hline
 $\mu$ & $L$ & $N$ & $T$ & constraint & observation\\
 \hline \hline

  	0 & 24 & 10 & 0.2 & none & Divergence of $R^2$ after about $1000$ MC steps, not enough stable MC steps for reliable measurements. \\ \hline

        0 & 24 & 16 & 0.2  & none & Divergence of $R^2$ after about $1000$ MC steps, not enough stable MC steps for reliable measurements. \\ \hline

       0 & 30 & 10 & 0.2 & none & See Section~\ref{sec:BFSSLimit}. \\ \hline

        0 & 30 & 12 & 0.2  & none & See Section~\ref{sec:BFSSLimit}. \\ \hline

        0 & 30 & 16 & 0.2 & none & See Section~\ref{sec:BFSSLimit}. \\ \hline

        0 & 36 & 10 & 0.2 & none & See Section~\ref{sec:BFSSLimit}. \\ \hline

        0 & 36 & 12 & 0.2  & none & See Section~\ref{sec:BFSSLimit}. \\ \hline

        0 & 36 & 16 & 0.2 & none & See Section~\ref{sec:BFSSLimit}. \\ \hline

        0 & 48 & 10 & 0.2 & none & See Section~\ref{sec:BFSSLimit}. \\ \hline

        0 & 48 & 12 & 0.2, 0,24, 0.25, 0.26, 0.28 & none & Starting from a confined configuration: stable confined signal for $T\leq 0.24$; divergence of $R^2$ after metastable confined signal for $T=0.25$; immediate divergence of $R^2$ for $T \geq 0.26$ and $P$ increasing to $\sim 0.5$. See also Section~\ref{sec:BFSSLimit}. \\ \hline

        0 & 48 & 16 & 0.2 & none & See Section~\ref{sec:BFSSLimit}. \\ \hline

	0 & 72 & 10 & 0.2 & none & See Section~\ref{sec:BFSSLimit}. \\ \hline
 \hline
\end{tabular}
}

\section{MC histories}\label{sec:MC_histories}

All MC histories not shown here but used (for $\mu<2$ and $\mu > 3$ as well for higher N and S) are quite short, e.g. about 3000-5000 Monte Carlo steps.
 
\section{Detail of the lattice regularization}\label{sec:lattice_regularization}

Below, we explain the details of the lattice regularization. The action is the same as the one used in Ref.~\cite{Berkowitz:2016jlq}, except that the Myers term is added.
\subsection{Gauge fixing}
\hspace{0.51cm}
The action of the BMN matrix model given in Sec.~\ref{sec:BMN_definition} is invariant under the SU($N$) gauge transformation.
For numerical efficiency, we take the static diagonal gauge,
\begin{eqnarray}
A_t=\frac{1}{\beta}\cdot{\rm diag}(\alpha_1,\cdots,\alpha_N),
\qquad
-\pi<\alpha_i\le\pi.
\end{eqnarray}
Associated with this gauge fixing, we add the Faddeev-Popov term
\begin{eqnarray}
S_{F.P.}
&= &
-
\sum_{i<j}2\log\left|\sin\left(\frac{\alpha_i-\alpha_j}{2}\right)\right|
\label{eq:Faddeev-Popov}
\end{eqnarray}
to the action.
\subsection{Lattice action}

We regularize the gauge-fixed continuum theory by introducing a lattice with $L$ sites. Our lattice action is
\begin{eqnarray}
S_{b}
&= &
\frac{N}{2a}\sum_{t=1}^L\sum_{I=1}^9{\rm Tr}\left(D_+X_I(t)\right)^2
-
\frac{Na}{4}\sum_{t=1}^L\sum_{I,J=1}^9{\rm Tr}[X_I(t),X_J(t)]^2.
\end{eqnarray}
\begin{eqnarray}
S_{f}
=
iN\sum_{t=1}^L{\rm Tr}\bar{\psi}(t)
\left(
\begin{array}{cc}
0 & D_+\\
D_- & 0
\end{array}
\right)
\psi(t)
-
aN\sum_{t=1}^L\sum_{I=1}^9\bar{\psi}(t)\Gamma^I[X_I(t),\psi(t)],
\end{eqnarray}
\begin{eqnarray}
\Delta S_b
=
aN\sum_{t=1}^L\ {\rm Tr}\left\{
\frac{\mu^2}{2}\sum_{i=1}^3X_i(t)^2
+
\frac{\mu^2}{8}\sum_{a=4}^9X_a(t)^2
+
i\sum_{i,j,k=1}^3\mu\epsilon^{ijk}X_i(t)X_j(t)X_k(t)
\right\}
\nonumber\\
\end{eqnarray}
and
\begin{eqnarray}
\Delta S_f
=
\frac{3i\mu}{4}\cdot aN\sum_{t=1}^L\ {\rm Tr}\left(
\bar{\psi}(t)\gamma^{123}\psi(t)
\right),
\end{eqnarray}
where
\begin{eqnarray}
D_\pm\psi(t)
\equiv
-\mp\frac{1}{2}U^2\psi(t\pm 2a)\left(U^\dagger\right)^2
\pm 2U\psi(t\pm a)U^\dagger
\mp\frac{3}{2}\psi(t)
=
aD_t\psi(t) + O(a^3).
\nonumber\\
\end{eqnarray}
Here, $U={\rm diag}(e^{i\alpha_1/L},e^{i\alpha_2/L}\cdots,e^{i\alpha_N/L})$,
$-\pi\le \alpha_i<\pi$.
The Faddeev-Popov term $S_{F.P.}$ is simply the same as the original form, \eqref{eq:Faddeev-Popov}
 
\section{Polyakov loop}\label{appendix:Polyakov}

Let us briefly review the Polyakov loop and the Polyakov line phases.
In gauge theories, the Polyakov loop $P$ is defined by
\begin{align}
	P
	&=
	\frac{1}{N}{\rm Tr}\left(\mathcal{P}\exp\left(i\int_{0}^{\beta}dt A_t\right)\right)\\
	&=
	\frac{1}{N}\sum_{j=1}^{N} e^{i\theta_j},
	\label{eq:Polyakov_loop_operator}
\end{align}
where $\mathcal{P}$ stands for path ordering.
The $N \times N$ unitary matrix inside the trace is called the Polyakov line.
Because the Polyakov line is a unitary matrix, its eigenvalues can be written as $e^{i\theta_1},\cdots, e^{i\theta_N}$,
where the phases $\theta_1,\cdots, \theta_N$ lie between $-\pi$ and $\pi$.

In the large-$N$ limit, we can introduce the distribution function of the Polyakov line phases $\rho(\theta)$ which is continuous and non-negative on $[-\pi, \pi)$.
We normalized it as
\begin{equation}
	\int_{-\pi}^{\pi} d\theta \rho(\theta) = 1.
\end{equation}
In terms of the distribution function, the Polyakov loop becomes
\begin{equation}
	P = \int_{-\pi}^{\pi} d\theta\; \rho(\theta)\,e^{i\theta},
\end{equation}
which is also useful for investigating the confinement/deconfinement transition.

Due to the $\mathbb{Z}_N$ center symmetry that shifts all the phases simultaneously by a multiple of $\frac{2\pi}{N}$, there is an ambiguity regarding $P$ and $\rho(\theta)$. In this paper, we fixed the center symmetry configuration-by-configuration in such a way that $P$ becomes real and positive.
 
\section{Canonical ensemble and microcanonical ensemble}\label{appendix:canonical-vs-microcanonical}

In the microcanonical ensemble, the energy $E$ is restricted to a small range $[E,E+dE]$, and all states contribute with the same weight.
The entropy $S(E)$ is related to the density of states $\Omega(E)$ as $S(E)=\log\Omega(E)$.
Dual gravity solutions such as 11d Schwarzschild black hole correspond to typical configurations dominating the entropy.
The microcanonical temperature $T_{\rm micro}$ is obtained from the entropy $S$ as
\begin{align}
\frac{1}{T_{\rm micro}}
=
\frac{dS}{dE}.
\end{align}
Therefore, the specific heat capacity $\frac{dE}{dT_{\rm micro}}$ is
\begin{align}
\frac{dE}{dT_{\rm micro}}
=
-\frac{1}{T_{\rm micro}^2}
\left(\frac{d^2S}{dE^2}\right)^{-1}.
\end{align}

The Euclidean path integral of the matrix model describes the canonical thermodynamics.
In the canonical ensemble, temperature $T$ is a controllable parameter.
The partition function is given by
\begin{align}
Z(T)
=
\int dE \Omega(E) e^{-E/T}
=
\int dE e^{-F(E,T)/T},
\label{def:canonical-partition-function}
\end{align}
where $F$ is the free energy defined by
\begin{align}
F(E,T)
=
E-TS(E).
\end{align}
For simplicity, let us assume that the maximum of entropy is uniquely determined at each $E$. (It is not the case if there are two or more separate local maxima and the first-order phase transition in the microcanonical ensemble takes place. We will consider such a case later.)
Then, by taking the derivative with respect to the energy, we obtain
\begin{align}
\frac{\partial F(E,T)}{\partial E}
=
1-T\frac{dS(E)}{dE}
=
1-\frac{T}{T_{\rm micro}(E)}.
\label{maximize-F}
\end{align}
Therefore, free energy is extremized at the value of $E$ which corresponds to $T_{\rm micro}(E)=T$.
The second derivative is
\begin{align}
\frac{\partial^2 F(E,T)}{\partial E^2}
=
\frac{T}{T_{\rm micro}^2}
\left(\frac{dE}{dT_{\rm micro}}\right)^{-1}.
\end{align}
Therefore, if the heat capacity in the microcanonical ensemble is positive (resp., negative), the free energy is minimized (resp., maximized) in the canonical ensemble.

\subsection{The case of first order transition in the microcanonical ensemble}
\begin{figure}
	\centering
	\scalebox{0.2}{
		\includegraphics{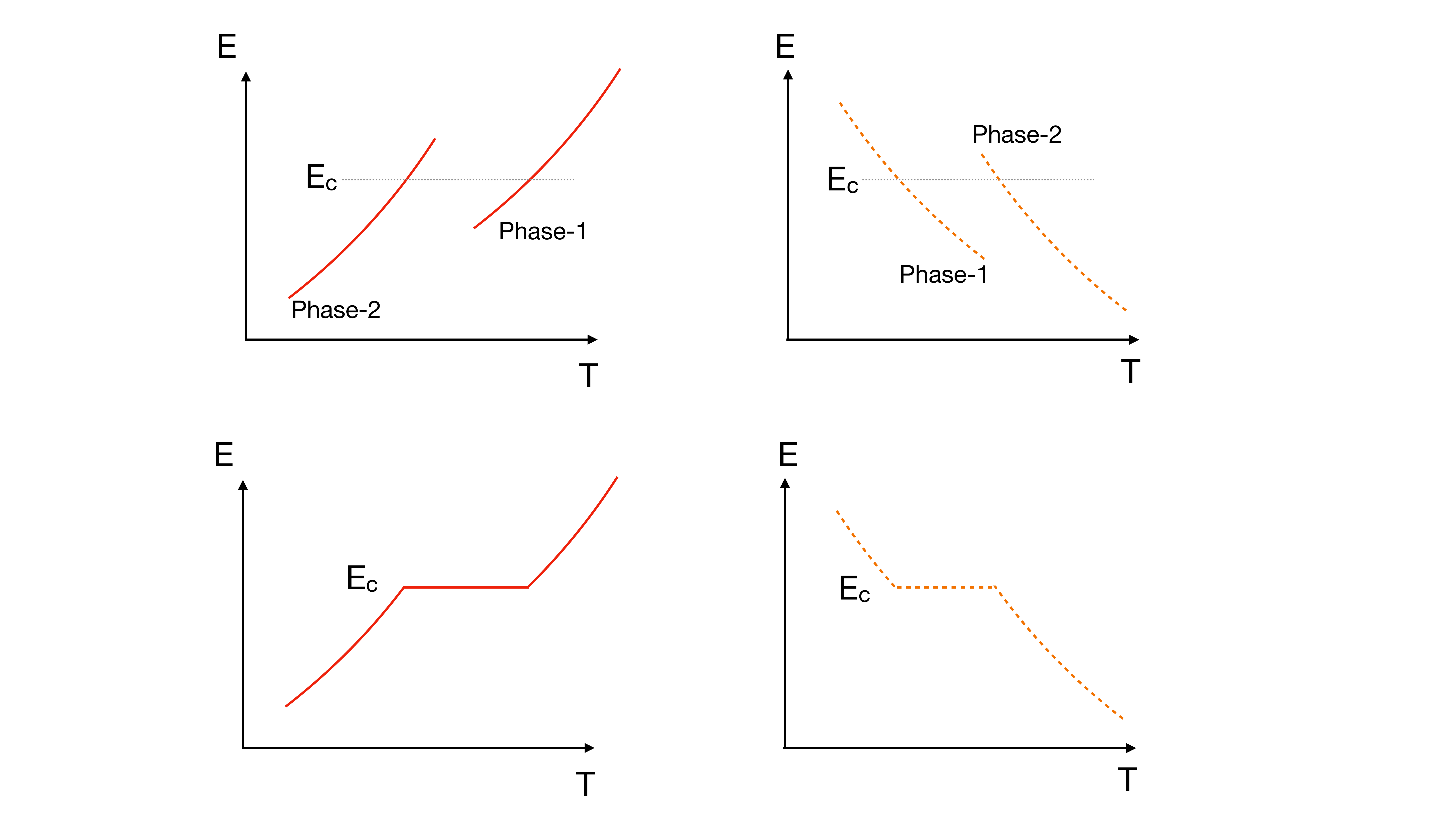}}
	\caption{[Top] Coexistence of two local maxima of entropy $S_1(E)$ and $S_2(E)$ corresponding to phase-1 and phase-2 in the microcanonical ensemble.
	We assume that $S_1(E)>S_2(E)$ at $E>E_c$ and $S_1(E)<S_2(E)$ at $E<E_c$.
	[Bottom] Corresponding canonical phase diagram; free energy minimum (left) and maximum (right).
	}
	\label{fig:two-entropy-maxima}
\end{figure}

Suppose there are two local maxima of the entropy $S_1(E)$ and $S_2(E)$ corresponding to phase-1 and phase-2, as shown in the top row of Fig.~\ref{fig:two-entropy-maxima}.
We assume that $S_1(E)>S_2(E)$ at $E>E_c$ and $S_1(E)<S_2(E)$ at $E<E_c$, and hence a first-order transition takes place at $E=E_c$.
The entropy $S(E)$ in \eqref{def:canonical-partition-function} becomes
$S=\log(e^{S_1}+e^{S_2})$.
This is approximated well by $S_1$ and $S_2$ at $E>E_c$ and $E<E_c$, respectively.
When $E$ is infinitesimally close to $E_c$, we have to take into account both phases.

We have
\begin{align}
\frac{dS(E)}{dE}
=
\frac{1}{e^{S_1}+e^{S_2}}\times
\left(
\frac{e^{S_1}}{T_{\rm micro,1}}
+
\frac{e^{S_2}}{T_{\rm micro,2}}
\right),
\end{align}
where $\frac{1}{T_{{\rm micro},i}}=\frac{dS_{i}}{dE}$ is the microcanonical temperature of phase-$i$ ($i=1,2$).
When $E$ is varied from slightly below $E_c$ to slightly above $E_c$,
it moves from $\frac{1}{T_{\rm micro,2}}$ to $\frac{1}{T_{\rm micro,1}}$.
Therefore, the canonical phase diagram becomes like the bottom row of Fig.~\ref{fig:two-entropy-maxima}.

\begin{figure}
	\centering
	\scalebox{0.2}{
		\includegraphics{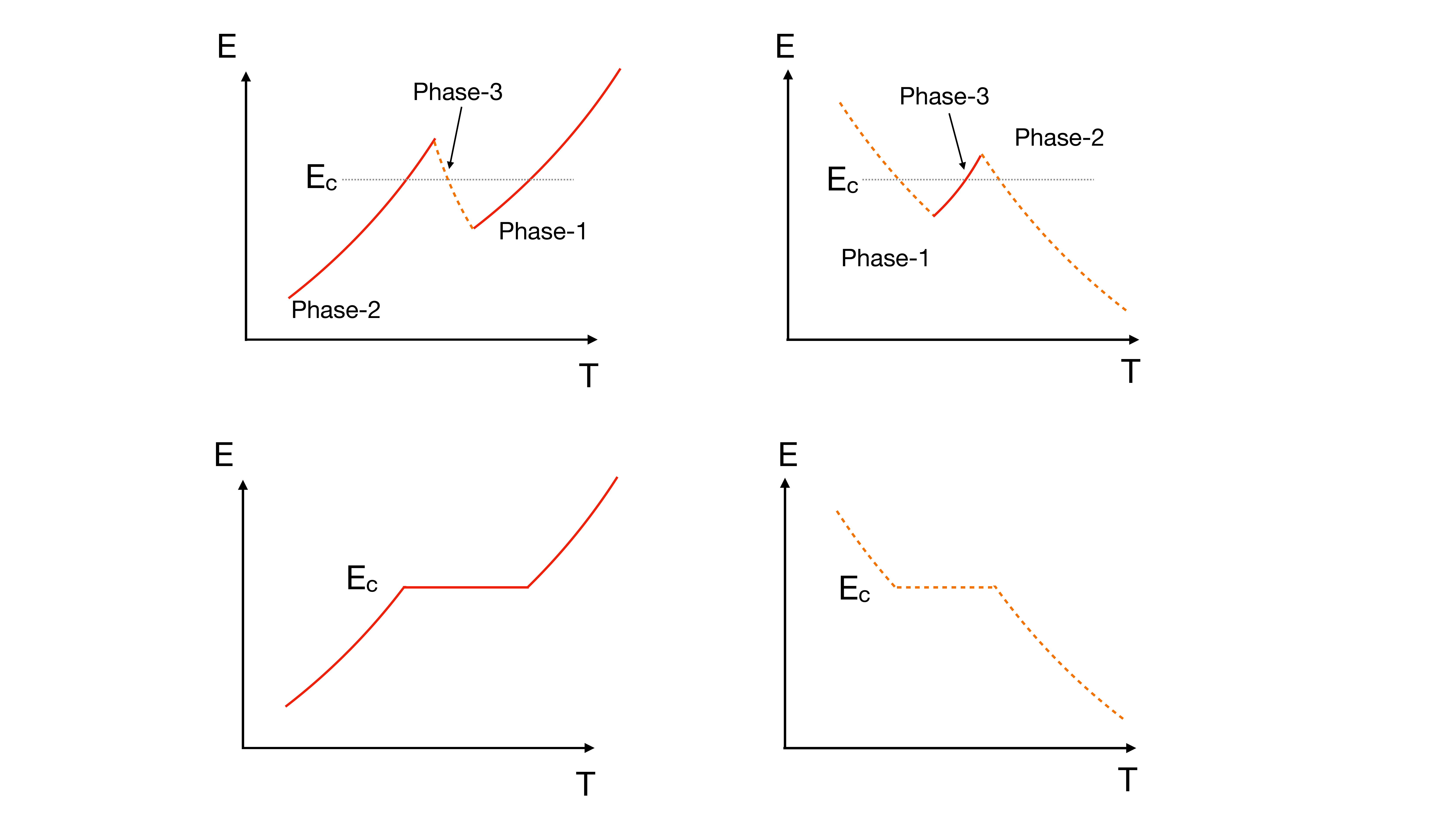}}
	\caption{[Top] Coexistence of three local maxima of entropy $S_1(E), S_2(E)$, and $S_3(E)$ in the microcanonical ensemble.
	We assume that $S_1(E)>S_2(E)$ at $E>E_c$ and $S_1(E)<S_2(E)$ at $E<E_c$, and $S_3(E)$ is always smaller than $S_1(E)$ or $S_2(E)$.
	[Bottom] Corresponding canonical phase diagram; free energy minimum (left) and maximum (right).
	}
	\label{fig:three-entropy-maxima}
\end{figure}

Suppose yet another maximum of the entropy $S_3(E)$ corresponding to phase-3 exists as shown in Fig.~\ref{fig:three-entropy-maxima}, and $S_3(E)$ is always smaller than $S_1(E)$ or $S_2(E)$.
Then phase-3 does not affect the canonical phase diagram at all; see the bottom row of Fig.~\ref{fig:three-entropy-maxima}. Still, if other parameters than energy are taken into account, phase-3 may be visible in the canonical simulation.

\section{Run-away behavior for BFSS with $L=30$, $N=16$}\label{sec:appendix_L30N16}

\begin{figure}
	\centering
	\scalebox{0.25}{
		\includegraphics[trim={0 0cm 0 0},clip]{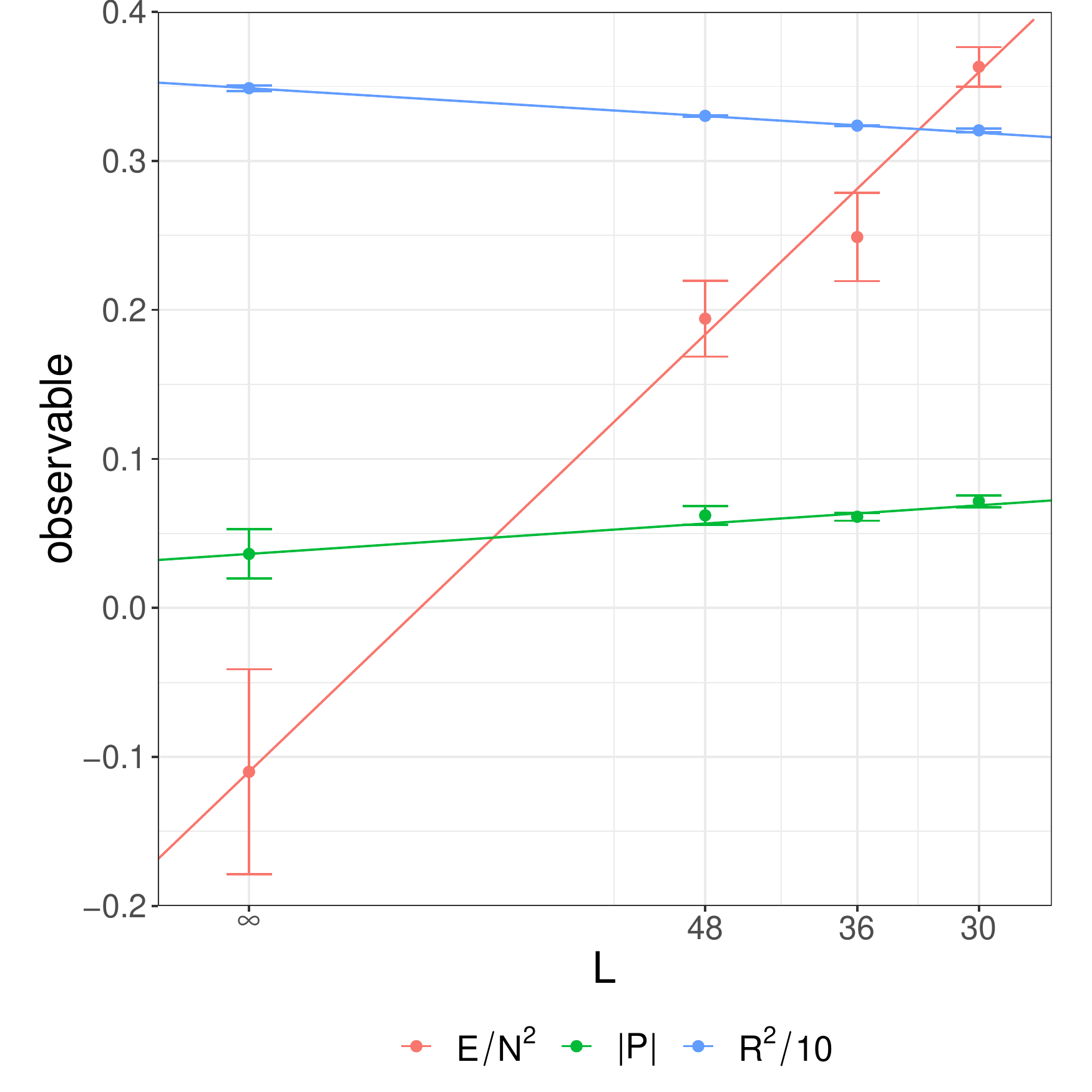}}
	\scalebox{0.25}{
		\includegraphics[trim={0 0cm 0 0},clip]{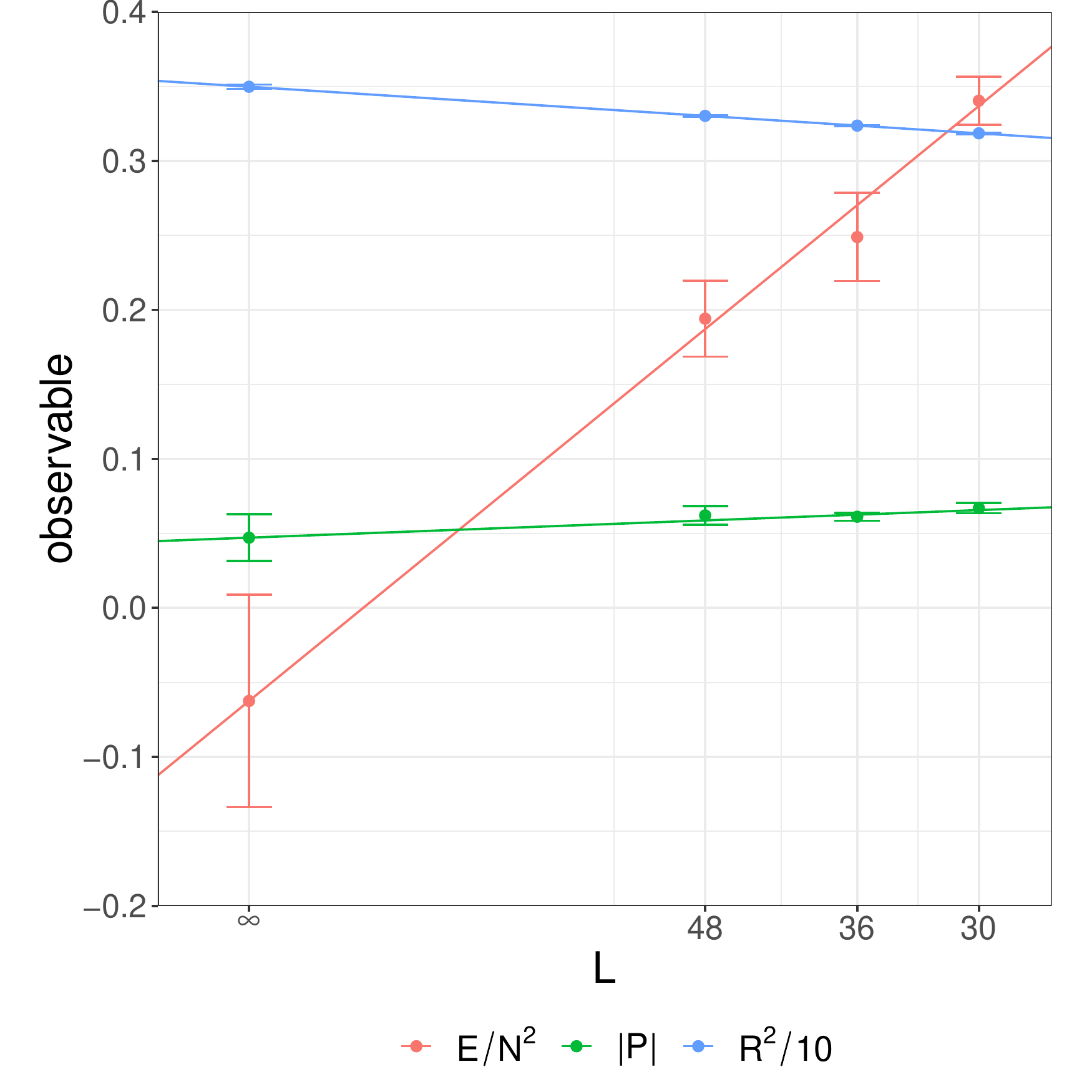}}
	\scalebox{0.25}{
		\includegraphics[trim={0 0cm 0 0},clip]{observable_large_S_GN16T02M0D9_F_321.pdf}}
\caption{BFSS at $T=0.2$, continuum extrapolations in the confined phase for $N=16$. Trajectories for $L=30$, $N=16$ were discarded if $R^2$ was greater than a certain cutoff value. From left to right: no cutoff, $R^2 \leq 3.23$, $R^2 \leq 3.21$. We observe that the continuum value of the energy approaches zero as the cutoff is lowered.}
		\label{fig:BFSSlargeL_cutoff}
\end{figure}

\begin{figure}
	\centering
	\scalebox{0.25}{
		\includegraphics[trim={0 0cm 0 0},clip]{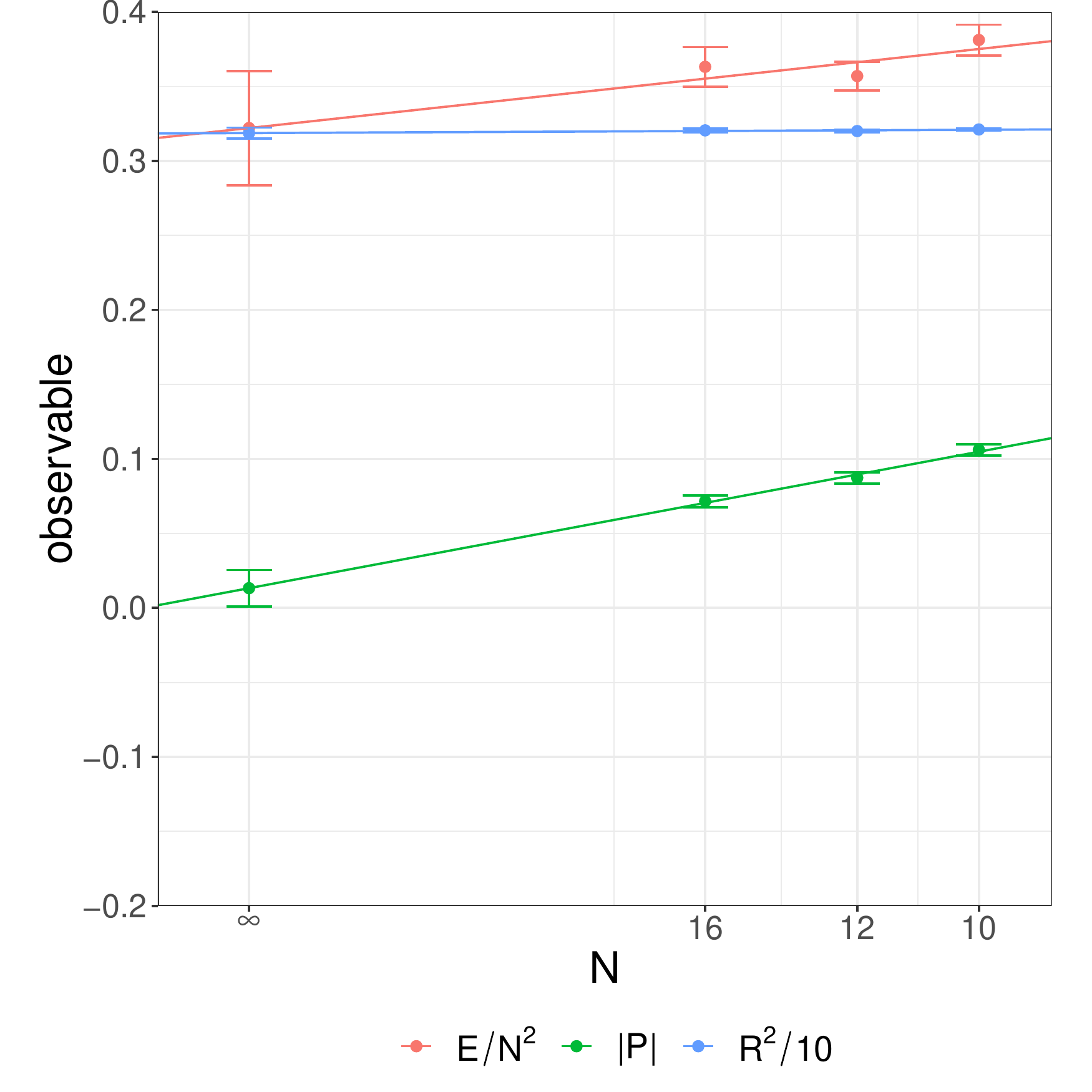}}
	\scalebox{0.25}{
		\includegraphics[trim={0 0cm 0 0},clip]{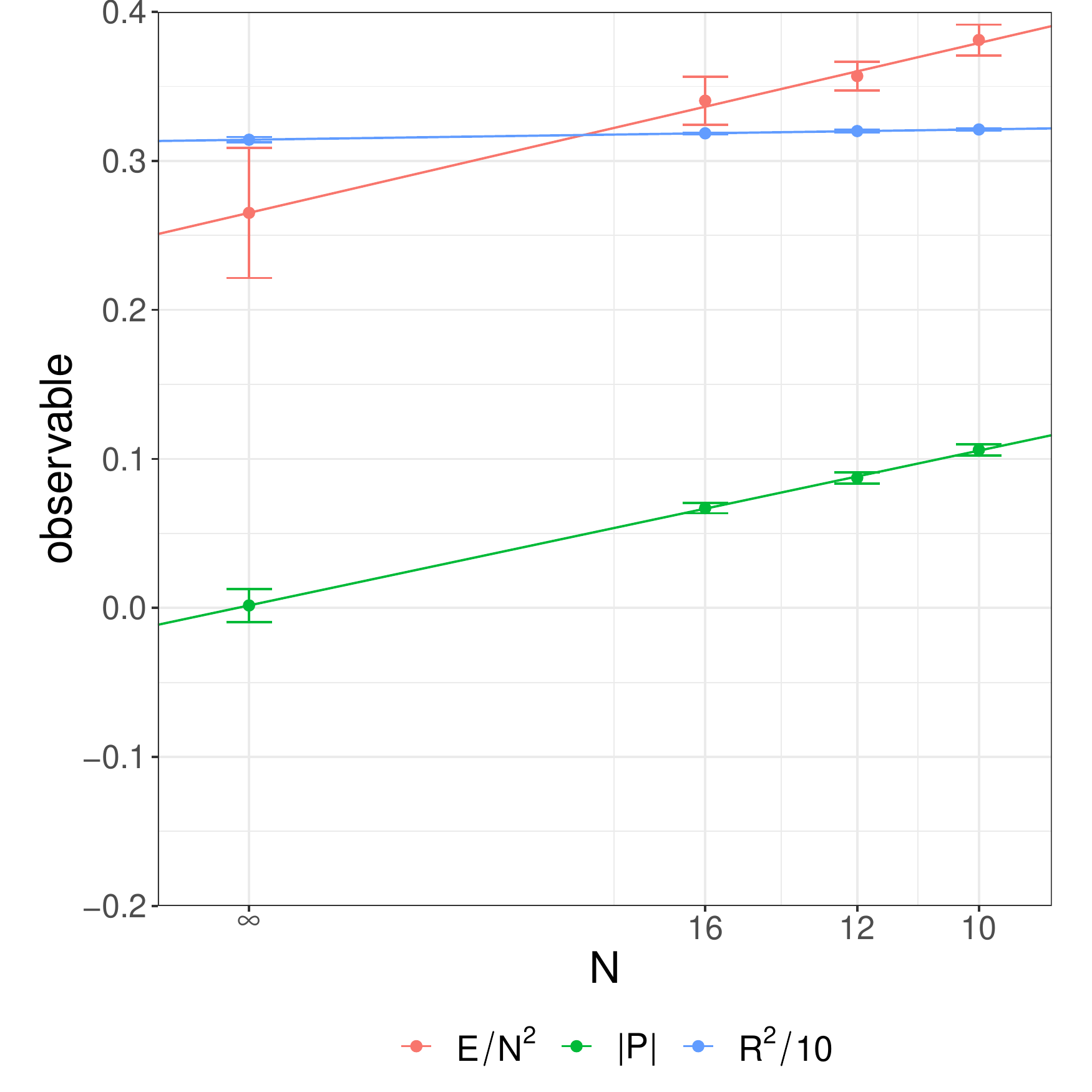}}
	\scalebox{0.25}{
		\includegraphics[trim={0 0cm 0 0},clip]{observable_large_N_GS30T02M0D9_F_321.pdf}}
\caption{BFSS at $T=0.2$, large $N$ extrapolations in the confined phase for $L=30$. Trajectories for $L=30$, $N=16$ were discarded if $R^2$ was greater than a certain cutoff value. From left to right: no cutoff, $R^2 \leq 3.23$, $R^2 \leq 3.21$. We observe that the large $N$ value of $P$ approaches zero as the cutoff is lowered.}
		\label{fig:BFSSlargeN_cutoff}
\end{figure}

As noted in the main text, we had to discard some configurations in the BFSS ($\mu=0$) simulations at $T=0.2$ with parameters $L=30$, $N=16$. The purpose of this appendix is to discuss this issue in more detail. In figure \ref{fig:BFSSlargeL_cutoff}, we show three different versions of the continuum extrapolation at $N=16$, one without discarding any trajectories, and two versions when discarding trajectories with $R^2 > 3.21$ and $R^2 > 3.23$, respectively. We observe that by making the cutoff smaller, the data fits better to the expected continuum  extrapolation. The same is true for the large $N$ extrapolation, see figure \ref{fig:BFSSlargeN_cutoff}.

The reasoning to obtain these cutoff values is as follows. Comparing the simulation results of $L=30$, $N=16$ to $L=36$, $N=16$ and $L=48$, $N=16$, we observe that $L=30$, $N=16$ has much more pronounced pre-run-away behavior, meaning that $R^2$ is increased for short phases of simulation time. By extrapolating the expectation value of $R^2$ from $L=36$, $N=16$ and $L=48$, $N=16$ to $L=30$, $N=16$, we find an expected value of $3.184$ and observe that by putting the cutoff to $3.23$, this expectation value is roughly achieved. Since $L=36$, $N=16$ has somewhat more pre-run-away behavior than  $L=48$, $N=16$, the cutoff value should be somewhat lower than this estimate, and we find that $3.21$ gives reasonable fits. We stress that due to this uncertainty, $3.21$ is only a rough estimate for a proper cutoff. 

This begs the question why a similar cutoff had not to be imposed on the data of the other simulations. 
It first has to be said that it is very hard to quantify the precise amount of run-away behavior in the simulations due to a lack of knowledge about the actual expected value of $R^2$. 
Nevertheless, a tentative answer may be that for $N<16$, the pre-run-away behavior is dominated by finite $N$ effects as opposed to finite $L$ effects. Hence, the extrapolations may still show the expected behavior, as we have roughly the same amount of pre-run-away behavior in all simulations. For $N=16$, on the other hand, the data points have a different amount of pre-run-away behavior, resulting in poor fits. Strong pre-run-away behavior for $L=30$ is also consistent with the observation that we did not achieve any stable simulation with $L=24$ in the BFSS limit due to immediate run-away behavior.

\end{document}